\documentclass[12pt]{article}
\pdfoutput=1

\usepackage{putex}
\usepackage{graphicx}
\usepackage{caption}
\usepackage{amsmath, amsbsy, mathtools}
\usepackage{changepage}   
\usepackage{empheq}       
\usepackage{array}
\usepackage{subcaption}
\usepackage{epstopdf}
\usepackage{enumerate}
\usepackage{cite}
\usepackage{youngtab}
\usepackage{tensor}
\usepackage{slashed}
\usepackage[aligntableaux=center]{ytableau}
\usepackage[utf8]{inputenc}
\usepackage[
      colorlinks=true,
      linkcolor=blue,
      urlcolor=blue,
      filecolor=black,
      citecolor=red
      ]{hyperref}
\usepackage{dsfont}
\usepackage{slashed}
\usepackage{bm}
\usepackage{bbm}
\usepackage{booktabs}
\usepackage{colortbl}
\usepackage{comment}
\usepackage{xcolor}
\usepackage{tikz}

\definecolor{99_1}{rgb}{0.65,0.,0.}
\definecolor{99_2}{rgb}{0.0504678,0.526626,0.627561}
\definecolor{99_3}{rgb}{0.752461,0.362306,0.125339}
\definecolor{99_4}{rgb}{0.435888,0.259065,0.71028}

\newcommand {\be} {\begin {equation}}
\newcommand {\ee} {\end {equation}}

\newcommand {\bes} {\begin {equation*}}
\newcommand {\ees} {\end {equation*}}

\newcommand{\es}[2] {\begin{equation} \label{#1} \begin{split} #2 \end{split} \end{equation}}

\newcommand{\Z}{\mathbb{Z}}

\newcommand{\beq}{\begin{equation}}
\newcommand{\eeq}{\end{equation}}

\def\ie{\begin{equation}\begin{aligned}}
\def\fe{\end{aligned}\end{equation}}

\numberwithin{equation}{section}

\def\<{\langle}
\def\>{\rangle}

\newcommand{\bra}[1]{\langle #1|}
\newcommand{\ket}[1]{|#1\rangle}
\newcommand{\braket}[1]{\langle#1\rangle}

\let\tr\relax
\DeclareMathOperator{\tr}{tr}

\renewcommand{\bar}[1]{\overline{#1}}

\begin{document}

\preprint{PUPT-2664}

\institution{PU}{Joseph Henry Laboratories and Leinweber Forum for Theoretical Physics, Princeton University, Princeton, NJ 08544, USA}
\institution{PCTS}{Princeton Center for Theoretical Science, Princeton University, Princeton, NJ 08544, USA}

\title{The Quasiparticle Ansatz for\\ Relativistic Continuous Matrix Product States}

\authors{Silviu S.~Pufu,\worksat{\PU,\PCTS} Benjamin T.~S\o{}gaard,\worksat{\PU} and Spyros Terlemes\worksat{\PU}}

\abstract{The Relativistic Continuous Matrix Product State (RCMPS) ansatz introduced in~\cite{Tilloy:2021yre} is a translation-invariant variational ansatz intended to approximate the ground states of gapped one-dimensional relativistic Quantum Field Theories directly in the continuum limit and in infinite volume.  In this work, we develop an ansatz, referred to as the Relativistic Quasiparticle (RQP) ansatz, for the low-lying one-particle excitations of the RCMPS vacuum, as well as for solitons interpolating between two distinct degenerate vacua.  The corresponding energies of these states can be obtained as the eigenvalues of an effective Hamiltonian.  The RQP ansatz generalizes constructions used in the Matrix Product State and Continuous Matrix Product State literature.  We apply the RQP ansatz to the sine-Gordon model and find good agreement with results from integrability for both the soliton and breather masses. We also apply the RQP ansatz to the $\phi^4$ theory, in both its $\mathbb Z_2$-preserving and $\mathbb{Z}_2$-broken phases, and determine the spectrum of particles and solitons. This enables us to determine the parameter regime for which multiple stable particles exist.}

\date{October 2026}

\maketitle

\tableofcontents
\section{Introduction}
Matrix Product States (MPSs) \cite{Ostlund:1995zn} and Hamiltonian lattice methods have become increasingly important tools in understanding strongly coupled dynamics in $(1+1)$d quantum field theories (QFTs) (see \cite{Banuls:2018jag,Banuls:2019rao,Banuls:2019bmf} for reviews). MPSs are ansätze for many-body wave functions of lattice systems that exhibit area law entanglement entropy, which make them efficient approximations of gapped ground states. MPS methods have been used to study gauge theories with fermionic matter \cite{Banuls:2013jaa,Buyens:2013yza, Dempsey:2025wia}, scalar field theories \cite{Milsted:2013rxa}, conformal field theories \cite{Stojevic:2014zta,ArguelloCruz:2024xzi}, and, importantly, an array of real-time observables \cite{Buyens:2016hhu,Jha:2024jan}. Furthermore, the MPS ansatz can be extended to model excitations using the quasiparticle ansatz \cite{Haegeman:2011lcd} and estimate bound state masses in the aforementioned models \cite{Dempsey:2025wia,Cuomo:2026kmy}. Since MPSs are formulated on a spatial lattice, the quantum dynamics is governed by a lattice Hamiltonian that approximates the continuum one as the lattice spacing is taken to zero. Fermionic theories are especially well-suited for such numerical lattice studies because the fermion statistics naturally yields a finite local Hilbert space at each lattice site.  Moreover, in $(1+1)$d, one can avoid some of the fermion doubling  that plagues higher dimensional lattice theories. On the other hand, lattice realizations of bosonic theories exhibit infinite-dimensional local Hilbert spaces that must be truncated in numerical implementations. This truncation introduces an additional cutoff, and one needs to make sure that the low-energy observables of interest converge as one increases the truncation parameter \cite{Milsted:2013rxa}.

An elegant way of dispensing with the lattice is the Continuous Matrix Product State (CMPS) introduced in \cite{Verstraete:2010ft}, which formulates an ansatz analogous to the discrete MPS for a non-relativistic QFT directly in the continuum limit. Using the CMPS ansatz, the issue of finiteness of local Hilbert spaces in bosonic theories does not arise since, in continuous space, multiple occupancy is equivalent to single occupancy of infinitesimally neighboring points \cite{Verstraete:2010ft}. CMPSs have been applied to both fermionic and bosonic relativistic QFTs, but they require the introduction of explicit counterterms for the ground state search to converge \cite{Haegeman:2010zz,Stojevic:2014zta}. For the case of scalar field theories with non-derivative interactions, this situation is somewhat unsatisfying, as it is well known that such theories can be rendered finite by normal-ordering \cite{Coleman:1974bu}. This issue was addressed more recently in \cite{Tilloy:2021yre} with the introduction of Relativistic Continuous Matrix Product States (RCMPSs), which adapt CMPSs to relativistic scalar field theories making the ultraviolet (UV) finiteness manifest. The key insight of RCMPSs is to employ a basis of states derived from the energy eigenstates of the free massive scalar field theory. RCMPSs automatically encode properties of relativistic QFT such as the short-distance behavior of two-point functions, making them better starting points for approximating interacting vacua. The RCMPS framework allows for efficient simulation of ground state properties directly in the continuum limit and at infinite volume. The method was initially used to study $\phi^4$ theory \cite{Tilloy:2021yre} and later the sine-Gordon and sinh-Gordon models \cite{Tilloy:2022kcn}, defect QFTs \cite{Tiwana:2025ptw}, the $O(2)$ model \cite{Tiwana:2025zhj}.  This framework was also used to constrain the spectral density through vacuum correlators \cite{Mutzel:2025veg}.

The RCMPS ansatz has the same structure as the CMPS ansatz, so it is natural to attempt to adapt CMPS constructions to the language of RCMPS\@. One such construction is the quasiparticle ansatz \cite{Draxler:2013oek}, which models a single particle- or soliton-like excitation as a local perturbation of a vacuum state as encoded by a CMPS\@. The quasiparticle ansatz was used to determine the dispersion of particles in non-relativistic QFTs but has, to the best of our knowledge, not been applied to relativistic systems. On the other hand, the quasiparticle ansatz for MPSs has proved to be a powerful tool in determining excitation spectra in relativistic gauge theories \cite{Dempsey:2025wia,Dempsey:2026hlf,Cuomo:2026kmy}.

In this work, we adapt the quasiparticle ansatz to RCMPSs, which we will refer to as Relativistic Quasiparticle (RQP) ansatz. The RQP ansatz allows us to study both the spectrum of particle-like excitations as well as solitons of $(1+1)$d scalar QFTs. We first apply the RQP ansatz to the sine-Gordon model, which is integrable \cite{Zamolodchikov:1978xm} and exhibits a rich spectrum of particles and solitons whose masses are known exactly. The numerical results from the RQP ansatz are in good agreement with integrability results. Furthermore, we find that the RQP ansatz accurately reproduces the relativistic dispersion relation required by the Lorentz invariance of the QFT\@. This agreement demonstrates that RCMPSs and the RQP ansatz are good approximations of relativistic states despite neither of them being manifestly relativistic.

We also apply the RQP ansatz to the $\phi^4$ theory. As the coupling is taken from weak to strong, this theory undergoes a phase transition that belongs to the Ising universality class \cite{Simon:1973qz}. We observe that the mass gap closes at a value of the coupling that is in agreement with other estimates in the literature \cite{Vanhecke:2021noi}. The RQP ansatz gives full access to the spectrum of particles and solitons, and it can be used to determine for which parameters stable excitations can exist. In particular, we find that deep in the strong coupling/$\mathbb Z_2$-broken phase, the spectrum contains two stable particles and a soliton (and associated anti-soliton), in agreement with semi-classical predictions and earlier numerical evidence \cite{Rychkov:2015vap, Mussardo:2006iv,Bajnok:2015bgw}.

The rest of the paper is organized as follows.  In Section~\ref{sec:rcmps}, we review the RCMPS ansatz and how to use it to compute expectation values in scalar field theories.  In Section~\ref{subsec:rqp}, we define the RQP ansatz and derive the necessary formulas to determine the spectrum of particles and solitons. The formalism is applied in Section~\ref{sec:application} to the sine-Gordon and $\phi^4$ theories.  Technical details are contained in the Appendices.

\section{Relativistic Continuous Matrix Product States}\label{sec:rcmps}

We begin by reviewing the RCMPS ansatz and establishing notation and formalism that will serve as the basis for the derivation of the RQP ansatz.

\subsection{The RCMPS ansatz}\label{subsec:rcmps}
RCMPSs were first introduced in \cite{Tilloy:2021yre} for studying relativistic $(1+1)$-dimensional field theories with a single scalar field $\phi$ and non-derivative interactions.  Such theories have Lagrangian and Hamiltonian densities of the form\footnote{We denote the coordinates by $x^\mu = (t, x)$, and we use the Minkowski metric $\eta_{\mu\nu} = \text{diag}\, \{1, -1\}$ to raise and lower space-time indices.}
\begin{equation}\label{eq:scalar_action}
    {\cal L} = \frac{1}{2} \dot{\phi}^2 - \frac 12 (\partial_x \phi)^2  - V(\phi) \,, \qquad 
    {\cal H} = \frac{1}{2} \pi^2 + \frac 12 (\partial_x \phi)^2 + V(\phi) \,,
\end{equation}
where $V(\phi)$ is the scalar potential and $\pi = \frac{\partial {\cal L}}{\partial \dot{\phi}}$ is the canonical momentum conjugate to $\phi$.  We may assume that $V(\phi)$ has a global minimum at $\phi = 0$ where $V(0) = 0$, so that
 \begin{equation} \label{eq:VtoVint}
   V(\phi) = \frac{1}{2} m^2 \phi^2 + V_\text{int}(\phi) \,,
 \end{equation}
where $m$ is the mass of the field $\phi$ around the $\phi =0$ vacuum and $V_\text{int}(\phi)$ is the interaction potential, whose Taylor series expansion around $\phi = 0$ is assumed to start at cubic order. The methods we will develop have general applicability to potentials that can be expressed as sums of polynomials and exponentials in $\phi$, with the concrete applications presented in Section~\ref{sec:application} being restricted to the $\phi^4$ and sine-Gordon theories, for which
 \begin{equation} \label{eq:potentials}
  \begin{split}
    \text{$\phi^4$ theory:} &\qquad
          V_{\phi^4} (\phi) =  \frac{1}{2} m^2 \phi^2 + g \phi^4 \,, \\
    \text{sine-Gordon theory:} &\qquad
          V_{\text{sG}} (\phi) =\frac{m^2}{\beta^2} 
           \left[ 1 - \cos (\beta \phi)  \right] \,.  
  \end{split}
\end{equation}
The two examples \eqref{eq:potentials} consist of families of theories parameterized by the dimensionless parameters $g/m^2$ in the $\phi^4$ case and $\beta$ in the sine-Gordon case\footnote{The mass parameter $m$ serves only as a renormalization scale as will be explained in Section \ref{subsec:sG}}.

At the quantum level, the precise definition of the field theory \eqref{eq:scalar_action} requires more care.  To be specific, we will work in the Schr\"odinger picture, where both $\phi$ and $\pi$  are time-independent quantum operators obeying
\begin{equation}\label{eq:commutation1}
    [\phi(x),\pi(y)] = i\delta(x-y)\,.
\end{equation}
Since $V(\phi)$ is a composite operator, it requires regularization.  A standard procedure in $(1+1)$-dimensional QFTs without irrelevant interactions is to use a normal ordering prescription with respect to a free scalar field of some mass $\mu$.\footnote{The parameter $\mu$ used for normal ordering is not necessarily equal to the mass $m$ mentioned above.}  (Different choices of $\mu$ correspond to different regularization prescriptions.)   As usual, the normal order product $N_\mu[\cdots]$ of any number of $\phi$ and $\pi$ operators is defined by expanding these operators in Fourier modes of a free scalar field of mass $\mu$,
\begin{equation}
\begin{aligned}
\label{eq:phi,pi}
    \phi(x) &= \int_{-\infty}^\infty \frac{dk}{2\pi}\frac{1}{\sqrt{2\omega_k}}(e^{ikx}a_k+e^{-ikx}a_k^\dagger)\,,\\
    \pi(x) &= -i\int_{-\infty}^\infty \frac{dk}{2\pi}\sqrt{\frac{\omega_k}{2}}(e^{ikx}a_k-e^{-ikx}a_k^\dagger) \,,
\end{aligned}
\end{equation}
with $\omega_k \equiv \sqrt{\mu^2+k^2}$, and then ordering the $a_k$ and $a_k^\dagger$ operators under the $N_\mu$ sign such that all $a_k$ appear to the right of all $a_k^\dagger$.  Normal ordering then provides a precise definition of our theory whereby the dynamics is given by the Hamiltonian
\begin{equation}\label{eq:scalar_ham}
    H=\int dx \,\mathcal H\,, \qquad {\cal H} \equiv N_\mu\left[\frac{1}{2}\pi^2+\frac{1}{2}(\partial_x\phi)^2 + V(\phi)\right]\,.
\end{equation}
Note the normal ordering $N_\mu[\cdots]$ distinguishes the Fock vacuum $\ket{0}$ as the state annihilated by all of the annihilation operators, $a_k\ket 0=0$.  This state is the ground state for the specific case $V(\phi)=\frac 12 \mu^2 \phi^2$.

Regardless of the value of $\mu$, the commutation relations \eqref{eq:commutation1} imply $[a_k,a_{k'}] = [a_k^\dagger,a_{k'}^\dagger] = 0$ and $[a_k,a_{k'}^\dagger] = 2\pi \delta(k-k')$, which, in turn, imply that the Fourier transforms of $a_k$ and $a_k^\dagger$,
\begin{equation}\label{eq:defa}
    a(x) \equiv \int_{-\infty}^\infty \frac{dk}{2\pi}\, e^{ikx}a_k\,, \qquad a^\dagger(x) \equiv \int_{-\infty}^\infty \frac{dk}{2\pi}\, e^{-ikx}a_k^\dagger\,,
\end{equation}
obey the commutation relations
\begin{equation}\label{eq:commutation_x}
    [a(x),a(y)] = [a^\dagger(x), a^\dagger(y)] = 0\,, \qquad [a(x),a^\dagger(y)] = \delta(x-y)\,.
\end{equation}
The commutation relations \eqref{eq:commutation_x} are identical to those of a non-relativistic (complex) scalar field $\psi(x)$:
\begin{equation}\label{eq:commutation_xNonRel}
    [\psi(x),\psi(y)] = [\psi^\dagger(x), \psi^\dagger(y)] = 0\,, \qquad [\psi(x),\psi^\dagger(y)] = \delta(x-y)\,.
\end{equation}
Of course, $a(x)$ defined in \eqref{eq:defa} should not be interpreted as a non-relativistic scalar field.\footnote{The interpretation of $a(x)$ and $a^\dagger(x)$ is that, in the free theory $V(\phi) = \frac 12 \mu^2 \phi^2$, the operator $a^\dagger(x)$ creates a single quantum of the field $\phi$ localized within a region centered at $x$ whose size is comparable to the Compton wavelength $\sim \mu^{-1}$, and the operator $a(x)$ annihilates such a quantum.}   Nevertheless, the fact that the equations \eqref{eq:commutation_x} and \eqref{eq:commutation_xNonRel} are identical allows one to consider an analog of the CMPS ansatz introduced in \cite{Verstraete:2010ft} for non-relativistic field theories by simply replacing $\psi \to a$ in many of the equations in \cite{Verstraete:2010ft}.  The result is the RCMPS ansatz introduced in \cite{Tilloy:2021yre}, which we now review.

For QFTs defined on the entire real line, the RCMPS is a translation-invariant ansatz that can be used to approximate the ground state.  It is parameterized by two complex $\chi \times \chi$ constant matrices $Q,R$, and two complex $\chi$-dimensional vectors $V_{+\infty}, V_{-\infty}$:  
\begin{equation}\label{eq:RCMPS}
    \ket{Q,R} = V_{+\infty}^\dagger \mathcal P \exp \left[\int_{-\infty}^\infty dx\, \Bigl( Q\otimes \mathds{1}+R\otimes a^\dagger(x) \Bigr) \right]V_{-\infty}\ket0 \,.
\end{equation}
Here, the tensor product is between the $\chi$-dimensional virtual space on which $Q,R$ act and the physical Hilbert space of the QFT,  the path ordered exponential $\mathcal P e^{\int_{-\infty}^\infty dx\,(\cdots)}$ is defined such that the operators with larger $x$ labels appear to the left of those with smaller $x$ labels, and the vectors $V_{+\infty},V_{-\infty}$ serve as boundary conditions for the virtual space of the path-ordered exponential.   The parameter $\chi$ is known as the bond dimension, and it serves as the truncation parameter of the ansatz. 
As we will explain shortly, the matrices $Q$, $R$ and the boundary vectors $V_{+\infty},V_{-\infty}$ must obey certain conditions in order for the state \eqref{eq:RCMPS} to be normalized. Furthermore, as we will explain, the precise choices of $V_{+\infty},V_{-\infty}$ do not affect any physical quantities computed in infinite volume.

Note that one can also construct other variants of RCMPSs.  One such variant that will be very useful in the following discussion consists of the more general family of states
\begin{equation}\label{eq:RCMPSxDep}
    \ket{{\bf Q},{\bf R}} = V_{+\infty}^\dagger \mathcal P \exp \left[\int_{-\infty}^\infty dx\, \Bigl( {\bf Q}(x) \otimes \mathds{1}+ {\bf R}(x)\otimes a^\dagger(x) \Bigr) \right]V_{-\infty}\ket0 \,,
\end{equation}
where ${\bf Q}(x)$ and ${\bf R}(x)$ are non-trivial functions of $x$.  Other generalizations of RCMPSs can be obtained by considering compact spaces and boundaries \cite{Haegeman:2012uk, Tiwana:2025ptw}.

\subsection{Normalization conditions}

To determine the conditions obeyed by $Q$, $R$, $V_{+\infty}$, $V_{-\infty}$ for the RCMPS \eqref{eq:RCMPS} to be normalized, we use the fact that the more general states \eqref{eq:RCMPSxDep} have the overlap (see Appendix~\ref{app:Norm} for a proof) 
\begin{equation}\label{eq:norm_fin}
 \begin{split}
    \braket{\mathbf Q',\mathbf R'|\mathbf Q,\mathbf R} &= (V_{+\infty}^\dagger\otimes V_{+\infty}^{\prime T})
     \mathcal P\exp\left[\int_{-\infty}^\infty dx \, 
     \Bigl( \mathbf Q(x)\otimes \mathds 1+\mathds 1 \otimes \bar{\mathbf Q}'(x)+\mathbf R(x)\otimes \bar{\mathbf R}'(x)  \Bigr) \right] \\
     &{}\times (V_{-\infty}\otimes \bar V'_{-\infty})\,,
 \end{split}
\end{equation}
where the tensor product is between $\mathbb{C}^\chi$ and $\bar{\mathbb{C}}^\chi$.
Taking $\mathbf Q(x)=\mathbf Q'(x)=Q$, $\mathbf R(x)=\mathbf R'(x)=R$, and $V'_{\pm \infty} = V_{\pm \infty}$ gives
\begin{equation}\label{eq:RCMPS_norm}
    \braket{Q,R|Q,R} = \left(v_{+\infty} \middle|  \exp \left( \int_{-\infty}^\infty dx \,\mathbb T \right) \middle| v_{-\infty} \right)  \,,
\end{equation}
where we defined the transfer matrix $\mathbb{T}$ and the boundary vectors $|v_{-\infty})$, $(v_{+\infty}|$:
\begin{equation}\label{eq:transfer_mat}
    \mathbb T \equiv Q\otimes\mathds 1+\mathds 1\otimes \bar Q +R\otimes\bar R\,, \qquad 
    (v_{+ \infty}| \equiv V_{+\infty}^\dagger \otimes  V_{+\infty}^T
    \qquad
     |v_{- \infty}) \equiv V_{- \infty}\otimes \bar V_{- \infty} \,,
\end{equation}
and we removed the path ordering symbol since the matrix appearing in the integrand of the exponent is $x$-independent.

The factor $e^{\int_{-\infty}^\infty dx \,\mathbb T}$ can be interpreted as the limit
 \begin{equation} \label{eq:limit}
     \exp \left( \int_{-\infty}^\infty dx \,\mathbb T \right)
      = \lim_{L \to \infty} e^{L \mathbb{T}} \,,
 \end{equation}
and it can be evaluated using the spectral decomposition of the $\chi^2 \times \chi^2$ matrix $\mathbb{T}$.  Let $\lambda_k$, $k = 1, \ldots, \chi^2$ be the eigenvalues, and $|r_k)$ and $(l_k|$ the right- and left-eigenvectors, obeying the eigenvalue equations
 \begin{equation}
   \mathbb{T} |r_k) = \lambda_k |r_k) \,, \qquad 
   (l_k | \mathbb{T}  = (l_k | \lambda_k  \,.
 \end{equation}
The eigenvectors of $\mathbb{T}$ can be chosen to be orthonormal, i.e.~$(l_k | r_{k'}) = \delta_{k k'}$, but the left- and right-eigenvectors are in general not related to each other by Hermitian conjugation.  As shown in \cite{Haegeman:2012uk} in the case of CMPSs\footnote{Technically, the (R)CMPS has to be ``injective'' for this result to hold (see \cite{Haegeman:2012uk}). However, a generic (R)CMPS is injective, so it will have this property.}, for matrices $\mathbb{T}$ of the form \eqref{eq:transfer_mat}, the eigenvalue with the largest real part (also referred to as the dominant eigenvalue) is unique and purely real.  Let us take this eigenvalue to correspond to $k=1$, and simply denote it by $\lambda$ and its corresponding eigenvectors by $|r)$ and $(l|$.  Using the spectral decomposition $\mathbb T = \sum_{k} \lambda_k|r_k)(l_k|$, it then follows that $e^{L \mathbb{T}} = \sum_k e^{L \lambda_k} |r_k) (l_k|  \approx e^{L \lambda} |r) (l|$ for $L \to \infty$.
As we can see, the factor $e^{L \mathbb{T}}$ that we need to evaluate does not have a finite limit as $L \to \infty$ unless $\lambda = 0$.  We therefore choose the $Q, R$ matrices such that $\lambda = 0$,\footnote{Such a choice can be easily made:  from any given pair of matrices $(Q, R)$ for which the dominant eigenvalue $\lambda$ does not necessarily vanish, one can construct the new pair  $(Q', R') = (Q - \frac{\lambda}{2} \mathds{1}, R)$ for which the dominant eigenvalue vanishes.} in which case $e^{L \mathbb{T}}$ approaches the projector $|r) (l|$ as $L \to \infty$,
 \begin{equation} \label{eq:eLT}
     e^{L \mathbb{T}} \approx |r) (l| \qquad \text{for $L \to \infty$.}
 \end{equation}
Thus, from \eqref{eq:RCMPS_norm}, we infer that the RCMPS is normalized provided that
 \begin{equation} \label{eq:norm_cond}
  \langle Q, R| Q, R \rangle = ({v}_{+\infty} |r) \times (l| v_{-\infty}) = 1 \,.
 \end{equation}
This is a condition that must be obeyed by the boundary vectors $V_{\pm \infty}$.  For future convenience, we will further restrict these boundary vectors such that each factor in \eqref{eq:norm_cond} equals one separately:
\begin{equation}\label{eq:norm_cond_2}
    \begin{aligned}
        \left(v_{+\infty} \middle| r \right)=1\,, \qquad
        \left(l \middle| v_{-\infty} \right) = 1.
    \end{aligned}
\end{equation}

To summarize, we restrict the parameters $Q$, $R$ such that the largest eigenvalue of the transfer matrix \eqref{eq:transfer_mat} is $\lambda = 0$, which implies that the corresponding eigenvectors obey
 \begin{equation} \label{eq:evalEqs}
  \mathbb{T} |r) = (l|\mathbb{T}= 0 \,, \qquad (l|r) = 1 \,,
 \end{equation}
and we furthermore restrict $V_{\pm \infty}$ such that the conditions \eqref{eq:norm_cond_2} are satisfied.  These conditions ensure that the RCMPS is normalized.

\subsection{Matrix notation and RCMPS gauge redundancy}
\label{sec:MatrixNotation}

To further characterize the zero-eigenvectors $(l|$ and $|r)$, it is convenient to interpret these eigenvectors as $\chi \times \chi$ matrices.  Such an identification is possible because the matrix $\mathbb{T}$ acts (on the right) on vectors $|u) \in \mathbb C^\chi\otimes \overline{\mathbb C}^\chi$ and on the left on co-vectors $(w| \in \overline{\mathbb C}^\chi\otimes \mathbb C^\chi$.  From the components of $|u)$ and $(w|$ we can build the components $u_i{}^j$ and $w_i{}^j$ of the matrices $u$ and $w$ using the standard rule for Kronecker products
\begin{equation} \label{eq:vecToMatrix}
    |u)_{(i-1)\chi+j} \  \cong \  u_i{}^j\,, \qquad (w|_{(i-1)\chi+j} \  \cong \   [w^\dagger]_j{}^i\,, \qquad 
    i,j=1,\ldots,\chi \,.
\end{equation}
In this notation, the inner product and the action of an operator of the form $A \otimes B$ become
 \begin{equation}\label{eq:mat_vec}
  \begin{split}
     (w | u )\  &\cong \ 
     \tr (w^\dagger u) \,, \\
    [A \otimes \overline{B}  |u)]_{(i-1)\chi+j}  \  &\cong \ 
     [A u B^\dagger]_i{}^j \,, \\
      [(w| A \otimes \overline{B}]_{(i-1)\chi+j}  \  &\cong \ 
     [B^\dagger w^\dagger A]_j{}^i \,.  
  \end{split}
 \end{equation}
We can then write the eigenvalue equations \eqref{eq:evalEqs} as
\begin{equation}
\begin{aligned} \label{eq:FPMatrices}
    \mathbb{T}|r) &=0 \qquad \Longleftrightarrow  \qquad Qr+rQ^\dagger+RrR^\dagger=0\,,\\
    (l|\mathbb{T} &=0 \qquad \Longleftrightarrow \qquad Q^\dagger l+lQ+R^\dagger l R=0 \,.
\end{aligned}
\end{equation}
The structure of the transfer matrix \eqref{eq:FPMatrices} furthermore guarantees that the fixed points $r$ and $l$ are positive definite matrices \cite{Haegeman:2012uk} and, in particular, Hermitian.

The RCMPS ansatz is overparameterized in the sense that the transformation
\begin{equation}\label{eq:rcmps_gauge}
    Q\rightarrow X QX^{-1}\,,\qquad R \rightarrow XRX^{-1}
\end{equation}
does not change the physical state for any invertible matrix $X$.\footnote{From the definition \eqref{eq:RCMPS}, it is clear that this gauge transformation should be accompanied by a transformation of the boundary vectors $V_{\pm \infty}$ for the state to be invariant.  However, since the boundary vectors do not change bulk observables, we can safely ignore their transformation.} We can use this redundancy to trivialize the left or right fixed point \cite{Haegeman:2012uk} i.e.~set $l=\mathds1$ or $r = \mathds 1$, respectively. In this work, we will do the former and use 
 \begin{equation} \label{eq:leftgauge}
     \text{{\em left-gauge}:} \qquad l = \mathds{1} \,,
 \end{equation}
which implies that the right fixed point satisfies $\tr(r)=1$, as per the normalization condition $(l|r) = \tr (l r) =1$. The choice $l = \mathds{1}$ further constrains the matrices $Q,R$ through \eqref{eq:FPMatrices}
\begin{equation}\label{eq:left_gauge}
    Q^\dagger+Q + R^\dagger R= 0\,,
\end{equation}
which fixes the Hermitian part of $Q$ in terms of $R^\dagger R$.

\subsection{Generating functional for correlators of \texorpdfstring{$a(x)$}{a(x)} and \texorpdfstring{$a^\dagger(x)$}{a†(x)}}
\label{sec:GenFuncRCMPS}

Having defined the states $\ket{Q, R}$, we would now like to compute expectation values of various (normal-ordered) operators in these states, most importantly of the Hamiltonian density ${\cal H}$.  All normal-ordered operators can be written in terms of normal-ordered products of $a(x)$ and $a^\dagger(x)$, so the first step is to compute a generating functional for such products:
\begin{equation} \label{eq:ZDef}
    \mathcal{Z}[\tilde\jmath,j] \equiv \left \langle Q,R \middle| N_\mu \left[ e^{\int_{-\infty}^\infty dx\,  \left( \tilde\jmath(x)a^\dagger(x)+ j(x)a(x) \right)} \right] \middle| Q,R \right \rangle \,,
\end{equation}
where $\tilde \jmath$ and $j$ are classical sources, which, for simplicity, we assume to have compact support.  Since
\begin{equation}
    N_\mu \left[ e^{\int_{-\infty}^\infty dx\,  \left( \tilde\jmath(x)a^\dagger(x)+ j(x)a(x) \right)} \right]  = e^{-\int_{-\infty}^\infty dx\,\tilde\jmath(x)j(x)} e^{\int_{-\infty}^\infty dx\, j(x)a(x)} e^{\int_{-\infty}^\infty dx\, \tilde\jmath(x)a^\dagger(x)} \,,
\end{equation}
we can relate \eqref{eq:ZDef} to the overlap of two generalized RCMPSs that were defined in \eqref{eq:RCMPSxDep}:
\begin{equation}\label{eq:Z_inter}
\begin{aligned}
    \mathcal{Z}[\tilde\jmath,j] 
    &= e^{-\int_{-\infty}^\infty dx\,\tilde\jmath(x)j(x)}\braket{{\bf Q}' ,{\bf R}' | {\bf Q}, {\bf R}} \,,
\end{aligned}
\end{equation}
with ${\bf Q}'(x) = {\bf Q}(x) = Q$, ${\bf R}'(x) = R+\bar j(x) \mathds{1}$, and ${\bf R}(x) = R+ \tilde \jmath(x) \mathds 1$, and with identical boundary vectors at both ends, $V_{\pm \infty}' = V_{\pm\infty}$. Now the expression for the generating functional follows directly from \eqref{eq:norm_fin}:
\begin{equation}\label{eq:gen_a}
    \mathcal{Z}[\tilde\jmath,j] = \left(v_{+\infty} \middle| \mathcal P\exp\left(\int_{-\infty}^\infty dx\, [\mathbb T+j(x)R\otimes\mathds 1+\tilde\jmath(x)\mathds 1\otimes \bar R]\right) \middle| v_{-\infty} \right)\,,
\end{equation}
where we used the definitions \eqref{eq:transfer_mat} and where the $\tilde\jmath j$ arising from product of $\mathbf R$s in \eqref{eq:norm_fin} cancels against the prefactor in \eqref{eq:Z_inter}.

We are now in a position to understand why the precise choice of boundary vectors $V_{\pm \infty}$ does not affect the bulk observables computed from the generating functional \eqref{eq:gen_a}.  Any observable of interest can be derived from $\mathcal{Z}$ using functions $j$ and $\tilde \jmath$ that have compact support, or that vanish sufficiently fast at infinity.  Suppose both $j$ and $\tilde \jmath$ vanish outside some interval $[x_1,x_2]$.  We can then split the path-ordered exponential in \eqref{eq:gen_a} into three factors, corresponding to the intervals $(-\infty, x_1]$, $[x_1, x_2]$, and $[x_2, \infty]$, where for the former and latter we can use \eqref{eq:eLT}:
 \begin{equation}
  \begin{split} \label{eq:PExp}
     &\mathcal P\exp\left(\int_{-\infty}^\infty dx\, [\mathbb T+j(x)R\otimes\mathds 1+\tilde\jmath(x)\mathds 1\otimes \bar R]\right) \\
      &\qquad\qquad = \left|r\middle) \middle(l \middle| \mathcal P\exp\left(\int_{x_1}^{x_2} dx\, [\mathbb T+j(x)R\otimes\mathds 1+\tilde\jmath(x)\mathds 1\otimes \bar R]\right)  \middle| r \middle) \middle(l\right| \,.
      \end{split}
 \end{equation}
In this expression, one can use \eqref{eq:eLT} to further extend the integration limits on the RHS to the entire real line.  Using the normalization condition \eqref{eq:norm_cond}, we arrive at
\begin{align}\label{eq:decouple}
 \mathcal{Z}[\tilde\jmath,j] 
   = \left(l \middle| \mathcal P\exp\left(\int_{-\infty}^{\infty} dx\, [\mathbb T+j(x)R\otimes\mathds 1+\tilde\jmath(x)\mathds 1\otimes \bar R]\right) \middle| r \right)\,,
\end{align}
which is manifestly independent of the choice of boundary vectors.   It follows that the precise choices of these vectors do not affect any bulk observables.  Note that the expression \eqref{eq:decouple} was derived for sources $j, \tilde \jmath$ with compact support, but the precise interval $[x_1, x_2]$ on which these functions are supported does not enter in the final answer.  It can be shown that \eqref{eq:decouple} continues to hold even if $j, \tilde \jmath$ do not have compact support but instead decay at $\pm \infty$ sufficiently fast.

The expression \eqref{eq:decouple} is the final answer for the generating functional, but it is still onerous to evaluate it as written.  An efficient strategy is to recast ${\cal Z}[\tilde \jmath, j]$ as the solution to a differential equation.  In particular, define the vector
 \begin{equation}\label{eq:rjjtDef}
   \bigl|r_{\tilde \jmath, j}(x) \bigr)  \equiv \left. \mathcal P\exp\left(\int_{-\infty}^{x} dy\, [\mathbb T+j(y)R\otimes\mathds 1+\tilde\jmath(y)\mathds 1\otimes \bar R]\right) \middle| r \right) \,,
 \end{equation}
in terms of which the generating functional is simply
 \begin{equation} \label{eq:Zright}
   \mathcal{Z}[\tilde\jmath,j] 
      = \bigl( l \big| r_{\tilde \jmath, j}(\infty) \bigr)  \,.
 \end{equation}
The advantage of defining \eqref{eq:rjjtDef} is that $ \bigl|r_{\tilde \jmath, j}(x) \bigr)$ obeys the differential equation
 \begin{equation} \label{eq:ODEGenrx}
   \frac{d \bigl|r_{\tilde \jmath, j}(x) \bigr)}{dx}
      = [\mathbb T+j(x)R\otimes\mathds 1+\tilde\jmath(x)\mathds 1\otimes \bar R] \bigl|r_{\tilde \jmath, j}(x) \bigr) \,, \qquad 
    \text{with } \bigl|r_{\tilde \jmath, j}(-\infty) \bigr) = |r) \,.
 \end{equation}
For given source functions $j, \tilde \jmath$, this is just a system of $\chi^2$ ODEs that can be solved numerically.  

Note that if $j, \tilde \jmath$ are compactly supported on the interval $[x_1, x_2]$, then outside of this interval, the differential equation \eqref{eq:ODEGenrx} takes the form
 \begin{equation} \label{eq:approxEq}
    \frac{d \bigl|r_{\tilde \jmath, j}(x) \bigr)}{dx}
      = \mathbb T \bigl|r_{\tilde \jmath, j}(x) \bigr) \,, \qquad \text{for $x<x_1$ or $x>x_2$} \,.
 \end{equation}
In terms of the right-eigenvectors $|r_k)$ of $\mathbb{T}$ (with $|r_1) = |r)$), the general solution of the equation \eqref{eq:approxEq} is
\begin{equation} \label{eq:approxSol}
    \bigl|r_{\tilde \jmath, j}(x) \bigr)
      = c_1 |r) + \sum_{k>1} c_k e^{\lambda_k x} |r_k) \,, \qquad \text{for $x<x_1$ or $x>x_2$} \,,
 \end{equation}
for some constants $c_k$. (We have a different set of $c_k$ for $x<x_1$ and $x>x_2$.)  Since $\Re \lambda_k<0$ for all $k>1$, the exponentials in \eqref{eq:approxSol} all grow as $x \to -\infty$, and thus the boundary condition $\bigl|r_{\tilde \jmath, j}(-\infty) \bigr) = |r)$ means that $\bigl|r_{\tilde \jmath, j}(x) \bigr) = |r)$ for all $x<x_1$.  It follows that we can start solving the differential equation \eqref{eq:ODEGenrx} at $x = x_1$.  For $x>x_2$, the exponentials in \eqref{eq:approxSol} all decay as $x \to \infty$, and thus the solution approaches $c_1|r)$ exponentially fast.  Thus, a good approximation can be obtained by solving the differential equation \eqref{eq:ODEGenrx} on a finite interval $[-L, L]$ instead of the entire real line.

Alternatively, one can also integrate from $+\infty$ to $-\infty$ instead.  Indeed, one can define the co-vector
 \begin{equation}\label{eq:ljjtDef}
   \bigl(l_{\tilde \jmath, j}(x) \bigr|  \equiv \left( l \middle|  \mathcal P\exp\left(\int_{x}^{\infty} dy\, [\mathbb T+j(y)R\otimes\mathds 1+\tilde\jmath(y)\mathds 1\otimes \bar R]\right) \right. \,,
 \end{equation}
in terms of which the generating functional is 
 \begin{equation} \label{eq:Zleft}
   \mathcal{Z}[\tilde\jmath,j] 
      = \bigl( l_{\tilde \jmath, j}(-\infty) \big| r \bigr)  \,.
 \end{equation}
The vector \eqref{eq:ljjtDef} obeys the differential equation
 \begin{equation} \label{eq:ODEGenlx}
   \frac{d \bigl(l_{\tilde \jmath, j}(x) \bigr|}{dx}
      = -\bigl(l_{\tilde \jmath, j}(x) \bigr| [\mathbb T+j(x)R\otimes\mathds 1+\tilde\jmath(x)\mathds 1\otimes \bar R]  \,, \qquad 
    \text{with } \bigl(l_{\tilde \jmath, j}(\infty) \bigr| = (l| \,.
 \end{equation}
This is again a system of $\chi^2$ ODEs that can be solved numerically.

\subsection{Expectation values in RCMPS}
\label{sec:expvalue}

Let us now move on to the expectation values of operators of interest, which are the Hamiltonian density \eqref{eq:scalar_ham} and potentially powers of the field operator $\phi$.  We will primarily compute the expectation values of three types of operators:  the Hamiltonian density ${\cal H}_\text{free}(x)$ of a free scalar field of the same mass $\mu$ as the mass parameter used for normal ordering, a vertex operator $V_b$ with parameter $b$, and powers of $\phi$:
 \begin{equation} \label{eq:opList}
  \begin{split} 
  {\cal H}_\text{free}(x) &\equiv N_\mu \left[ \frac{1}{2} \pi^2 + \frac 12 (\partial_x \phi)^2 + \frac 12 \mu^2 \phi^2 \right] \,, \\
  V_b(x) &\equiv N_\mu \left[ e^{b \phi(x)} \right] \,, \\
  {\cal O}_n(x) &\equiv N_\mu \left[ \phi(x)^n \right] \,.
  \end{split}
 \end{equation}
These operators will serve as building blocks for all local operators that we will be interested in.  For instance, for the two examples mentioned in \eqref{eq:potentials}, the Hamiltonian density is
 \begin{equation}
  \begin{split} \label{eq:HamExamples}
    \text{$\phi^4$ theory:} &\qquad
     {\cal H}_{\phi^4} =  {\cal H}_\text{free} 
        + \frac 12 (m^2 - \mu^2) {\cal O}_2 + g {\cal O}_4 \,, \\
     \text{sine-Gordon theory:} &\qquad
     {\cal H}_{\text{sG}} =  {\cal H}_\text{free} 
      - \frac{1}{2} \mu^2 {\cal O}_2 - \frac{m^2}{2\beta^2} (V_{i \beta} + V_{-i \beta}) + \frac{m^2}{\beta^2} \,.    
  \end{split}   
 \end{equation}

To compute the expectation values of the operators \eqref{eq:opList}, we should express $\phi(x)$ and $\pi(x)$ in terms of $a(x), a^\dagger(x)$ by combining the definitions \eqref{eq:phi,pi} and \eqref{eq:defa}.  For $\phi(x)$, this relation is
\begin{equation}\label{eq:phi_inv}
 \begin{split}
    \phi(x) &= \int_{-\infty}^\infty dy \,J(x-y)\left[a(y)+a^\dagger(y) \right]\,,
 \end{split}
\end{equation}
where we defined the kernel
\begin{equation}\label{eq:J_kernel}
    J(x) \equiv \int_{-\infty}^\infty \frac{dk}{2\pi}\frac{e^{ikx}}{\sqrt{2\omega_k}}=\frac{\mu^{1/4}}{2^{1/4}\sqrt{\pi}\Gamma(\frac{1}{4})}\frac{K_{1/4}(\mu|x|)}{|x|^{1/4}}\,.
\end{equation}
(We will not need the analogous relation for $\pi(x)$.)  Note that the function $J(x)$ has an integrable singularity at $x=0$ and decays as $\sim \frac{e^{-\mu|x|}}{(\mu|x|)^{3/4}}$ for $|x|\gg \mu^{-1}$. The non-locality of the relation \eqref{eq:phi_inv} makes computations using RCMPSs more involved than those using their non-relativistic counterparts, where the field operators are fully local.  Note that due to translation invariance of the RCMPS, the expectation values of the operators in \eqref{eq:opList} are $x$-independent, so we can set $x=0$.

One notable feature of RCMPSs is that they encode the expected UV behavior of two-point functions. To see this, we note that expectation values in the RCMPS ansatz \eqref{eq:RCMPS} of normal-ordered operators are finite as they can be evaluated using the generating function~\eqref{eq:decouple}. In particular, this implies that the two-point function of $\phi(x)$ satisfies
\begin{equation}\label{eq:RCMPStwopoint}
\begin{aligned}
    \braket{Q,R|\phi(x)\phi(y)|Q,R} &= G_\mu(x-y)+ \braket{Q,R|N_\mu[\phi(x)\phi(y)]|Q,R}\\
    &= -\frac{1}{4\pi}\log \left[ \mu^2(x-y)^2 \right] +\text{regular terms}\,,
\end{aligned}
\end{equation}
where the singular behavior derives from the free equal-time two-point function $G_\mu(x-y)=\braket{0|\phi(x)\phi(y)|0}$. This singular behavior is exactly what is expected for the interacting theory~\eqref{eq:scalar_ham}, because all interactions are relevant and, accordingly, the theory is free in the~UV\@.

\subsubsection{Expectation value of the vertex operator \texorpdfstring{$V_b$}{Vb}}
\label{sec:VbExp}

Of the operators in \eqref{eq:opList}, the simplest expectation value is that of $V_b$ because using \eqref{eq:phi_inv} we see that
 \begin{equation} \label{eq:expVb}
     \left\langle Q, R \middle| V_b(0) \middle|Q, R \right\rangle = {\cal Z}[b J, b J] \,,
 \end{equation}
with ${\cal Z}$ being the generating functional defined in \eqref{eq:ZDef} that we computed in the previous section.  The expectation value \eqref{eq:expVb} can be computed either from \eqref{eq:Zright} or from \eqref{eq:Zleft} after solving numerically the differential equations \eqref{eq:ODEGenrx} or \eqref{eq:ODEGenlx}, respectively, in the particular case $j = \tilde \jmath = b J$.  Due to the fact that the kernel $J(x)$ decays exponentially at large distances, one can obtain a good approximation by solving the differential equation on the interval $[-L, L]$ for a sufficiently large $L$---see the discussion after \eqref{eq:ODEGenrx}.

\subsubsection{Expectation value of \texorpdfstring{${\cal O}_n$}{On}}
\label{sec:OnExp}

We now turn to the expectation values of the normal-ordered monomials ${\cal O}_n = N_\mu[\phi^n(x)]$. 
As explained in \cite{Tilloy:2021yre}, the most efficient way of computing $\langle {\cal O}_n(x) \rangle$ is to use $\braket{V_b}$ as a generating function for monomials:
\begin{equation}
    \braket{Q,R|{\cal O}_n(x)|Q,R} = \left.\frac{\partial^n}{\partial b^n}\braket{Q, R | V_b(x) | Q, R}\right |_{b=0} \,.
\end{equation}
As for $V_b(0)$ itself, we evaluate the expectation value of ${\cal O}_n(0)$ using ODEs. These ODEs can be obtained by taking $n$ derivatives with respect to $b$ of the equations \eqref{eq:rjjtDef}--\eqref{eq:ODEGenlx}, with $j = \tilde \jmath = bJ$, and then setting $b =0$.

Thus, we define the $n$th derivative of $|r_{bJ, bJ}(x))$
\begin{equation}\label{eq:phin_VEV}
     |r^{(n)}(x))\equiv\left.\frac{\partial^n|r_{bJ, bJ} (x))}{\partial b^n} \right|_{b=0} \,,
\end{equation}
which, from \eqref{eq:Zright} and \eqref{eq:expVb}, allows us to write
 \begin{equation}
    \braket{Q,R|{\cal O}_n(0)|Q,R} 
     = \bigl(l\big|r^{(n)}(\infty) \bigr) \,.
 \end{equation}
Note that $\big|r^{(0)}(x) \bigr) = |r)$ is $x$-independent. In practice, the vectors $\bigl|r^{(n)}(x) \bigr)$, with $n>0$, can be computed efficiently  by solving a separate system of ODEs derived from \eqref{eq:ODEGenrx} by repeated differentiation:
\begin{equation}\label{eq:rnODE}
        \frac{d\bigl|r^{(n)}(x) \bigr)}{dx} = \mathbb{T} \bigl|r^{(n)}(x))+nJ(x) (R \otimes \mathds 1 + \mathds 1 \otimes \bar R) |r^{(n-1)}(x) \bigr) \,,
\end{equation}
with the boundary condition
 \begin{equation} \label{eq:rnBC}
      \bigl|r^{(n)}(-\infty) \bigr) = 
             0 \qquad \text{for $n>0$} \,.
 \end{equation}
To compute $\bigl|r^{(n)}(x) \bigr)$, one has to solve the equations for $\bigl|r^{(n')}(x) \bigr)$ with all $1 \leq n' \leq n$.  

One can also derive an analogous system of ODEs using $\bigl(l_{b J, bJ}(x) \bigr|$ by taking derivatives of~\eqref{eq:Zleft} and~\eqref{eq:ODEGenlx}.  In particular, we have
 \begin{equation}
     \braket{Q,R|{\cal O}_n(0)|Q,R} 
     = \bigl(l^{(n)}(-\infty)\big|r \bigr) \,,
 \end{equation}
where 
 \begin{equation} \label{eq:lnDef}
    (l^{(n)}(x)|\equiv\left.\frac{\partial^n(l_{bJ, bJ} (x)|}{\partial b^n} \right|_{b=0} \,,  
 \end{equation}
with $(l^{(0)}| = (l|$, and $(l^{(n)}(x)|$ with $n>0$ obeys
 \begin{equation} \label{eq:lnODE}
   \frac{d \bigl(l^{(n)}(x) \bigr|}{dx}
      = -\bigl(l^{(n)}(x)\bigr| \mathbb{T} -n J(x) \bigl(l^{(n-1)}(x) \bigr| \left( R\otimes\mathds 1+\mathds 1\otimes \bar R\right)  \,, \qquad 
    \text{with } \bigl(l^{(n)}(\infty) \bigr| = 0 \,.
 \end{equation}

This method of evaluating VEVs of ${\cal O}_n$ by solving a system of ODEs rather than by performing a large number of iterated integrals is one of the main technical insights of \cite{Tilloy:2021yre}. We will adapt this method to the RQP ansatz in the following section. 

\subsubsection{Expectation value of \texorpdfstring{${\cal H}_{\text{free}}$}{Hfree}}
\label{sec:expvalueHfree}

Lastly, let us now discuss the computation of the normal ordered free Hamiltonian density ${\cal H}_{\text{free}}$ defined in \eqref{eq:opList}.  Writing the operators $\phi$ and $\pi$ in terms of the creation and annihilation operators $a_k$ and $a_k^\dagger$, it is straightforward to show that in any translation-invariant state we have 
  \begin{equation}  \label{eq:HfreeExp}
   \langle \mathcal{H}_{\text{free}}(0) \rangle 
    = \int \frac{dk\,dk'}{(2 \pi)^2}\, \frac{(\mu + i k) (\mu - i k')}{\sqrt{\omega_k \omega_{k'}}}
        \langle a_{k'}^\dagger a_k \rangle \,.
 \end{equation}
(See Appendix~\ref{app:free_term} for a detailed proof.)  Using the definitions \eqref{eq:defa}, this expression can be written in terms of $a(x)$ and $a^\dagger(x)$:
 \begin{equation} \label{eq:HFreeToa}
     \langle {\cal H}_\text{free}(0) \rangle = 2 \int dy\, dz\, H(y) H(z) \langle a^\dagger(y) a(z) \rangle \,, \qquad  H(x) \equiv  (\mu - \partial_x) J(x) \,.
 \end{equation}
In an RCMPS $\ket{Q, R}$, the expression \eqref{eq:HFreeToa} implies that the expectation value of ${\cal H}_\text{free}(0)$ can be written in terms of the generating functional \eqref{eq:ZDef} as
 \begin{equation} \label{eq:HfreeGen}
   \langle Q, R| {\cal H}_\text{free}(0) | Q, R\rangle = 2 \frac{\partial^2}{\partial \alpha \partial \beta} 
   {\cal Z}[\alpha H, \beta H]
   \bigg|_{\alpha = \beta = 0} \,. 
 \end{equation}
Note that this approach for computing $\braket{{\cal H}_\text{free}(0)}$ differs from that presented in \cite{Tilloy:2021yre} and requires one to solve only half the number of ODEs.
The generating functional ${\cal Z}[\alpha H, \beta H]$ can be computed from \eqref{eq:Zright} by solving \eqref{eq:ODEGenrx} with $j = \beta H$ and $\tilde \jmath = \alpha H$.
However, this differential equation is complicated to solve because $H(x) \sim 1/|x|^{3/2}$ at small $x$, which is not a classically integrable singularity and would have to be treated as a distribution.  Since the $1/|x|^{3/2}$ behavior of $H(x)$ comes solely from terms involving derivatives of $J(x)$, it is possible to get rid of these derivatives (and hence of the non-integrable divergence) by making the definition
 \begin{equation}\label{eq:rtilde}
     \bigl|r_{\alpha H, \beta H}(x) \bigr)
      = e^{-F(x)} \bigl|\tilde r_{\alpha H, \beta H}(x) \bigr) \,, \qquad F(x) \equiv \beta J(x) R \otimes \mathds 1 
       + \alpha J(x) \mathds 1 \otimes \bar R \,,
 \end{equation}
where $e^{-F}$ is an integrating factor.  Using \eqref{eq:ODEGenrx}, one can then show that the quantity $\bigl|\tilde r_{\alpha H, \beta H}(x) \bigr)$ obeys the differential equation
 \begin{equation} \label{eq:ODEGenrxParticular2}
   \frac{d \bigl|\tilde r_{\alpha H, \beta H}(x) \bigr)}{dx}
      =  e^{F(x)} \bigl(\mathbb T+\beta \mu J(x) R\otimes\mathds 1+\alpha \mu J(x) \mathds 1\otimes \bar R
        \bigr) e^{-F(x)} \bigl|\tilde r_{\alpha H, \beta H}(x) \bigr) \,,
 \end{equation}
with the boundary condition $\bigl|\tilde r_{\alpha H, \beta H}(-\infty) \bigr) = |r)$ that follows from \eqref{eq:ODEGenrx} and the fact that $F \to 0$ as $|x| \to \infty$.  That $F$ vanishes at infinity also implies
\begin{equation} \label{eq:ZH}
   {\cal Z}[\alpha H, \beta H] = (l | \tilde r_{\alpha H, \beta H}(\infty) ) \,.
 \end{equation}
Note that the differential equation \eqref{eq:ODEGenrxParticular2} involves only $J(x)$, which has an integrable singularity at $x = 0$, and it is therefore simpler to solve than \eqref{eq:ODEGenrx}.

We can now go back to \eqref{eq:HfreeGen}, and, using \eqref{eq:ZH}, write the expectation value of the free Hamiltonian as 
 \begin{equation} \label{eq:HFreeFromrtilde}
     \langle Q, R| {\cal H}_\text{free}(0) | Q, R\rangle = 2 \bigl(l \big| \tilde r^{(1, 1)}(\infty) \bigr) \,,
 \end{equation}
where we defined 
\begin{equation}
     \big|\tilde r^{(n, m)}(x)\big)\equiv\left.\frac{\partial^{n+m}|\tilde r_{\alpha H, \beta H} (x))}{\partial \alpha^n \partial \beta^m} \right|_{\alpha = \beta =0} \,.
\end{equation}
Note that $\big|\tilde r^{(0, 0)}(x)\big) = |r)$.   By taking derivatives of \eqref{eq:ODEGenrxParticular2}, we obtain a system of equations for $\big|\tilde  r^{(n, m)}(x) \bigr)$ with $n, m \leq 1$:
 \begin{equation}  \label{eq:rmnSecondEqs}
  \begin{split}
   \frac{d \bigl|\tilde r^{(1,1)}(x) \bigr)}{dx}
      &= \mathbb T \bigl|\tilde r^{(1, 1)}(x) \bigr)
      + J(x)\left(\mu R + [R, Q] \right) 
      \otimes\mathds 1 \bigl|\tilde r^{(1, 0)}(x) \bigr) \\
      &{}+ 
      J(x) \mathds 1\otimes \left( \mu \bar R+ [\bar R, \bar Q] \right) \bigl|\tilde r^{(0, 1)} (x) \bigr)
    \,, \\
     \frac{d \bigl|\tilde r^{(1,0)}(x) \bigr)}{dx}
      &= 
      \mathbb T \bigl|\tilde r^{(1, 0)}(x) \bigr)
       + J(x)  \mathds 1\otimes \left(\mu \bar R + [\bar R, \bar Q] \right)\bigl|r \bigr)
    \,, \\
   \frac{d \bigl|\tilde r^{(0,1)}(x) \bigr)}{dx}
      &= \mathbb T \bigl|\tilde r^{(0, 1)}(x) \bigr)
       +J(x) \left(\mu R + [R, Q] \right)  \otimes \mathds{1} \bigl|r \bigr) 
    \,,
  \end{split}
 \end{equation}
with the boundary conditions $\bigl|\tilde r^{(m,n)}(-\infty) \bigr) = 0$ for $(m, n) \neq (0, 0)$.  The system of equations \eqref{eq:rmnSecondEqs} can be solved simultaneously, and from the solution we can then determine $\langle Q, R| {\cal H}_\text{free}(0) | Q, R\rangle$ via \eqref{eq:HFreeFromrtilde}. 

Finally, following \eqref{eq:Zleft}, we can repeat this procedure while integrating in the other direction by considering $\big(\tilde l_{\alpha H, \beta H}\big|$ defined through
 \begin{equation}\label{eq:ltilde}
     \bigl(l_{\alpha H, \beta H}(x) \bigr|
      =  \bigl(\tilde l_{\alpha H, \beta H}(x) \bigr| e^{F(x)}\,.
 \end{equation}
One can then utilize \eqref{eq:ODEGenlx} to find
 \begin{equation} \label{eq:ODEGenlxParticular2}
   \frac{d \bigl(\tilde l_{\alpha H, \beta H}(x) \bigr|}{dx}
      =  -\bigl(\tilde l_{\alpha H, \beta H}(x) \bigr| e^{F(x)} \bigl(\mathbb T+\beta \mu J(x) R\otimes\mathds 1+\alpha \mu J(x) \mathds 1\otimes \bar R
        \bigr) e^{-F(x)} \,.
 \end{equation}
With this, the expectation value of the free Hamiltonian density takes the form
\begin{equation}
     \langle Q, R| {\cal H}_\text{free}(0) | Q, R\rangle=2\big(\tilde{l}^{(1,1)}(-\infty)\big|r\big)\,,
\end{equation}
where
\begin{equation}
     \big(\tilde l^{(n, m)}(x)\big|\equiv\left.\frac{\partial^{n+m}\big(\tilde l_{\alpha H, \beta H} (x)\big|}{\partial \alpha^n \partial \beta^m} \right|_{\alpha = \beta =0} \,.
\end{equation}
The system of differential equations needed to be solved is
\begin{equation}  \label{eq:lmnSecondEqs}
  \begin{split}
   \frac{d \bigl(\tilde l^{(1,1)}(x) \bigr|}{dx}
      &= -\bigl(\tilde l^{(1, 1)}(x) \bigr| \mathbb T
      - J(x) \bigl(\tilde l^{(1, 0)}(x) \bigr| \left(\mu R + [R, Q] \right)
      \otimes\mathds 1 \\
      &{}-
      J(x) \bigl(\tilde l^{(0, 1)} (x) \bigr| \mathds 1\otimes \left( \mu \bar R+ [\bar R, \bar Q] \right)
    \,, \\
     \frac{d \bigl(\tilde l^{(1,0)}(x) \bigr|}{dx}
      &=
      -\bigl(\tilde l^{(1, 0)}(x) \bigr| \mathbb T
       - J(x) \bigl(l \bigr| \mathds 1\otimes \left(\mu \bar R + [\bar R, \bar Q] \right)
    \,, \\
   \frac{d \bigl(\tilde l^{(0,1)}(x) \bigr|}{dx}
      &= -\bigl(\tilde l^{(0, 1)}(x) \bigr| \mathbb T
       - J(x) \bigl(l \bigr| \left(\mu R + [R, Q] \right)  \otimes \mathds{1}
    \,,
  \end{split}
 \end{equation}
accompanied by the boundary conditions $\big(\tilde{l}^{(n,m)}(\infty)\big|=0$ for $(n,m) \neq (0,0)$.

\section{Relativistic Quasiparticle ansatz}\label{subsec:rqp}
We will now define the RQP ansatz and derive the formulas needed to determine the spectrum of excitations. As explained in Section \ref{subsec:rcmps}, RCMPSs differ from CMPSs only in that the physical operators of interest are non-local when written in terms of the operators $a(x)$ appearing in the ansatz.  Thus, we can straightforwardly adapt the CMPS quasiparticle ansatz \cite{Draxler:2013oek} to the RCMPS framework. For simplicity, in this section we will first focus on particle-like excitations and postpone the generalizations to solitonic excitations to Section~\ref{sec:soliton}.

\subsection{Definition of the RQP ansatz}

The RQP ansatz is a state that resembles the vacuum state $\ket{Q,R}$ for $|x|\rightarrow \infty$, with a local perturbation at a position $x$ parameterized by two $\chi \times \chi$ matrices $V,W$, further smeared into a plane wave of momentum~$k$:
\begin{equation}\label{eq:rqp}
\begin{aligned}
    \ket{V,W;k}&\equiv 
      \int_{-\infty}^\infty dx\,e^{ikx} 
      V_{+\infty}^\dagger 
      \mathcal P \left[ e^{\int_{-\infty}^{\infty} dy\,  \left( Q\otimes \mathds{1}+ R\otimes a^\dagger(y) \right) }
      \left(V\otimes \mathds{1}+W\otimes a^\dagger(x)\right)   \right]
    V_{-\infty}\ket0\,.
\end{aligned}
\end{equation}
Here, the path ordering refers to everything inside the square bracket, so it can also be written as a path ordered exponential from $x$ to $\infty$, $\mathcal P  e^{\int_{x}^{\infty} dy\,  \left( Q\otimes \mathds{1}+ R\otimes a^\dagger(y) \right) }$,  followed by the factor $\left(V\otimes \mathds{1}+W\otimes a^\dagger(x)\right)$, followed by a path ordered exponential from $-\infty$ to $x$, $\mathcal P e^{\int_{-\infty}^{x} dy\,  \left( Q\otimes \mathds{1}+ R\otimes a^\dagger(y) \right) }$. 
Also note that the spatial profile of the RQP is not that of a delta function because the local perturbation can influence a finite neighborhood around $x$ through the virtual indices of $V,W$.

It is useful to view the state $\ket{V, W; k}$ above as a linearized variation of the generalized RCMPS $\ket{{\bf Q}, {\bf R}}$ defined in \eqref{eq:RCMPSxDep}, expanded around the state $\ket{Q, R}$, with a specific plane-wave profile for $\delta \mathbf{Q}$ and $\delta \mathbf{R}$:
\begin{equation}\label{eq:var_rule}
\ket{V,W;k} = \delta\ket{\mathbf{Q},\mathbf{R}}\Big|_{\substack{\mathbf{Q}(x)=Q \\ \mathbf{R}(x)=R}}\qquad\text{with}\qquad \begin{aligned}
\delta \mathbf{Q}(x)&=e^{ikx}V\,,\\
\delta \mathbf{R}(x)&=e^{ikx}W\,.
\end{aligned}
\end{equation}
All properties of the RQP states $\ket{V, W; k}$ can thus be derived from the properties of the generalized RCMPSs $\ket{{\bf Q}, {\bf R}}$ by taking a variation as in \eqref{eq:var_rule}.

The first requirement we should impose is that the states $\ket{V,W;k}$ should be orthogonal to the RCMPS  $\ket{Q, R}$.  The inner product $\langle Q, R | V, W; k \rangle$ can be computed as the variation of the inner product \eqref{eq:norm_fin} of two generalized RCMPSs, where we only vary the ket $| {\bf Q}, {\bf R} \rangle$, and afterwards we set ${\bf Q}' = {\bf Q} = Q$, ${\bf R}' = {\bf R} = R$, and $V_{\pm}' = V_{\pm}$.  The result can be written as
 \begin{equation}
     \langle Q, R | V, W; k \rangle
      = \int dx\, e^{i k x} \left(v_{+\infty} \middle| e^{\int_{x}^\infty dy\, \mathbb{T}} \left( V \otimes \mathds{1} + W \otimes \bar R \right) 
      e^{\int_{-\infty}^x dy\, \mathbb{T}}
       \middle| v_{-\infty}  \right) \,.
 \end{equation}
Further, using the normalization conditions \eqref{eq:norm_cond_2} of the RCMPS and the fact that the exponential of the $\mathbb{T}$ matrix with a large coefficient amounts to  a projection operator, as in \eqref{eq:eLT}, one can simplify the inner product above to
\begin{equation}
     \langle Q, R | V, W; k \rangle
      = 2 \pi \delta(k)  \left(l |  V \otimes \mathds{1} + W \otimes \bar R  
       | r \right) \,.
 \end{equation}
As expected, RQP states with non-zero momentum are  orthogonal to the (zero-momentum) RCMPS $|Q, R \rangle$, but this is not necessarily the case for $k=0$.   Thus, we restrict the RQP with any~$k$ to be orthogonal to the RCMPS by requiring $V, W$ to obey
\begin{equation}\label{eq:tangent_ortho}
    \left(l \middle |(V \otimes \mathds1 +W \otimes \bar{R}) \middle| r \right)=0\,.
\end{equation} 
A similar calculation gives the inner product of two RQP excitations over the same $\ket{Q, R}$ RCMPS:
 \begin{align} \label{eq:dot_gen}
   \langle V',W'; k'|V,W; k\rangle &= 2\pi\delta(k-k')
     \Biggl[
    \, \left(l \middle|\,W \otimes \bar{W}'\,\middle|r\right) \\
    &{}+ \left(l\middle|\left(\mathds1 \otimes \bar{V}' + R \otimes \bar{W}' \right) 
    \left( \int_0^\infty d\xi\, e^{(\mathbb{T} - i k) \xi} \right)
    \left(V \otimes \mathds1 + W \otimes \bar{R} \right)\middle|r\right) \notag\\
    &{}+\left(l \middle|\left(V \otimes \mathds{1} + W \otimes \bar{R}\right) 
    \left( \int_{0}^\infty d\xi\, e^{(\mathbb{T} + i k) \xi} \right)
    \left(\mathds{1} \otimes \bar{V}' + R \otimes \bar{W}'\right) \middle|r \right)
    \Biggr] \,. \notag
 \end{align}
The second and third lines of this expression can be made to vanish by an appropriate choice of gauge. In particular, the RQP ansatz \eqref{eq:rqp} introduces its own redundancy
\begin{equation}\label{eq:tangent_gauge}
    V\rightarrow V +[X,Q] +ikX\,,\qquad W \rightarrow W + [X,R]\,,
\end{equation}
which is the analog of~\eqref{eq:rcmps_gauge}. Under \eqref{eq:tangent_gauge}, the integrand in \eqref{eq:rqp} changes by a total derivative. Combined with the choice \eqref{eq:tangent_ortho}, we will use this freedom to set 
\begin{equation}\label{eq:tangent_left}
    \text{{\em tangent left-gauge}:} \qquad (l|(V \otimes \mathds{1} + W \otimes \bar{R}) =0 \,,
\end{equation}
which we refer to as \textit{tangent left-gauge}.\footnote{The name \textit{tangent} is appropriate as the RQP ansatz \eqref{eq:rqp} can be interpreted as a ``boosted'' tangent vector to the RCMPS manifold.  We use the term {\em tangent left gauge} as opposed to simply {\em left gauge} because the latter refers to the gauge choice made for the base RCMPS.}${}^{,}$\footnote{Alternatively, we can use the gauge freedom to set $\left(V \otimes \mathds1 + W \otimes \bar{R} \right)|r)=0$, which corresponds to tangent right-gauge.}  In tangent left-gauge, the inner product between two RQP states is
 \begin{equation} \label{eq:dot_left}
  \begin{split}
   \langle V',W'; k'|V,W; k\rangle &= 2\pi\delta(k-k')
    \, \left(l \middle|\,W \otimes \bar{W}'\,\middle|r\right)  \,.
  \end{split}
 \end{equation}

In the matrix notation introduced, the tangent left-gauge condition reads $V^\dagger l + W^\dagger l R = 0$, which allows us to solve for $V$: 
\begin{equation}
    \text{{\em tangent left-gauge}:}  \qquad V = - l^{-1} R^\dagger l W\,.
\end{equation}
In all subsequent equations, the tangent left-gauge should be assumed (unless stated otherwise) and all occurrences of $V$ should be understood as $V = - l^{-1} R^\dagger l  W$.  In the tangent left-gauge, the inner product of two RQP states in \eqref{eq:dot_left} reads
 \begin{equation}\label{eq:inner_simple}
     \braket{V',W';k'|V,W;k} = 2\pi \delta(k-k') \tr(W'^\dagger l W r) \,.
 \end{equation}

\subsection{Computing the excitation energy}

To estimate the spectrum, we apply the Rayleigh-Ritz method to restrict the Hamiltonian to the basis of RQP $\ket{V, W; k}$ over the RCMPS $\ket{Q, R}$. 
Thus, we consider the projected time-independent Schr\"odinger equation
\begin{equation}\label{eq:eigval1}
    \braket{V',W';k'|H|V,W;k} = \left(\omega(k) +E_{\text{RCMPS}} \right) \braket{V',W';k'|V,W;k}\,,
\end{equation}
where $\omega(k)$ is the excitation energy of the RQP and $E_\text{RCMPS} \equiv \langle Q, R | H | Q, R \rangle$ is the (formally infinite) energy of the RCMPS, which we can also write as
 \begin{equation} \label{eq:ERCMPS}
     E_\text{RCMPS} = \langle Q, R | {\cal H}(0) | Q, R \rangle \, \frac{\braket{V',W';k|V,W;k}}{\left(l \middle|\,W \otimes \bar{W}'\,\middle|r\right)} \,,
 \end{equation}
where ${\cal H}(x)$ is the Hamiltonian density. The inner product on the RHS of \eqref{eq:eigval1} was already evaluated in \eqref{eq:dot_left}.  
For the LHS, we write $H = \int_{-\infty}^\infty dx\, {\cal H}(x)$ and use the fact that ${\cal H}(x) = e^{-i P x} {\cal H}(0) e^{i P x}$ together with $P \ket{V, W; k} = k\ket{V, W; k} $, where $P$ is the momentum operator,\footnote{We can write $P = \int \frac{dk}{2\pi} k a_k^\dagger a_k$, as usual, even though $a_k$ and $a_k^\dagger$ are the ladder operators corresponding to only the free part of the Hamiltonian.} to simplify \eqref{eq:eigval1} to
\begin{equation} \label{eq:HamDensity}
     2\pi \delta(k-k')\braket{V',W';k'|\mathcal H(0)|V,W;k} = 2\pi \delta(k-k') \left(\omega(k) +E_\text{RCMPS} \right) \left(l \middle|\,W \otimes \bar{W}'\,\middle|r\right) \,.
\end{equation}
As RCMPSs are cluster-decomposing, the matrix element of a local operator such as the Hamiltonian density decomposes into a connected and disconnected contribution
\begin{equation}\label{eq:cluster}
    \braket{V',W';k'|\mathcal H(0)|V,W;k} = \braket{Q,R|\mathcal H(0)|Q,R}\braket{V',W';k'|V,W;k}+\braket{V',W';k'|\mathcal H(0)|V,W;k}_\text{conn}\,,
\end{equation}
where the disconnected piece contains a delta function (see \eqref{eq:dot_left}) and the connected piece is regular as $k' \to k$.  Plugging \eqref{eq:cluster} and \eqref{eq:ERCMPS} into \eqref{eq:HamDensity}, we notice that the disconnected piece of \eqref{eq:cluster} precisely cancels the $E_\text{RCMPS}$ term on the RHS, giving
 \begin{equation}
  2\pi \delta(k-k')\braket{V',W';k'|\mathcal H(0)|V,W;k}_\text{conn} = 2\pi \delta(k-k') \omega(k)  \left(l \middle|\,W \otimes \bar{W}'\,\middle|r\right) \,. 
 \end{equation}
We can then cancel the delta functions from both sides of this equation and set $k' = k$ in the remaining expression.  Accordingly, the eigenvalue problem is
 \begin{equation} \label{eq:effeval}
   \braket{V',W';k|\mathcal H(0)|V,W;k}_\text{conn} =  \omega(k) \left(l \middle|\,W \otimes \bar{W}'\,\middle|r\right) \,.
 \end{equation}

The LHS of \eqref{eq:effeval} determines the matrix elements of an effective Hamiltonian $H_\text{eff}$ that acts  linearly on the space of matrices $W$, defined implicitly by
\begin{equation}\label{eq:Heff}
    \tr(W'^\dagger l H_{\text{eff},k}(W)r) \equiv \braket{V',W';k|\mathcal H(0)|V,W;k}_\text{conn}\,,
\end{equation}
As determining the connected part of the matrix element is a linear operation, we can compute the effective RQP Hamiltonian for each term appearing in the Hamiltonian individually. For the operators \eqref{eq:opList}, we define the effective operators
\begin{equation}\label{eq:HeffSystem}
    \begin{split}
    \tr \left[W'^\dagger l {\mathbf{H}}_{k}(W)r\right]&\equiv \braket{V',W';k|\mathcal H_{\text{free}}(0)|V,W;k}_\text{conn}
    \\
    \tr \left[W'^\dagger l {\bf{V}}_{b,k}(W)r\right]&\equiv \braket{V',W';k|V_b(0)|V,W;k}_\text{conn}\\
        \tr \left[W'^\dagger l {\bf{O}}_{n,k}(W)r\right]&\equiv \braket{V',W';k|\mathcal O_n(0)|V,W;k}_\text{conn}
    \end{split}
\end{equation}
Then the full effective Hamiltonians take the form
 \begin{equation}\label{eq:HeffEx}
  \begin{split}
    \text{$\phi^4$ theory:} &\qquad
     {H}^{\phi^4}_{\text{eff},k}(W) =  \mathbf{H}_{k}(W)
        + \frac 12 (m^2 - \mu^2)\, \mathbf{O}_{2,k}(W) + g\, \mathbf{O}_{4,k}(W) \,, \\
     \text{sine-Gordon:} &\qquad
     {H}^{\text{sG}}_{\text{eff},k}(W) = \mathbf{H}_{k}(W)
      - \frac{1}{2} \mu^2\, \mathbf{O}_{2,k}(W) - \frac{m^2}{2\beta^2} \left(\mathbf{V}_{i \beta,k}(W) + \mathbf{V}_{-i \beta,k}(W) \right) \,.    
  \end{split}   
 \end{equation}
Given the effective Hamiltonian, the dispersion relation $\omega(k)$ of the RQP ansatz can then be extracted by solving the eigenvalue problem\footnote{In the tangent left-gauge, this is more naturally a generalized eigenvalue problem as one has a non-trivial metric as implemented by the multiplication of $l$ and $r$ in \eqref{eq:dot_left}. In \eqref{eq:eigval}, this metric has been removed by inverting $l$ and $r$ when defining \eqref{eq:Heff}.} involving finite matrices
\begin{equation}\label{eq:eigval}
    H_{\text{eff},k}(W) = \omega(k)W\,.
\end{equation}
One can determine up to $\chi^2$ eigenvalues from \eqref{eq:eigval}, but typically only the lowest eigenvalues can be interpreted as asymptotic states. Most eigenvalues of \eqref{eq:eigval} will fall in the two-particle continuum, and such two-body states are poorly approximated by the RQP ansatz \eqref{eq:rqp} and thus the associated energies should not be trusted.

\subsection{Generating functional for matrix elements of \texorpdfstring{$a(x)$}{a(x)} and \texorpdfstring{$a^\dagger(x)$}{a†(x)}}
\label{sec:GenFunsec}

To compute matrix elements of local operators in the RQP states, let us define the generating functional for matrix elements of products of $a(x)$ and $a^\dagger(x)$:
 \begin{equation} \label{eq:GenFuncMatrixElems}
     \widetilde{\cal Z}[\tilde \jmath, j] \equiv
      \left\langle V', W'; k' \middle| 
      N_\mu \left[ e^{ \int_{-\infty}^\infty
       dx\, \left(\tilde \jmath(x) a^\dagger(x)
        + j(x) a(x) \right)} \right]
        \middle| V, W;k \right\rangle \,.
 \end{equation}
As in Section~\ref{sec:GenFuncRCMPS}, we assume that the source functions $j(x)$ and $\tilde \jmath(x)$ are of compact support. Given the interpretation of the RQP states as variations of generalized RCMPSs, as in \eqref{eq:var_rule}, the expression \eqref{eq:GenFuncMatrixElems} can be recast as
 \begin{equation}
   \widetilde{\cal Z}[\tilde \jmath, j] 
      = e^{- \int_{-\infty}^\infty dx\, \tilde \jmath(x) j(x)} \delta \delta' \langle {\bf Q}', {\bf R}' | {\bf Q}, {\bf R} \rangle \,,
 \end{equation}
where, after taking the variations, one sets ${\bf Q} = {\bf Q}' = Q$, ${\bf R} = R + \tilde \jmath \mathds{1}$ and ${\bf R}' = R + \bar j \mathds{1}$.  We then use \eqref{eq:norm_fin} with $V_{\pm \infty}' = V_{\pm \infty}$ and replace the boundary vectors with the left and right eigenvectors of $\mathbb{T}$, as per the projection property \eqref{eq:eLT}.  The result is
 \begin{equation} \label{eq:GenFuncRQP}
   \begin{split}
    \widetilde{\cal Z}[\tilde \jmath, j] 
     &= \int dx\, e^{i (k-k') x} 
     \left( l \middle | 
      {\cal P} \left[ e^{\int_{-\infty}^\infty dy\, \left( \mathbb{T}
       + j(y) R \otimes \mathds 1 
       + \tilde \jmath(y) \mathds{1} \otimes \bar R \right) }
       W \otimes 
       \bar W'
       \right]
       \middle| r \right)  \\
    &{}+\int dx\, dz\,  e^{i k z - i k' x}
     \left( l \middle | 
     {\cal P} \left[ e^{\int_{-\infty}^\infty dy\, \left( \mathbb{T}
       + j(y) R \otimes \mathds 1 
       + \tilde \jmath(y) \mathds{1} \otimes \bar R \right)}
      U(z) \widetilde{U}(x)
       \right]
     \middle| r \right) \,,
   \end{split}    
 \end{equation}
where, in the first line, the $W \otimes \bar W'$ insertion is assigned the position label $x$ in the path ordering, and 
 \begin{equation}\label{eq:Udef}
  \begin{split}
     U(x) &\equiv V \otimes \mathds{1} 
       + W \otimes \bar R + j(x) W \otimes \mathds 1 \,,  \\
     \widetilde{U}(x) &\equiv \mathds{1} \otimes \bar V' + R \otimes \bar W' + \tilde \jmath(x) \mathds{1} \otimes \bar W' \,.  
   \end{split}          
 \end{equation}
Using the definitions \eqref{eq:rjjtDef} and \eqref{eq:ljjtDef} we can write the first line of \eqref{eq:GenFuncRQP} as
 \begin{equation}\label{eq:Z1stline}
   \widetilde{\cal Z}[\tilde \jmath, j] 
    \bigg|_{\text{1st line}}
     = \int dx\, e^{i (k-k') x} 
     \left( l_{\tilde \jmath, j}(x) \middle| 
     W \otimes \bar W' 
     \middle| r_{\tilde \jmath, j}(x) \right) \,.
 \end{equation}
This expression can be easily computed once we have solved for the vectors $|r_{\tilde \jmath, j}(x) )$ and $(l_{\tilde \jmath, j}(x)|$ using the differential equations \eqref{eq:ODEGenrx} and \eqref{eq:ODEGenlx}, respectively.  We write the second line of \eqref{eq:GenFuncRQP} by splitting the path ordered expression into two halves at the location $x$:
 \begin{equation} \label{eq:Z2ndLine}
     \widetilde{\cal Z}[\tilde \jmath, j] 
    \bigg|_{\text{2nd line}}
     = \int dx\,  e^{i (k-k') x}
     \left[ \left( l_{\tilde \jmath, j}(x) \middle| 
    \widetilde{U}(x) 
     \middle| \widehat{r}_{\tilde \jmath, j}(x) \right)
     +
     \left( \widehat{l}_{\tilde \jmath, j}(x) \middle| 
    \widetilde{U}(x) 
     \middle| r_{\tilde \jmath, j}(x) \right)
     \right] 
 \end{equation}
where
 \begin{equation} \label{eq:rlhatDef}
  \begin{split}
     \left| \widehat{r}_{\tilde \jmath, j}(x) \right)
      &\equiv
    e^{-i k x} \int_{-\infty}^x dz\, e^{ikz} \left. {\cal P} \left[ e^{\int_{-\infty}^x dy\, \left( \mathbb{T}
       + j(y) R \otimes \mathds 1 
       + \tilde \jmath(y) \mathds{1} \otimes \bar R \right)}
      U(z) 
       \right]
     \middle| r \right) \,, \\
   \left( \widehat{l}_{\tilde \jmath, j}(x) \right|
      &\equiv
    e^{-i k x} \int_{x}^\infty dz\, e^{ikz} \left( l \middle|  {\cal P} \left[ e^{\int_{x}^\infty dy\, \left( \mathbb{T}
       + j(y) R \otimes \mathds 1 
       + \tilde \jmath(y) \mathds{1} \otimes \bar R \right)}
      U(z) 
       \right]
     \right. \,.
  \end{split}
 \end{equation}
The two terms in \eqref{eq:Z2ndLine} correspond to the cases $z<x$ and $z>x$, respectively.  

Just like $|r_{\tilde \jmath, j}(x) )$ and $(l_{\tilde \jmath, j}(x)|$, the vectors $\left| \widehat{r}_{\tilde \jmath, j}(x) \right)$ and $\left( \widehat{l}_{\tilde \jmath, j}(x) \right|$ can be found by solving systems of ODEs that can be derived by taking $x$-derivatives of \eqref{eq:rlhatDef}:
 \begin{equation} \label{eq:rlhateqs}
  \begin{split}
     \frac{d \left| \widehat{r}_{\tilde \jmath, j}(x) \right)}{dx}
      &=  U(x) \left| r_{\tilde \jmath, j}(x) \right)
       + \left(-ik + \mathbb{T}
       + j(x) R \otimes \mathds 1 
       + \tilde \jmath(x) \mathds{1} \otimes \bar R \right) \left| \widehat{r}_{\tilde \jmath, j}(x) \right) \,, \\
    \frac{d \left( \widehat{l}_{\tilde \jmath, j}(x) \right|}{dx}
      &=  -\left( l_{\tilde \jmath, j}(x) \right| U(x) 
       -\left( \widehat{l}_{\tilde \jmath, j}(x) \right|  \left(ik + \mathbb{T}
       + j(x) R \otimes \mathds 1 
       + \tilde \jmath(x) \mathds{1} \otimes \bar R \right)  \,.
  \end{split}
 \end{equation}
The boundary conditions are as follows.  For $\left| \widehat{r}_{\tilde \jmath, j}(x) \right)$, we start solving at $x = -\infty$.  If $j$ and~$\tilde \jmath$ have compact support, then for very negative $x$, we have from the definition \eqref{eq:rlhatDef} that
 \begin{equation} \label{eq:hateqs}
      \left| \widehat{r}_{\tilde \jmath, j}(x) \right)
      \approx
    \int_{-\infty}^x dz\,  \left. e^{(x-z)(\mathbb{T} - i k)} ( V \otimes \mathds{1} + W \otimes \bar R) 
     \middle| r \right)
 \end{equation}
We are free to add $-\lambda |r ) (l|(x-z)$ in the exponent for any $\lambda > 0$, provided that we work in tangent left-gauge.  We can then do the $z$ integral and take $x \to -\infty$ to infer
 \begin{equation} \label{eq:BCrhat}
     \left| \widehat{r}_{\tilde \jmath, j}(-\infty) \right)
      = -\frac{1}{\mathbb{T} - i k - \lambda |r)(l|} \left. ( V \otimes \mathds{1} + W \otimes \bar R) 
     \middle| r \right) \,.
 \end{equation}
A similar calculation gives the boundary condition for $\left( \widehat{l}_{\tilde \jmath, j}(x) \right|$ at $x = +\infty$:
 \begin{equation} \label{eq:BClhat}
     \left( \widehat{l}_{\tilde \jmath, j}(\infty) \right|
      = - \left(l \middle| ( V \otimes \mathds{1} + W \otimes \bar R) 
     \right.  \frac{1}{\mathbb{T}+ i k - \lambda |r)(l|}  = 0 \,,
 \end{equation}
where this expression vanishes in tangent left-gauge.

Lastly, let us discuss the generating functional for {\em connected} correlators of $a(x), a^\dagger(x)$:
 \begin{equation}
     \widetilde{\cal Z}_\text{conn}[\tilde \jmath, j] \equiv
      \left\langle V', W'; k \middle| 
      N_\mu \left[ e^{ \int_{-\infty}^\infty
       dx\, \left(\tilde \jmath(x) a^\dagger(x)
        + j(x) a(x) \right)} \right]
        \middle| V, W;k \right\rangle_\text{conn} \,,
 \end{equation}
where for simplicity we set $k'= k$.  Due to cluster decomposition, we have
 \begin{equation}\label{eq:Zconn}
   \widetilde{\cal Z}_\text{conn}[\tilde \jmath, j]
    = \lim_{k' \to k} \left( \widetilde{\cal Z}[\tilde \jmath, j]
     - \langle V', W', k' | V, W;k \rangle {\cal Z}[\tilde \jmath, j] \right) \,,
 \end{equation}
where ${\cal Z}[\tilde \jmath, j]$ is the generating functional \eqref{eq:ZDef} in the underlying RCMPS, and the inner product of RQP states was given in \eqref{eq:dot_left} in tangent left-gauge.  These equations imply that 
 \begin{equation} \label{eq:ZconnFinal}
     \begin{split}
     \widetilde{\cal Z}_\text{conn}[\tilde \jmath, j]
     &= \int dx\,  
     \biggl[ \left( l_{\tilde \jmath, j}(x) \middle| 
     W \otimes \bar W' 
     \middle| r_{\tilde \jmath, j}(x) \right)
      -  \left(l \middle|\,W \otimes \bar{W}'\,\middle|r\right) (l | r_{\tilde \jmath, j}(\infty) ) \\
     &{}+ \left( l_{\tilde \jmath, j}(x) \middle| 
    \widetilde{U}(x) 
     \middle| \widehat{r}_{\tilde \jmath, j}(x) \right)
     +
     \left( \widehat{l}_{\tilde \jmath, j}(x) \middle| 
    \widetilde{U}(x) 
     \middle| r_{\tilde \jmath, j}(x) \right)
       \biggr] \,.
    \end{split}
 \end{equation}
This is the expression we will use in the following subsection, with $|r_{\tilde \jmath, j}(x) )$ and $(l_{\tilde \jmath, j}(x)|$ computed by solving \eqref{eq:ODEGenrx} and \eqref{eq:ODEGenlx}, respectively, and  $\left| \widehat{r}_{\tilde \jmath, j}(x) \right)$ and $\left( \widehat{l}_{\tilde \jmath, j}(x) \right|$ determined by solving \eqref{eq:rlhateqs}.

Note that the integrals corresponding to the terms on the first line of \eqref{eq:ZconnFinal} are not absolutely convergent, but their sum is.  Indeed, as we will show shortly, one can argue that if $j$ and $\tilde \jmath$ are compactly supported in the interval $[-x_*, x_*]$, then for any $|x| > x_*$ we have
 \begin{equation} \label{eq:approach}
     \left( l_{\tilde \jmath, j}(x) \middle| 
     W \otimes \bar W' 
     \middle| r_{\tilde \jmath, j}(x) \right)
       =  \left(l \middle|\,W \otimes \bar{W}'\,\middle|r\right) (l | r_{\tilde \jmath, j}(\infty) )+\mathcal O(e^{\Re(\lambda_2)|x|}) \,, \qquad \text{for $|x|>x_*$} \,,
 \end{equation}
where $\lambda_2$ is the eigenvalue of $\mathbb T$ with the second largest real part. That the integrals on the first line of \eqref{eq:ZconnFinal} are not absolutely convergent then follows, since the second term is manifestly $x$-independent at large $x$;  moreover, because of the equality \eqref{eq:approach}, the sum of these two terms decays exponentially and hence yields an absolutely convergent integral.  To show \eqref{eq:approach}, note that for $|x|>x_*$ the differential equations \eqref{eq:ODEGenrx} and \eqref{eq:ODEGenlx} simplify to
 \begin{equation} 
  \frac{d \bigl|r_{\tilde \jmath, j}(x) \bigr)}{dx}
      = \mathbb T \bigl|r_{\tilde \jmath, j}(x) \bigr) \,, \qquad 
   \frac{d \bigl(l_{\tilde \jmath, j}(x) \bigr|}{dx}
      = - \bigl(l_{\tilde \jmath, j}(x) \bigr| \mathbb T   \,, \qquad \text{for $|x|>x_*$} \,.
 \end{equation}
Because of the projection property \eqref{eq:eLT}, these equations imply exponential approach $\bigl|r_{\tilde \jmath, j}(x) \bigr) \rightarrow \alpha_r|r)$ and $\bigl(l_{\tilde \jmath, j}(x) \bigr| \rightarrow \alpha_l(l|$ for $|x|>x_*$ with $\alpha_r,\alpha_l$ constants.  Let us examine the region $x > x_*$ first.  Since $\bigl(l_{\tilde \jmath, j}(\infty) \bigr| = (l|$ we have $\bigl(l_{\tilde \jmath, j}(x) \bigr| = (l|$.  Meanwhile, the proportionality $\bigl|r_{\tilde \jmath, j}(x) \bigr) = \alpha_r|r)+\mathcal O(e^{\Re(\lambda_2)x})$ implies that $\bigl|r_{\tilde \jmath, j}(x) \bigr) = |r) \bigl(l \big|r_{\tilde \jmath, j}(x) \bigr)+\mathcal O(e^{\Re(\lambda_2)x})$, since multiplying a vector proportional to $|r)$ by the projector $|r)(l|$ should leave this vector unchanged.  Thus, 
 \begin{equation} \label{eq:limit1}
   \left( l_{\tilde \jmath, j}(x) \right|
    = (l | \,, \qquad \left| r_{\tilde \jmath, j}(x) \right)
    = |r )\bigl(l \big| r_{\tilde \jmath, j}(\infty) \bigr)+\mathcal O(e^{\Re(\lambda_2)x}) \,, \qquad \text{for $x >x_*$} \,.
 \end{equation}
A similar analysis shows that 
 \begin{equation} \label{eq:limit2}
    \bigl(l_{\tilde \jmath, j}(x) \bigr| = \bigl(l_{\tilde \jmath, j}(-\infty) \big| r \bigr) (l|+\mathcal O(e^{-\Re(\lambda_2)x}) \,, \qquad 
    \left| r_{\tilde \jmath, j}(x) \right) = |r)
    \qquad \text{for $x < - x_*$} \,. 
 \end{equation}
The equality \eqref{eq:approach} then follows from \eqref{eq:limit1}--\eqref{eq:limit2} and the fact that $\bigl(l_{\tilde \jmath, j}(x) \big| r_{\tilde \jmath, j}(x) \bigr)$ is $x$-independent, and hence $\bigl(l_{\tilde \jmath, j}(-\infty) \big| r \bigr) = \bigl(l \big| r_{\tilde \jmath, j}(\infty) \bigr)$.

If $j$ and $\tilde \jmath$ do not have compact support but instead decay sufficiently fast at infinity, then \eqref{eq:ZconnFinal} still holds.  The equality \eqref{eq:approach} is only approximate, but the same conclusions are true.

\subsection{Matrix elements of the Hamiltonian density}
\label{sec:effHamiltonian}

We can evaluate the connected matrix element of the Hamiltonian density ${\cal H}(0)$ in the state $\ket{V, W; k}$, which is the main ingredient in \eqref{eq:Heff} for determining the effective Hamiltonian.  We will use the same strategy as in Section~\ref{sec:expvalue} whereby we first evaluate the matrix elements of the operators $V_b(0)$, ${\cal O}_n(0)$, and ${\cal H}_\text{free}(0)$ defined in \eqref{eq:opList}, from which we can then construct the matrix elements of ${\cal H}(0)$ in the theories of interest  \eqref{eq:HamExamples}.

\subsubsection{Matrix elements of \texorpdfstring{$V_b(0)$}{Vb(0)}}

For the operator $V_b(0) = N_\mu \left[ e^{b \phi(0)} \right]$, from \eqref{eq:phi_inv}, we see that 
 \begin{equation} \label{eq:VbMatrix}
     \left\langle V', W';k
       \middle| V_b(0) 
       \middle| V, W;k  \right\rangle_\text{conn}
     = \widetilde{\mathcal Z}_\text{conn}[bJ, bJ] \,,
 \end{equation}
where the functional $\widetilde{\mathcal Z}_\text{conn}$ was computed in \eqref{eq:ZconnFinal}. In order to compute the excitation spectrum, we need to deduce the effective interaction term $\mathbf{V}_{b,k}(W)$ defined in \eqref{eq:HeffSystem}. This is straightforward to achieve by starting from \eqref{eq:VbMatrix} and using the vectorization properties to write the effective operator in matrix form
\begin{equation}
\begin{aligned}
    \mathbf{V}_{b,k}(W) &= l^{-1}\int dx\, \biggl[
    l^{\dagger}_{bJ,bJ}(x)\, W\, r_{bJ,bJ}(x)-\langle V_b\rangle_{Q,R}\, lWr
    + \\&[\widehat{l}_{bJ,bJ}^{\dagger}(x),R]\, r_{bJ,bJ}(x)
    + [l^{\dagger}_{bJ,bJ}(x),R]\, \widehat{r}_{bJ,bJ}(x) 
    + \\&bJ(x)\Bigl(\widehat{l}^{\dagger}_{bJ,bJ}(x) r_{bJ,bJ}(x)
    + l^{\dagger}_{bJ,bJ}(x)\widehat{r}_{bJ,bJ}(x)\Bigr)
    \biggr] r^{-1}\,.
\end{aligned}
\end{equation}
The effective operator for the sine-Gordon potential $\cos(\beta \phi)$ can then be computed as $\frac{1}{2}(\mathbf{V}_{i\beta,k}(W) +\mathbf{V}_{-i\beta,k}(W))$.

\subsubsection{Matrix elements of \texorpdfstring{${\cal O}_n(0)$}{On(0)}}

The matrix elements of ${\cal O}_n(0) = N_\mu \left[\phi(0)^n \right]$ can be obtained in a similar way to Section~\ref{sec:OnExp} by viewing \eqref{eq:VbMatrix} as a generating function.  In particular,
 \begin{equation} \label{eq:OnMatrix}
   \left\langle V', W';k
       \middle| {\cal O}_n(0) 
       \middle| V, W;k  \right\rangle_\text{conn}
    = \frac{\partial^n}{\partial b^n}
     \widetilde{\mathcal Z}_\text{conn}[bJ, bJ] \bigg|_{b=0} \,.
 \end{equation}
We can write a more explicit formula using the definitions for $|r^{(n)}(x))$ and $(l^{(n)}(x)|$ in \eqref{eq:phin_VEV} and \eqref{eq:lnDef}, respectively, as well as
 \begin{equation} \label{eq:allRLDefs}
  \begin{split}
    |\widehat{r}^{(n)}(x))
      \equiv\left.\frac{\partial^n|\widehat{r}_{bJ, bJ} (x))}{\partial b^n} \right|_{b=0} \,, \qquad
    (\widehat{l}^{(n)}(x)|
       \equiv\left.\frac{\partial^n(\widehat{l}_{bJ, bJ} (x)|}{\partial b^n} \right|_{b=0} \,.
  \end{split}   
 \end{equation}
From \eqref{eq:ZconnFinal}, the matrix element \eqref{eq:OnMatrix} becomes
 \begin{equation} \label{eq:OnMatrix2}
       \begin{split}
   &\left\langle V', W';k
       \middle| {\cal O}_n(0) 
       \middle| V, W;k  \right\rangle_\text{conn} \\
     &\quad= \int dx\, 
     \Biggl[ -  \left(l \middle|\,W \otimes \bar{W}'\,\middle|r\right) (l | r^{(n)}(\infty) ) 
    + \sum_{p=0}^n \begin{pmatrix} n \\ p \end{pmatrix} \biggl[ \left( l^{(p)}(x) \middle| 
     W \otimes \bar W' 
     \middle| r^{(n-p)}(x) \right) \\
     &\quad{}+ \left( l^{(p)} (x) \middle| 
    \mathds{1} \otimes  \bar V' + R \otimes \bar W'
     \middle| \widehat{r}^{(n-p)}(x) \right) 
     +
     \left( \widehat{l}^{(p)} (x) \middle| 
    \mathds{1} \otimes  \bar V' + R \otimes \bar W'
     \middle| r^{(n-p)}(x) \right)
       \biggr] \\
    &\quad{}+
    n J(x) \sum_{p=0}^{n-1} \begin{pmatrix} n-1 \\ p \end{pmatrix} \biggl[
    \left( l^{(p)} (x) \middle| 
    \mathds{1}  \otimes \bar W'
     \middle| \widehat{r}^{(n-1-p)}(x) \right)
     + 
     \left( \widehat{l}^{(p)} (x) \middle| 
     \mathds{1} \otimes \bar W'
     \middle| r^{(n-1-p)}(x) \right) \biggr]
    \Biggr] \,.
    \end{split}
 \end{equation}
The equations obeyed by the functions \eqref{eq:allRLDefs} are
 \begin{equation}
  \begin{split}
    \frac{d \left| \widehat{r}^{(n)}(x) \right)}{dx}
      &=  \left( V \otimes \mathds{1} + W \otimes \bar R \right) \left| r^{(n)}(x) \right)
      + n J(x)   W \otimes \mathds{1}  \left| r^{(n-1)}(x) \right) \\
      &{}+ \left(-ik + \mathbb{T} \right) \left| \widehat{r}^{(n)}(x) \right) 
       + n J(x) \left( R \otimes \mathds 1 
       +  \mathds{1} \otimes \bar R \right) \left| \widehat{r}^{(n-1)}(x) \right) \,, \\
   \frac{d \left( \widehat{l}^{(n)}(x) \right|}{dx}
      &=  -\left( l^{(n)}(x) \right| \left( V \otimes \mathds{1} + W \otimes \bar R \right)
      -n J(x) \left( l^{(n-1)}(x) \right| W \otimes \mathds{1} \\
      &{}-\left( \widehat{l}^{(n)}(x) \right|  \left(ik + \mathbb{T} \right)
      - n J(x) \left( \widehat{l}^{(n-1)}(x) \right|
       \left(  R \otimes \mathds 1 
       + \mathds{1} \otimes \bar R \right)  \,.
   \end{split}    
 \end{equation}
with boundary conditions obtained from \eqref{eq:BCrhat}--\eqref{eq:BClhat}:
 \begin{equation}
     \left| \widehat{r}^{(n)}(-\infty) \right)
      = \begin{cases} -\frac{1}{\mathbb{T} - i k - \lambda |r)(l|} \left. ( V \otimes \mathds{1} + W \otimes \bar R) 
     \middle| r \right) & n=0 \,, \\
     0 & n > 0 \,, 
     \end{cases} \qquad 
      \left( \widehat{l}^{(n)}(\infty) \right| = 0 \,.
 \end{equation}
Finally, the effective term of \eqref{eq:OnMatrix} becomes
\begin{equation}
\begin{aligned}
   \mathbf{O}_{n,k}(W)
   &= (l^{\dagger})^{-1}\int dx\,\,\Bigg[
   -\braket{\mathcal{O}_n}_{Q,R} \,l^{\dagger} W r 
   +\sum_{p=0}^n \binom{n}{p} \bigg[
      l^{\dagger(p)}(x) W r^{(n-p)}(x) 
      - R\,l^{\dagger(p)}(x) \widehat{r}^{(n-p)}(x)
       \\&+l^{\dagger(p)}(x) R \widehat{r}^{(n-p)}(x)
      - R\,\widehat{l}^{\dagger(p)}(x) r^{(n-p)}(x) 
      + \widehat{l}^{\dagger(p)}(x) R r^{(n-p)}(x) \bigg] \\
   &+ n J(x) \sum_{p=0}^{n-1} \binom{n-1}{p} \Big(
      l^{\dagger(p)}(x) \widehat{r}^{(n-1-p)} (x)
      + \widehat{l}^{\dagger(p)}(x) r^{(n-1-p)}(x) \Big)
   \Bigg] \, r^{-1}\,,
\end{aligned}
\label{HeffOn}
\end{equation}
where we denoted $\braket{\mathcal{O}_n}_{Q,R}\equiv \bra{Q,R}\mathcal{O}_n(0)\ket{Q,R}$ and again used \eqref{eq:vecToMatrix}.

\subsubsection{Matrix elements of \texorpdfstring{${\cal H}_\text{free}(0)$}{Hfree(0)}}\label{sec:HeffFree}

We now consider the matrix elements of the free part of the Hamiltonian ${\cal H}_\text{free}(0)$ defined in~\eqref{eq:opList}.
The statement in Section~\ref{sec:expvalueHfree} that the expectation value of ${\cal H}_\text{free}(0)$ can be computed using \eqref{eq:HfreeExp} generalizes to any matrix elements between momentum eigenstates of the same momenta (see Appendix~\ref{app:free_term}). This identity allows us to retrace the steps of Section~\ref{sec:expvalueHfree} and write the effective free Hamiltonian as
\begin{equation}\label{eq:rqp_Hfree}
     \bra{V',W';k}\mathcal{H}_\text{free}(0)\ket{V,W;k}_\text{conn}=2\frac{\partial^2}{\partial \alpha\, \partial \beta }\mathcal{\widetilde{Z}_\text{conn}}[\alpha H, \beta H] \Bigg|_{\alpha=\beta=0}\,,
\end{equation}
where $\mathcal{\widetilde{Z}}_{\text{conn}}$ is the connected generating functional defined in \eqref{eq:ZconnFinal}.
To evaluate \eqref{eq:ZconnFinal}, we need to solve the ODEs \eqref{eq:rlhateqs}. However, we encounter the same issue as for the RCMPS ansatz in Section \ref{sec:expvalueHfree}; namely, that $\partial_x J(x)$ has a non-integrable singularity. To resolve this, we define
\begin{equation}\label{eq:urul}
     \bigl|\widehat{r}_{\alpha H, \beta H}(x) \bigr)
      = e^{-F(x)} \bigl|u^r_{\alpha H, \beta H}(x) \bigr)\,,\qquad\bigl(\widehat{l}_{\alpha H, \beta H}(x) \bigr|
      = \bigl(u^l_{\alpha H, \beta H}(x) \bigr|e^{F(x)} \,,
\end{equation}
where $e^{-F(x)}$ is the same integrating factor used in \eqref{eq:ltilde} and \eqref{eq:rtilde}. It is straightforward then to show that the ODEs obeyed by $\bigl|u^r_{\alpha H, \beta H}(x) \bigr)$ and $\bigl(u^l_{\alpha H, \beta H}(x) \bigr|$ are
\begin{equation}\label{eq:ODEur}
\begin{aligned}
    \frac{d \bigl|u^r_{\alpha H, \beta H}(x) \bigr)}{dx}
      &= e^{F(x)} U(x)e^{-F(x)} \bigl|\widetilde r_{\alpha H, \beta H}(x) \bigr)  \\& \quad  +e^{F(x)} \bigl(-ik +\mathbb T+\beta \mu J(x) R\otimes\mathds 1+\alpha \mu J(x) \mathds 1\otimes \bar R
        \bigr) e^{-F(x)} \bigl|u^r_{\alpha H, \beta H}(x) \bigr) \,,
\end{aligned}
\end{equation}
\begin{equation}\label{eq:ODEul}
\begin{aligned}
    \frac{d \bigl(u^l_{\alpha H, \beta H}(x) \bigr|}{dx}
      &= -\bigl(\widetilde l_{\alpha H, \beta H}(x) \bigr|e^{F(x)} U(x)e^{-F(x)}   \\& \quad -\bigl(u^l_{\alpha H, \beta H}(x) \bigr| e^{F(x)} \bigl(ik +\mathbb T+\beta \mu J(x) R\otimes\mathds 1+\alpha \mu J(x) \mathds 1\otimes \bar R
        \bigr) e^{-F(x)}  \,,
\end{aligned}
\end{equation}
with the boundary conditions
\begin{equation}\label{eq:u_BC}
\begin{aligned}
     &\left| u^r_{\alpha H, \beta H}(-\infty) \right) \equiv\left| u^r(-\infty) \right)
      = -\frac{1}{\mathbb{T} - i k - \lambda |r)(l|} \left. ( V \otimes \mathds{1} + W \otimes \bar R) 
     \middle| r \right) \,, \\&
     \left( u^l_{\alpha H, \beta H}(\infty) \right|=0\,,
\end{aligned}
\end{equation}
where $\bigl|\widetilde{r}_{\alpha H, \beta H}(x)\bigr)$ and $\bigl(\widetilde{l}_{\alpha H, \beta H}(x)\bigr|$ are defined in \eqref{eq:rtilde} and \eqref{eq:ltilde}, respectively. According to \eqref{eq:rqp_Hfree}, we must take derivatives of $\mathcal{\widetilde{Z}}[\alpha H, \beta H]_{\text{conn}}$ with respect to $\alpha$ and $\beta$, which requires that we solve two ODE systems: one for $\bigl|u^{r\,(n,m)}(x)\bigr)$ and one for $\bigl(u^{l\,(n,m)}(x)\bigr|$, where we again use the notation $f^{(n,m)}\equiv\left .\frac{\partial^{n+m}f}{\partial \alpha^n\partial \beta^m}\right|_{\alpha=\beta=0}$.
Having solved for $\bigl|u^{r\,(n,m)}(x)\bigr),\bigl(u^{l\,(n,m)}(x)\bigr|$, we can deduce the derivatives of $\bigl|\widehat{r}_{\alpha H, \beta H}(x) \bigr)$ and $\bigl(\widehat{l}_{\alpha H, \beta H}(x) \bigr|$ using  \eqref{eq:urul}. We refer the reader to Appendix~\ref{urulAppendix} for detailed equations. 

This determines all quantities needed to evaluate \eqref{eq:rqp_Hfree} and we can combine them into an expression for the effective free Hamiltonian
\begin{equation}\label{eq:Hfree11}
\begin{aligned}
\mathbf{H}_{k}&(W)
    = 2{l}^{-1}\int dx\,\Bigg[-\frac{1}{2}\braket{\mathcal{H}_{\text{free}}}_{Q,R}\,l W r+
        \,[\widehat{l}^{\dagger\,(1,1)}(x),R]\,r
        +[\widehat{l}^{\dagger\,(0,1)}(x),R]\,r^{(1,0)}(x)
        \\&+[\widehat{l}^{\dagger\,(1,0)}(x),R]\,r^{(0,1)}(x)
        +[\widehat{l}^\dagger,R]\,r^{(1,1)}(x) \\[4pt]
        &+\,[l^{\dagger\,(1,1)}(x),R]\,\widehat r
        +[l^{\dagger\,(0,1)}(x),R]\,\widehat r^{(1,0)}(x)
        +[l^{\dagger\,(1,0)}(x),R]\,\widehat r^{(0,1)}(x)
        +[l^\dagger,R]\,\widehat r^{(1,1)}(x) \\[4pt]
        &+\,l^{\dagger\,(1,1)}(x)\,W\,r
        +l^{\dagger\,(0,1)}(x)\,W\,r^{(1,0)}(x)
        +l^{\dagger\,(1,0)}(x)\,W\,r^{(0,1)}(x)
        +l^\dagger\,W\,r^{(1,1)}(x) \\[4pt]
        &+\,J(x)\Big(\widehat{l}^{\dagger\,(1,0)}(x)\,r
            +l^{\dagger\,(1,0)}(x)\,\widehat r
            \Big)
    \Bigg]\,r^{-1}\,,
\end{aligned}
\end{equation}
where we defined $\langle \mathcal{H}_{\text{free}} \rangle_{Q,R} \equiv \bra{Q,R}\mathcal{H}_{\text{free}}(0)\ket{Q,R}$.

\subsection{Dispersion relation at large momentum}\label{sec:large_mom}
In Section~\ref{sec:expvalue}, we saw that the RCMPS ansatz encodes the correct relativistic short distance behavior of correlation functions. Analogously, we will now show that the RQP ansatz at large momentum recovers the Lorentz-covariant dispersion $\omega(k)\sim |k|$. To that end, we need to determine the large momentum limit of the connected matrix elements of the various ingredients \eqref{eq:HeffSystem} used for computing the effective Hamiltonian.

The connected Hamiltonian matrix elements can be computed in terms of the generating functional \eqref{eq:ZconnFinal}, which we find convenient to rewrite as
\begin{equation}\label{eq:largek_Zconn}
\begin{aligned}
    \widetilde{\cal Z}_\text{conn}[\tilde \jmath,j] 
     &= \int dx\, 
     \left( l \middle | 
      {\cal P} \left[ e^{\int_{-\infty}^\infty dz\, \left( \mathbb{T}
       + j(z) R \otimes \mathds 1 
       + \tilde\jmath(z) \mathds{1} \otimes \bar R \right) }
       \left\{W \otimes 
       \bar W'-(l|(W \otimes 
       \bar W')|r)\right\}
       \right]
       \middle| r \right)  \\
    &{}+\int dx\, dy\,  e^{ik(x-y)}
     \left( l \middle | 
     {\cal P} \left[ e^{\int_{-\infty}^\infty dz\, \left( \mathbb{T}
       + j(z) R \otimes \mathds 1 
       + \tilde j(z) \mathds{1} \otimes \bar R \right)}
      U(x) \widetilde{U}(y)
       \right]
     \middle| r \right) \,,
\end{aligned}
\end{equation}
where, in the first line,  the insertion in the path ordering is at $x$, and in the second line the quantities $U,\widetilde U$ were defined in \eqref{eq:Udef}. The first line is independent of $k$, while the second line takes the form of Fourier transforms with respect to $x,y$. For integrable sources such as $j,\tilde \jmath\propto J$, which are relevant for the interaction potentials, the Riemann-Lebesgue lemma ensures that the Fourier transforms decay. Thus, the interaction potential does not give rise to an asymptotically increasing energy at large momentum. For the free part of the Hamiltonian, the necessary source function is proportional to $H(x) = (\mu-\partial_x)J(x)$ which is not integrable due to $\partial_x J(x)$ and one has to take care in the application of the Riemann-Lebesgue lemma. The connected matrix element of the free Hamiltonian \eqref{eq:rqp_Hfree} can be written as
\begin{equation}\label{eq:largekapprox}
    \bra{V',W';k}\mathcal{H}_\text{free}(0)\ket{V,W;k}_\text{conn}
    =2\frac{\partial^2}{\partial \alpha\, \partial \beta }\mathcal{\widetilde{Z}_\text{conn}}[\alpha H, \beta H]\Bigg|_{\alpha=\beta=0}
    = I_1+I_2+I_3+I_4+I_5\,,
\end{equation}
where the integrals $I_i$ are defined as
 \es{eq:Is}{
    I_1 &= 2\int dx\, dy\,  e^{ik(x-y)}\,H(x)H(y)
     \left( l \middle | 
     {\cal P} \left[ e^{\int_{-\infty}^\infty dz\,  \mathbb{T}}
      (W\otimes\mathds 1)_x\,(\mathds 1\otimes \bar W')_y
       \right]
     \middle| r \right)\,, \\
    I_2 &= 2\int dx\, dy\, dz\,  e^{ik(x-y)}\,H(x)H(z)
     \left( l \middle | 
     {\cal P} \left[ e^{\int_{-\infty}^\infty dz'\,  \mathbb{T}}
      (W\otimes\mathds 1)_x\,(\mathds 1\otimes \bar R)_z\,\widetilde U_0(y)
       \right]
     \middle| r \right)\,, \\
    I_3 &= 2\int dx\, dy\, dz\,  e^{ik(x-y)}\,H(z)H(y)
     \left( l \middle | 
     {\cal P} \left[ e^{\int_{-\infty}^\infty dz'\,  \mathbb{T}}
      (R\otimes\mathds 1)_z\,U_0(x)\,(\mathds 1\otimes \bar W')_y
       \right]
     \middle| r \right)\,, \\
    I_4 &= 2\int dx\, dy\, dz\, dz'\,  e^{ik(x-y)}\,H(z)H(z') \\
     &{}\qquad\quad\times
     \left( l \middle | 
     {\cal P} \left[ e^{\int_{-\infty}^\infty dz''\,  \mathbb{T}}
      (R\otimes\mathds 1)_z\,(\mathds 1\otimes \bar R)_{z'}\,U_0(x)\,\widetilde U_0(y)
       \right]
     \middle| r \right)\,, \\
    I_5 &= 2\int dx\, dz\, dz'\,  H(z)H(z')
     \\
     &{}\qquad\quad\times
     \left( l \middle | 
     {\cal P} \left[ e^{\int_{-\infty}^\infty dz''\,  \mathbb{T}}
      (R\otimes\mathds 1)_z\,(\mathds 1\otimes \bar R)_{z'}\,
      \bigl\{W\otimes\bar W'-(l|(W\otimes\bar W')|r)\bigr\}_x
       \right]
     \middle| r \right)\,, 
 }
where we denoted $U$ and $\tilde{U}$ evaluated at $\alpha=\beta=0$ as $U_0$ and $\tilde{U}_0$, respectively, and indicated the insertion points in the path-ordering by subscripts, such as in $(W \otimes 1)_x$. We begin by examining the large momentum behavior of $I_1$. Since the path-ordered exponentials preserve the fixed points $(l|,|r)$, we can simplify the integral and insert $H(x)=(\mu-\partial_x)J(x)$
\begin{equation}\begin{split}\label{eq:largkint1}
I_1 &=2\int dx \, dy\, 
e^{ik(x-y)}\,
\Bigl(\mu^{2}J(x)J(y)-\mu J(y)\partial_x J(x)
  -\mu J(x)\partial_y J(y)+\partial_x J(x)\partial_y J(y)\Bigr)\\
&\times\left( l \middle| \theta(x-y)(W\otimes\mathds{1})\,e^{\mathbb{T}(x-y)}\,
  (\mathds{1}\otimes\bar W') +\theta(y-x) (\mathds{1}\otimes\bar W')\,e^{\mathbb{T}(y-x)}\,
  (W\otimes\mathds{1})\middle| r \right)\,.
\end{split}\end{equation}
The term containing $J(x)J(y)$ is integrable and decays at large momentum due to the Riemann-Lebesgue lemma. The terms containing $ J(y)\partial_xJ(x)$ and $ J(x)\partial_yJ(y)$ cancel exactly. To see this we use the identity
\begin{equation}
    \int dx\, dy\, f(y) \partial_xg(x)S(x-y)=-\int dx\, dy\, g(x)\partial_yf(y) S(x-y)\,,
\end{equation}
which holds for functions that decay at infinity and is easily proven by integration by parts. The remaining term in \eqref{eq:largkint1} contains $\partial_xJ(x)\partial_yJ(y)$ and thus the large momentum behavior of $I_1$ is 
\begin{equation}\label{eq:largkint2}
\begin{aligned}
    I_1&= 2\int_{-\infty}^\infty dx\int^x_{-\infty} dy\, e^{ik(x-y)}\partial_xJ(x)\partial_yJ(y)\, \bigl(l\bigr|(W\otimes\mathds 1)\,e^{\mathbb T (x-y)}(\mathds 1\otimes \bar W')\bigl|r\bigr)\\
    &\;{}+ 2\int_{-\infty}^\infty dy\int^y_{-\infty} dx\, e^{ik(x-y)}\partial_xJ(x)\partial_yJ(y)\, \bigl(l\bigr|(\mathds 1\otimes \bar W') e^{\mathbb T(y-x)}(W\otimes\mathds 1)\bigl|r\bigr) +o(|k|^0) \,,
\end{aligned}
\end{equation}
Upon integration by parts, we find
\begin{equation}\label{eq:largek1}
\begin{aligned}
    I_1&= 2k^2\int_{-\infty}^\infty dx\int^x_{-\infty} dy\, e^{ik(x-y)}J(x)J(y)\, \bigl(l\bigr|(W\otimes\mathds 1)\,e^{\mathbb T (x-y)}(\mathds 1\otimes \bar W')\bigl|r\bigr)\\
    &\;{}+ 2k^2\int_{-\infty}^\infty dy\int^y_{-\infty} dx\, e^{ik(x-y)}J(x)J(y)\, \bigl(l\bigr|(\mathds 1\otimes \bar W') e^{\mathbb T(y-x)}(W\otimes\mathds 1)\bigl|r\bigr)\\
    &\;{}-4ik\int_{-\infty}^\infty dx\int^x_{-\infty} dy\, e^{ik(x-y)}J(x)J(y)\, \bigl(l\bigr|(W\otimes\mathds 1)\,\mathbb T\,e^{\mathbb T (x-y)}(\mathds 1\otimes \bar W')\bigl|r\bigr)\\
    &\;{}+4ik\int_{-\infty}^\infty dy\int^y_{-\infty} dx\, e^{ik(x-y)}J(x)J(y)\, \bigl(l\bigr|(\mathds 1\otimes \bar W')\,\mathbb T\, e^{\mathbb T(y-x)}(W\otimes\mathds 1)\bigl|r\bigr) +\mathcal O(|k|^0) \,,
\end{aligned}
\end{equation}
where the remainder contains terms that decay according to the Riemann-Lebesgue lemma and a divergent term independent of $k$. However, this divergence must be canceled in \eqref{eq:largekapprox}, because the effective Hamiltonian is finite, as shown in \eqref{eq:Hfree11}.
The remaining integrals in \eqref{eq:largek1} can be computed using the spectral decomposition $e^{\mathbb T (x-y)}= \sum_{n}e^{\lambda_n (x-y)}|r_n)(l_n|$, which reduces the integrals in \eqref{eq:largek1} to the scalar integral
\begin{equation}\label{eq:largek_integral}
    \int_{-\infty}^\infty dx\int^x_{-\infty} dy\, e^{ik(x-y)}J(x)J(y)e^{\lambda(x-y)} = \frac{i}{4\pi k}\log\frac{4k^2}{\mu ^2}+\frac{1}{4|k|}+\mathcal O\left(\frac{1}{k^2}\log\frac{k^2}{\mu^2}\right )\,.
\end{equation}
Adding up the terms in \eqref{eq:largek1}, and noting that the last two lines contribute only at $\mathcal O\bigl(\log\frac{k^2}{\mu^2}\bigr)$, we get
\begin{equation}\label{eq:largek2First}
\begin{aligned}
   I_1&\approx \left(\frac{ik}{2\pi}\log\frac{4k^2}{\mu^2}+\frac{|k|}{2}\right)\sum_n\bigl(l\bigr|W\otimes\mathds 1\bigl|r_n\bigr)\bigl(l_n\bigr|\mathds 1\otimes \bar W'\bigl|r\bigr)\\
    &\;{}+ \left(-\frac{ik}{2\pi}\log\frac{4k^2}{\mu^2}+\frac{|k|}{2}\right)\sum_n\bigl(l\bigr|\mathds 1\otimes \bar W'\bigl|r_n\bigr)\bigl(l_n\bigr|W\otimes\mathds 1\bigl|r\bigr)+\mathcal O\left(\log\frac{k^2}{\mu^2}\right ) \,.
\end{aligned}
\end{equation}
In this expression we recognize the sum over $n$ as the resolution of identity, and therefore
\begin{equation}\label{eq:largek2}
\begin{aligned}
   I_1
    &\;\approx |k|\bigl(l\bigr|W\otimes \bar W'\bigl|r\bigr) +\mathcal O\left(\log\frac{k^2}{\mu^2}\right ) \,.
\end{aligned}
\end{equation}
A similar analysis for the remaining integrals in \eqref{eq:largekapprox} shows that $I_2 =\mathcal{O}\left(\log\frac{k^2}{\mu^2}\right)$, $I_3 = \mathcal{O}\left(\log\frac{k^2}{\mu^2}\right)$, $I_4 = \mathcal{O}\left(\frac{\mu}{|k|}\log\frac{k^2}{\mu^2}\right)$, and $I_5$ is $k$-independent. Consequently, the large $k$ behavior of the dispersion relation follows directly from \eqref{eq:effeval}
\begin{equation}
    \omega(k) \approx |k|+\mathcal{O}\left(\log \frac{k^2}{\mu^2} \right)\,, \quad |k|\rightarrow \infty\,.
\end{equation}
This shows that the large momentum leading contribution to the effective Hamiltonian of the RQP is $|k|$ regardless of the specifics of the state $\ket{V,W;k}$. Thus, the RQP ansatz naturally encodes the expected relativistic behavior. In particular, the derivation of the UV behavior of the dispersion $\omega(k)$ relied crucially on the use of $a^\dagger(x)$ in the construction of the ansatz \eqref{eq:rqp}, as it gave rise to the kernel $J(x)$ and in turn the $\frac{1}{|k|}$ behavior in \eqref{eq:largek_integral}.  On the other hand, since the ansatz \eqref{eq:rqp} is not Lorentz covariant, the convergence of $\omega(k)$ to the expected relativistic expression $\sqrt{M^2+k^2}$ at small momenta will serve as an important consistency check of our method.

\subsection{Solitonic excitations}\label{sec:soliton}
The RQP ansatz for the particle-like excitations \eqref{eq:rqp} can be generalized to solitonic excitations \cite{Draxler:2013oek} interpolating between two degenerate vacua described by $\ket{Q_1,R_1}$ at $x\rightarrow\infty$ and $\ket{Q_2, R_2}$ at $x\rightarrow-\infty$. In particular, solitonic excitations are implemented by substituting $(Q,R)\rightarrow( Q_1,R_1)$ to the left of the local perturbation in \eqref{eq:rqp} and $(Q,R)\rightarrow( Q_2,R_2)$ to the right of it, yielding 
\begin{equation}\label{eq:rqp_sol}
\begin{aligned}
    \ket{V,W;k}_{\text{sol}}&=\int_{-\infty}^\infty dx\,e^{ikx} V_{+\infty}^\dagger \mathcal Pe^{\int_x^{\infty} dy\, Q_1\otimes \mathds{1}+R_1\otimes a^\dagger(y)}\left(V\otimes \mathds{1}+W\otimes a^\dagger(x)\right)\\
    &\hspace{4cm}{}\times \mathcal Pe^{\int_{-\infty}^x dy\, Q_2\otimes \mathds{1}+R_2\otimes a^\dagger(y)}V_{-\infty}\ket0\,.
\end{aligned}
\end{equation}
For the solitonic excitations, the momentum $k$ in \eqref{eq:rqp_sol} is ambiguous as one can rephase, say, the vacuum parametrized by $(Q_2,R_2)$ by $Q_2\rightarrow Q_2+i\alpha \mathds 1$ for $\alpha\in\mathbb R$, which is equivalent to $k\rightarrow k+\alpha$ and an overall phase. This ambiguity does not appear for the particle-like excitations \eqref{eq:rqp} because rephasing both sides simultaneously contributes an overall phase to the state.

The solitonic RQP ansatz is also invariant under a gauge transformation analogous to~\eqref{eq:tangent_gauge}
\begin{equation}\label{eq:sol_gauge}
    V\rightarrow V +XQ_2-Q_1 X +ik X\,,\qquad W \rightarrow W + XR_2-R_1X\,.
\end{equation}
Similarly, one can go to the tangent left-gauge where $V$ can be eliminated as $V=- l_1^{-1}  R_1^\dagger  l_1 W$, where one uses $R_1$ as the gauge choice referring to the left vacuum $\ket{Q_1,R_1}$. The derivation of the effective Hamiltonian in Section~\ref{sec:effHamiltonian} still applies with only minor changes to keep track of the two vacua. These modifications are necessary because $\ket{V,W;k}_{\text{sol}}$ does not look like a tangent vector (cf.\ \eqref{eq:var_rule}) above either of the two vacuum states. 

To derive the effective Hamiltonian for the solitonic RQP ansatz \eqref{eq:rqp_sol}, it will be convenient to embed the two RCMPSs in a doubled RCMPS ansatz $\ket{\mathcal Q,\mathcal R}$ with
\begin{equation}\label{RQgen1}
  \mathcal{Q} \equiv \begin{pmatrix} Q_1& 0 \\ 0 & Q_2\end{pmatrix}\,, \qquad \mathcal{R} \equiv \begin{pmatrix} R_1 & 0 \\ 0 & R_2\end{pmatrix}\,,
\end{equation}
where $\mathcal{Q},\mathcal{R}$ are now $2\chi \times 2\chi$ matrices and the corresponding boundary vectors for the RCMPS are similarly constructed as direct sums. The solitonic ansatz \eqref{eq:rqp_sol} can then be considered a particle-like RQP over the doubled vacuum $\ket{\mathcal{Q},\mathcal{R}}$. In particular, the choice of tangent vector
\begin{equation}\label{VWgen}
  \mathcal{V} \equiv \begin{pmatrix} 0 & V \\ 0 & 0 \end{pmatrix}\,, \quad
  \mathcal{W} \equiv \begin{pmatrix} 0 & W \\ 0 & 0 \end{pmatrix}\,,
\end{equation}
immediately leads to the identification
\begin{equation}\label{eq:VWequality}
\ket{\mathcal{V},\mathcal{W};k}=\ket{V,W;k}_{\text{sol}}\,.
\end{equation}
The LHS of \eqref{eq:VWequality} can be treated as a non-topological RQP as in Section \ref{sec:GenFunsec} allowing us to reuse the framework derived for particle-like excitations. In particular, one has to evaluate the generating functional \eqref{eq:ZconnFinal} using \eqref{RQgen1} and \eqref{VWgen} and simplify the formulas using the block structure of the matrices. An important caveat is subtraction of the connected part of the matrix element in \eqref{eq:cluster}. As the two vacuum states are degenerate, they agree on the subtraction as it only depends on the vacuum energy density. However, we would like to compute the effective Hamiltonian for each term in the Hamiltonian separately, but this is not a well-defined operation if the two vacua do not agree on the expectation value of each term separately. Computing the effective Hamiltonian term by term will in general give rise to divergences that cancel only once added up to the full effective Hamiltonian. Below, we will use a symmetric subtraction of the connected part which will render each term of effective Hamiltonian finite but physically meaningless when the VEVs do not agree between the left and right vacua. This ambiguity cancels in the sum of the separate terms.

Upon carrying out the above program, equations are again expressed in terms of $\chi\times \chi$ matrices and the solitonic generalization of the generating functional for the connected matrix elements $\widetilde{\cal Z}_\text{conn}[\tilde \jmath, j]$ is
 \begin{align} \label{eq:ZconnSol}
     \widetilde{\cal Z}^{\text{sol}}_\text{conn}[\tilde \jmath, j]
     &= \int dx\,  
     \biggl[ \left( {l_1}_{\tilde \jmath, j}(x) \middle| 
     W \otimes \bar W' 
     \middle| {r_2}_{\tilde \jmath, j}(x) \right)
      -  \left(l_{1} \middle|\,W \otimes \bar{W}'\,\middle|r_{2}\right) [(l_1 | {r_{1}}_{\tilde \jmath, j}(\infty))+ (l_2 | {r_{2}}_{\tilde \jmath, j}(\infty)) ]/2\notag \\
     &{}+ \left( {l_1}_{\tilde \jmath, j}(x) \middle| 
    \tilde{U}_{1}(x)
     \middle| {\widehat{r}}^{\text{sol}}_{\tilde \jmath, j}(x) \right)
     +
     \left( {\widehat{l}}^{\text{sol}}_{\tilde \jmath, j}(x) \middle| 
   \tilde{U}_{2}(x)
     \middle| {r_2}_{\tilde \jmath, j}(x) \right)
       \biggr] \,,
 \end{align}
where $\big(l_{a,\tilde \jmath, j}|$ and $\big|r_{a,\tilde\jmath, j}\big)$ refer to \eqref{eq:ljjtDef} and \eqref{eq:rjjtDef}, respectively, as computed in the state $\ket{Q_a,R_a}$ for $a=1,2$. Further, we defined the matrices
 \begin{equation}\label{eq:Udefsol}
  \begin{split}
     U_{a}(x) &\equiv V \otimes \mathds{1} 
       + W \otimes \bar R_{a} + j(x) W \otimes \mathds 1 \,,  \\
     \widetilde{U}_{a}(x) &\equiv \mathds{1} \otimes \bar V' + R_{a}\otimes \bar W' + \tilde \jmath(x) \mathds{1} \otimes \bar W' \,.  
   \end{split}          
 \end{equation}
and the generalization of $\left( \widehat{l}_{\tilde \jmath, j}(x) \right|,\left| \widehat{r}_{\tilde \jmath, j}(x) \right)$ defined by the ODEs
 \begin{equation}
  \begin{split}
     \frac{d \left| \widehat{r}^\text{sol}_{\tilde \jmath, j}(x) \right)}{dx}
      &=  U_2(x) \left| {r_2}_{\tilde \jmath, j}(x) \right)
       + \left(-ik + \mathbb{T}_{1,2}
       + j(x) R_1 \otimes \mathds 1 
       + \tilde \jmath(x) \mathds{1} \otimes \bar R_2 \right) \left| \widehat{r}^\text{sol}_{\tilde \jmath, j}(x) \right) \,, \\
    \frac{d \left( \widehat{l}^\text{sol}_{\tilde \jmath, j}(x) \right|}{dx}
      &=  -\left( {l_1}_{\tilde \jmath, j}(x) \right| U_1(x) 
       -\left( \widehat{l}^\text{sol}_{\tilde \jmath, j}(x) \right|  \left(ik + \mathbb{T}_{2,1}
       + j(x) R_2 \otimes \mathds 1 
       + \tilde \jmath(x) \mathds{1} \otimes \bar R_1 \right)  \,.
  \end{split}
 \end{equation}
and the boundary condition\footnote{The regulator $-\lambda|r)(l|(x-z)$ we used in \eqref{eq:BCrhat} is now not needed, since the mixed transfer matrix has no zero eigenvalues, rendering $(\mathbb{T}_{1,2} - i k)$ invertible for generic $k$.}
 \begin{equation}
 \begin{aligned}
     \left| \widehat{r}^\text{sol}_{\tilde \jmath, j}(-\infty) \right)
      &= -\frac{1}{\mathbb{T}_{1,2} - i k} \left. ( V \otimes \mathds{1} + W \otimes \bar R_2) 
     \middle| r_2 \right) \,,\\
     \left( \widehat{l}^{\text{sol}}_{\tilde \jmath, j}(\infty) \right|
      &= - \left(l_1 \middle| ( V \otimes \mathds{1} + W \otimes \bar R_1) 
     \right.  \frac{1}{\mathbb{T}_{2,1} + i k} 
     = 0 \,,
    \end{aligned}
 \end{equation}
with the notation for the mixed transfer matrix
\begin{equation}
    \mathbb{T}_{a,b}=Q_a \otimes \mathds 1 + \mathds 1 \otimes  \bar{Q}_b+R_a \otimes \bar{R}_b\,.
\end{equation}
To see that the symmetric subtraction of the connected part in the first line of \eqref{eq:ZconnSol} makes the integral finite notice that if one repeats the derivation of \eqref{eq:approach} for $\left( {l_1}_{\tilde \jmath, j}(x) \middle| W \otimes \bar W' \middle| {r_2}_{\tilde \jmath, j}(x) \right)$ then one finds that it is proportional to the VEV of $\ket{Q_1,R_1}$ for large positive $x$ and to the VEV of $\ket{Q_2,R_2}$ for large negative $x$. Thus, subtracting the average of the VEVs in~\eqref{eq:ZconnSol} correctly cancels the divergence when the integral is done symmetrically about $x=0$.

From \eqref{eq:ZconnSol}, one can specialize to the necessary operators \eqref{eq:opList} as done for the particle-like RQP in Section \ref{sec:effHamiltonian}.

\section{Applications}\label{sec:application}
In this section, we will apply the RCMPS and RQP frameworks developed in the previous section to the sine-Gordon and $\phi^4$ models \eqref{eq:potentials}. To extract the excitation spectrum of a scalar field theory, we first need to solve for the RCMPS representing the vacuum and then  solve for the RQP energies by finding the low-lying eigenvalues of the effective Hamiltonian. The vacuum state is determined by minimizing the expectation value of the Hamiltonian density, as computed in Section \ref{sec:expvalue}, over the space of RCMPSs. This is a non-linear optimization problem, which we solve using a gradient-descent-type solver. Once the vacuum RCMPS is known, the effective Hamiltonian can be constructed as in Section~\ref{sec:effHamiltonian}, and we can then solve for its low-lying eigenvalues. We give further details on the numerical implementation in Appendix~\ref{app:num_impl}.

\subsection{Sine-Gordon model}\label{subsec:sG}

\subsubsection{Definition and excitation spectrum}

\begin{figure}[t]
        \centering
        \includegraphics[width=4in]{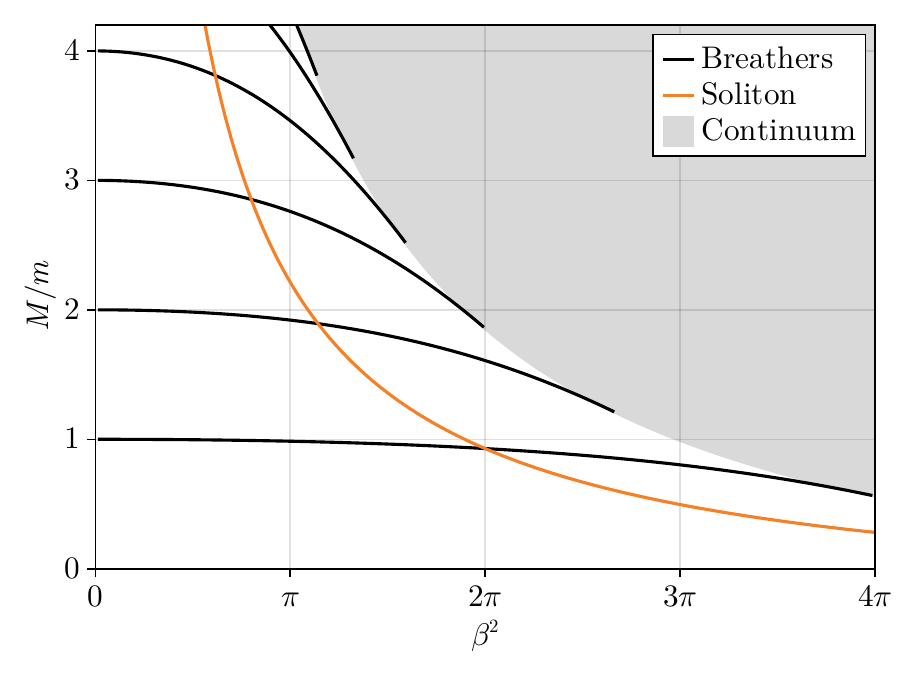}
        \caption{Exact spectrum of excitations for the sine-Gordon model obtained from integrability. The orange curve indicates the mass of the soliton \eqref{eq:soliton_mass}, and the shaded region denotes the 2-soliton continuum. The black curves indicate the breather masses in \eqref{eq:particle_mass}.  The breathers  cease to exist as asymptotic particles when their masses reach the soliton–antisoliton threshold.}
        \label{fig:sG_exact}
\end{figure}

The sine-Gordon model is a real scalar theory with a rich spectrum of particle and solitonic excitations, which furthermore is integrable \cite{Zamolodchikov:1978xm}. Hence, the sine-Gordon model serves as a highly non-trivial benchmark of our numerical setup. The Hamiltonian density is
\begin{equation}\label{eq:sG_ham}
    \mathcal H_\text{sG} = N_m\left[\frac{1}{2}\pi^2+\frac{1}{2}(\partial_x\phi)^2 +\frac{m^2}{\beta^2}(1-\cos(\beta\phi))\right]\,,
\end{equation}
where, relative to \eqref{eq:HamExamples}, we took the normal ordering mass to be $\mu=m$. Despite the interaction potential not being polynomial, the Hamiltonian \eqref{eq:sG_ham} is still UV finite for $\beta^2<4\pi$ \cite{Frohlich:1975rcn}; further renormalization is required for $ \beta^2\geq4\pi$~\cite{Dimock:1993qw}. In this work, we restrict ourselves to $\beta^2<4\pi$.
In this range, the spectrum of excitations consists of a soliton and an anti-soliton, also called kink and anti-kink, with equal masses given by \cite{Zamolodchikov:1995xk}
\begin{equation}\label{eq:soliton_mass}
    \frac{M_\text{sol}}{m} = \frac{2\pi^{\frac{\xi(\beta)}{2}}\Gamma\left(\frac{\xi(\beta)}{2}\right)}{\Gamma\left(\frac{1+\xi(\beta)}{2}\right)}\left(\frac{\mu(\beta)\Gamma\left(1-\frac{\beta^2}{8\pi}\right)}{\Gamma\left(\frac{\beta^2}{8\pi}\right)}\right)^{\frac{4\pi}{8\pi-\beta^2}}\,,\qquad\text{with}\qquad\begin{aligned}
        \xi(\beta)&=\frac{\beta^2}{8\pi-\beta^2}\,,\\
        \mu(\beta)&= \frac{\exp\left(-\frac{\gamma \beta^2}{4\pi}\right)}{2^{1-\frac{\beta^2}{4\pi}}\beta^2}\,,
    \end{aligned}
\end{equation}
where $\gamma$ is the Euler-Mascheroni constant, as well as a number of stable particles known as breathers.\footnote{The quantity $\mu(\beta)$ in \eqref{eq:soliton_mass} should not be confused with the normal ordering mass used in previous sections.}   The relationship between the masses of the breathers and the mass of the soliton was first determined by a semi-classical analysis in \cite{Dashen:1975hd} and later confirmed to be exact using integrability \cite{Zamolodchikov:1978xm}:
\begin{equation}\label{eq:particle_mass}
    M_n= 2\sin \left(\frac{\pi n \xi(\beta)}{2}\right)M_\text{sol}\,,\qquad n =1,2,\ldots,n_\text{max}\,,
\end{equation}
with $n_\text{max}$ the largest integer such that $n_\text{max}< \frac{1}{\xi(\beta)}$. Roughly speaking, the solitons can be interpreted as interpolating between two adjacent equivalent classical minima $\phi_\text{min} \in \frac{2\pi \mathbb{Z}}{\beta}$ of the interaction potential, and the breathers can be thought of as soliton-anti-soliton bound states.\footnote{Normal-ordering respects the shift symmetry $\phi \rightarrow\phi +\frac{2 \pi \mathbb{Z}}{\beta}$ of the bare Hamiltonian.} In the weak coupling limit $\beta^2\ll 1$, the particle spectrum contains arbitrarily many stable particle states with masses $M_n= nm+\mathcal O(\beta^2)$ that can be interpreted as loosely bound states of $n$ elementary quanta of the free $\beta^2=0$ theory. The exact spectrum of solitons and breathers is shown in Figure~\ref{fig:sG_exact}. 

\begin{figure}[t]
    \centering
    \begin{subfigure}[b]{0.48\textwidth}
        \centering
        \includegraphics[width=3in]{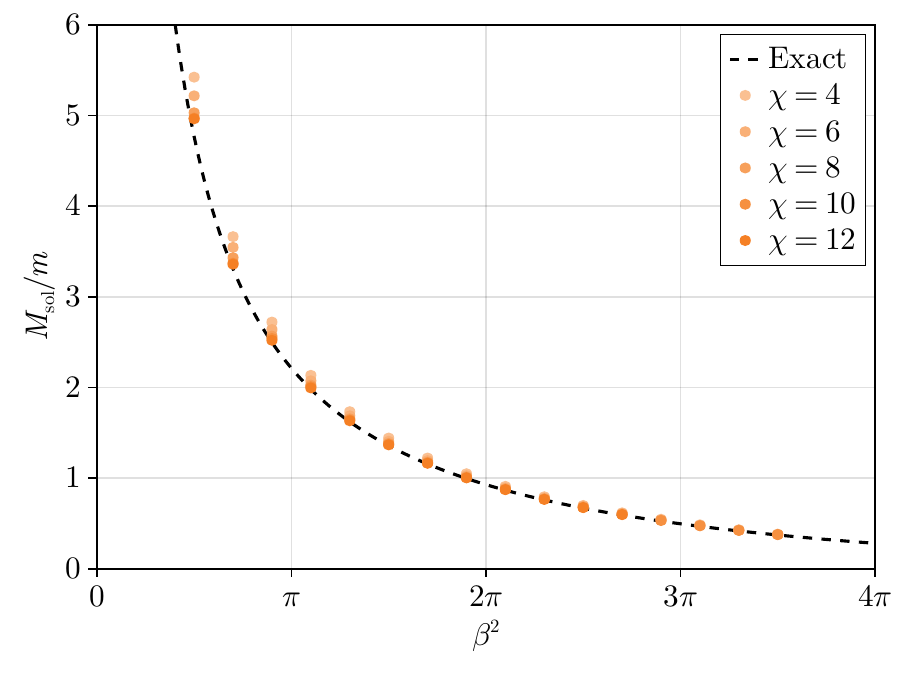}
        \caption{}
        \label{fig:kink_beta}
    \end{subfigure}
    \hfill 
    \begin{subfigure}[b]{0.48\textwidth}
        \centering
        \includegraphics[width=3in]{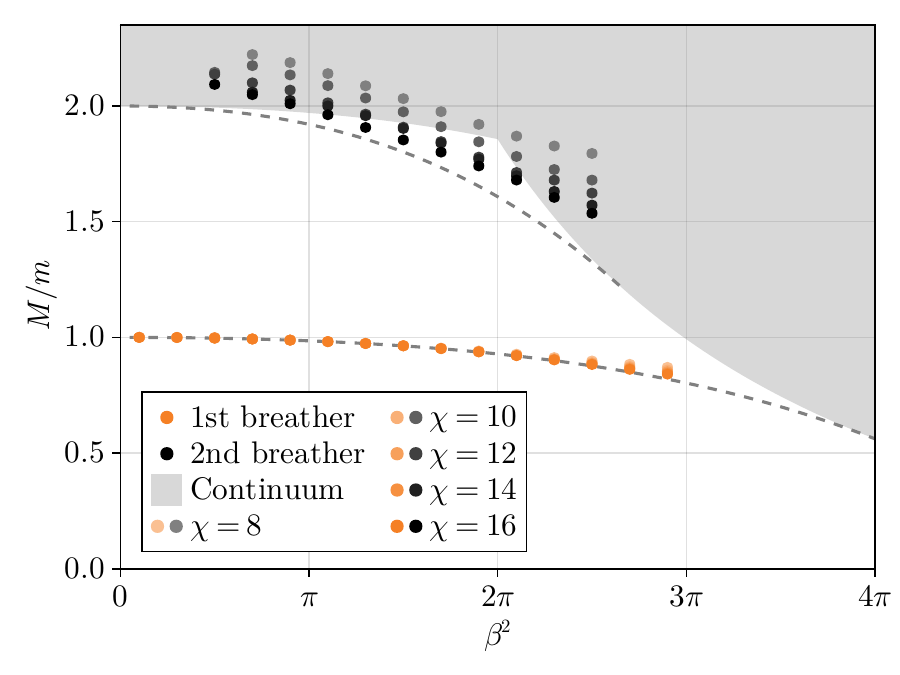}
        \caption{}
        \label{fig:particle_beta}
    \end{subfigure}
    \caption{Mass of the soliton (Figure \ref{fig:kink_beta}) and mass of the two lowest breathers (Figure~\ref{fig:particle_beta}) for the sine-Gordon model as a function of the coupling constant $\beta^2$. The exact results from integrability \eqref{eq:soliton_mass} and \eqref{eq:particle_mass} are plotted in dashed gray.  In either case, we observe that the numerical data approaches the exact results as the bond dimension $\chi$ is increased. The shaded area includes the two-particle continuum as well as the 2-soliton continuum also drawn in Figure~\ref{fig:sG_exact}.} 
    \label{fig:kink_fig}
\end{figure}

\subsubsection{Numerical results}
The ground state properties of the sine-Gordon model were studied using the RCMPS ansatz in \cite{Tilloy:2022kcn}, where excellent agreement was found with integrability results for $\beta^2\lesssim 2\pi$ already for modest bond dimensions. We now  extend this study to the excitations of the ground state using the RQP ansatz introduced in Section~\ref{subsec:rqp}. 

\begin{figure}
    \centering
    \includegraphics[width=4in]
    {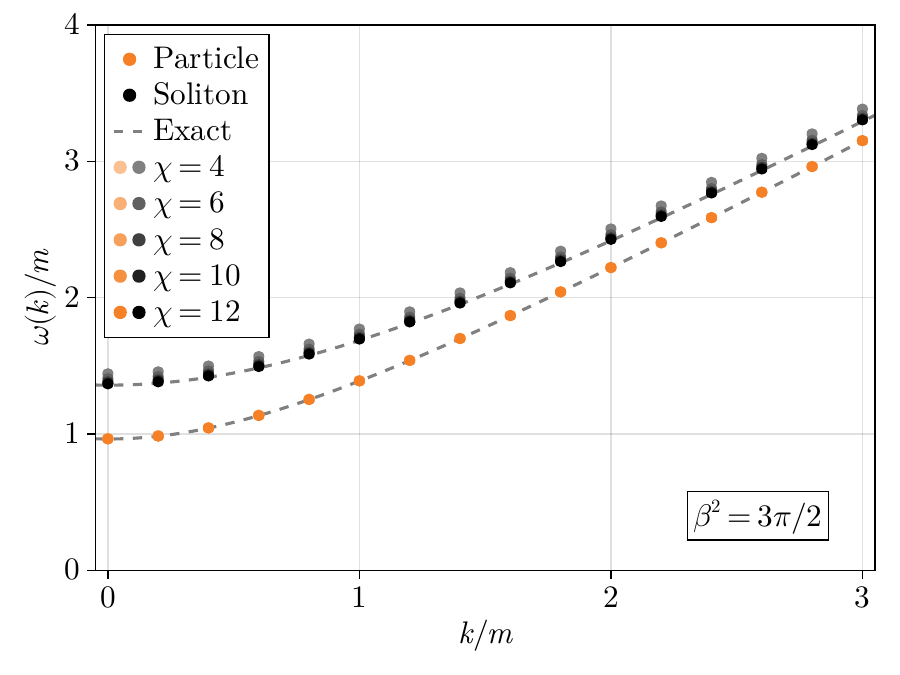}
    \caption{Dispersion relation of the soliton and of the lowest particle for the sine-Gordon model at $\beta^2=\frac{3\pi}{2}$. The exact results plotted in dashed gray come from combining the integrability results \eqref{eq:soliton_mass} and \eqref{eq:particle_mass} with the relativistic dispersion $\omega(k)=\sqrt{M^2+k^2}$. The numerical results converge to the exact ones as the bond dimension is increased.}
    \label{fig:kink_disp}
\end{figure}

In Figure \ref{fig:kink_fig}, we show the numerical data for the solitonic and particle-like excitations in the sine-Gordon model as determined by the RQP ansatz. In Figures~\ref{fig:kink_beta} and~\ref{fig:particle_beta}, we plot the soliton and particle masses, respectively, together with the analytical results \eqref{eq:soliton_mass} and \eqref{eq:particle_mass} for various values of the coupling. The data for the soliton and first breather show good agreement with the integrability results. The figure also demonstrates that the RQP ansatz approximates the exact excited states more accurately as the bond dimension is increased. The second breather is poorly captured at the bond dimensions considered here, and, for large ranges of $\beta^2$, it is indistinguishable from the two-particle continuum (the shaded region for $\beta^2<2\pi$). However, one can check that the second breather mass does tend to the exact result as the bond dimension is increased.

Figure~\ref{fig:kink_disp} shows the numerically determined dispersion relation for the soliton and the lowest breather at $\beta^2=3\pi/2$, along with the relativistic energy formula $\omega(k)=\sqrt{M^2+k^2}$ that follows from Lorentz invariance, with the masses computed from integrability using \eqref{eq:soliton_mass} and \eqref{eq:particle_mass}. In Section~\ref{sec:large_mom}, we showed that a generic RQP state satisfies the relativistic dispersion $\omega(k)\sim |k|$ for large momenta $k$. The data in Figure~\ref{fig:kink_disp} demonstrate that the RQP ansatz not only recovers the relativistic dispersion at large momenta, but also at small momenta as the bond dimension is increased. This relativistic dispersion emerges despite neither the RCMPS nor the RQP ansatz being Lorentz invariant/covariant constructs. Figure~\ref{fig:kink_disp} therefore serves as important evidence that the RCMPS framework is able to reliably model relativistic physics.

\begin{figure}
    \centering
    \includegraphics[width=4in]{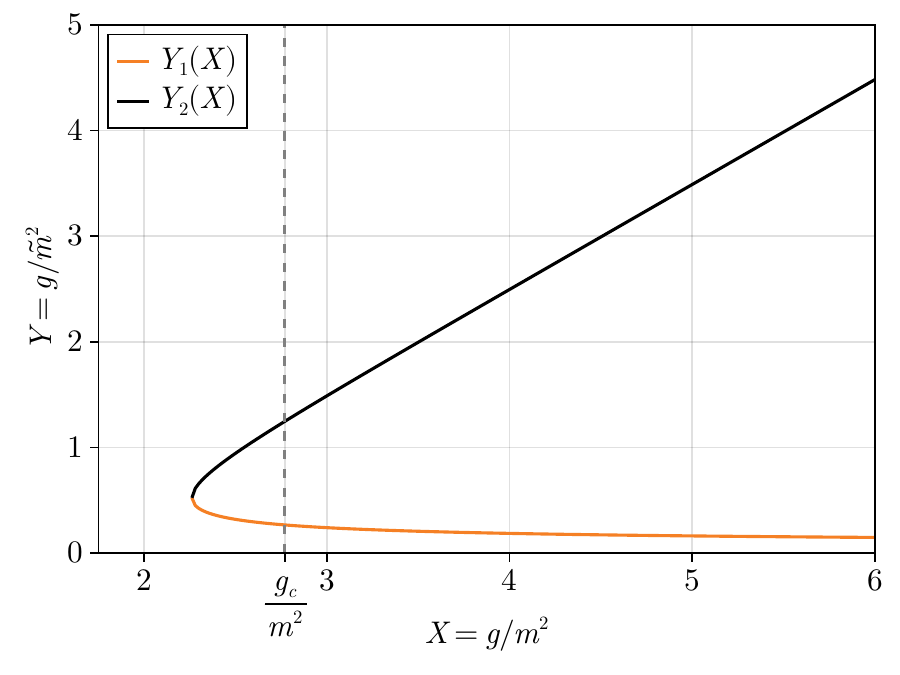}
    \caption{Plot of the relation \eqref{eq:logeq} with the strong/weak branch in orange and the strong/strong branch in black. The vertical gray dashed line indicates the critical value of the coupling above which there is spontaneous $\Z_2$ symmetry breaking.  No real solutions of \eqref{eq:logeq} exist for $X< X_*\approx2.26$.}
    \label{fig:changdual}
\end{figure}
\subsection{\texorpdfstring{$\phi^4$}{ϕ⁴} theory}\label{subsec:phi4rqp}
We now turn to the $\phi^4$ theory, which, despite its simplicity, is not integrable and exhibits an array of interesting physical phenomena. The $\phi^4$ theory is defined as a free massive scalar of mass $m$ deformed by a quartic self-coupling of strength $g$, with Hamiltonian density
\begin{equation}\label{eq:phi4_ham}
    \mathcal H_{\phi^4} = N_m\left[\frac{1}{2}\pi^2+\frac{1}{2}(\partial_x\phi)^2 + \frac{1}{2}m^2\phi^2+g\phi^4\right]\,,
\end{equation}
where we again have taken the normal ordering mass to be $\mu=m$ (see \eqref{eq:HamExamples}). The Hamiltonian density \eqref{eq:phi4_ham} has an internal $\mathbb Z_2$ symmetry $\phi \rightarrow -\phi$, which is preserved by the normal-ordering procedure. Even though we will only consider $m^2>0$, the symmetry breaks spontaneously at some $g_c>0$ because the normal-ordering of the quartic term effectively renormalizes the mass term. The critical value of the coupling has been studied in many works, such as \cite{Loinaz:1997az,Vanhecke:2021noi}, and it is approximately ${g_c}\approx 2.77{m^2}$ in the scheme defined by \eqref{eq:phi4_ham} (see \cite{Durr:2025hfu} for a recent study with a comparison of the estimates in the literature). The quantum phase transition at $g=g_c$ is in the Ising universality class.

In the perturbative regime $g\ll m^2$, the $\phi^4$ interaction is repulsive and the spectrum consists of only the fundamental particle created by $\phi$. This particle remains the only stable particle in the symmetry preserving phase $0\leq g < g_c$ and its mass decreases from $m$ at $g=0$ to zero at the critical coupling $g = g_c$.

To understand the spectrum in the $\mathbb Z_2$-broken phase, for $g>g_c$, it is convenient to make use of the strong/weak duality transformation known as Chang duality \cite{Chang:1976ek}. The transformation is derived by changing the normal-ordering scale in the Hamiltonian \eqref{eq:phi4_ham} to a mass scale $\widetilde m$ such that the classical potential exhibits spontaneous symmetry breaking. Furthermore, the mass $\widetilde m$ is chosen such that the excitations above these classical minima have classical mass equal to $\widetilde m$. As we will see, the correct Hamiltonian density to consider is \cite{Chang:1976ek}
\begin{equation}\label{eq:phi4_ham_M}
    \mathcal H_{\phi^4} = N_{\widetilde m}\left[\frac{1}{2}\pi^2+\frac{1}{2}(\partial_x\phi)^2 - \frac{1}{4}\widetilde m^2\phi^2+g\phi^4\right]\,,
\end{equation}
which for a proper choice of $\widetilde{m}$ equals the original Hamiltonian density \eqref{eq:phi4_ham} up to an additive constant, as we will see. Note that in \eqref{eq:phi4_ham_M}, the mass term is negative, giving rise to two symmetry-broken classical minima, and the precise coefficient of the $\phi^2$ term is chosen such that it gives the correct normalizations of the mass term for the fluctuations around these minima. Indeed, the two minima are at $\phi=\pm\frac{\widetilde m}{\sqrt{8g}}$, and, if we write $\phi = \tilde \phi \pm \frac{\widetilde m}{\sqrt{8g}}$, the Hamiltonian density \eqref{eq:phi4_ham_M} takes the form
\begin{equation}\label{eq:phi3}
    \mathcal{H}_{\phi^4}=N_{\widetilde m}\left[\frac{1}{2}\tilde\pi^2+\frac{1}{2}(\partial_x\tilde \phi)^2 +\frac{1}{2}\widetilde m^2\tilde \phi^2\pm \sqrt{2g}\,\widetilde m \tilde \phi^3+g\tilde\phi^4\right]\,,
\end{equation}
where we neglected an additive constant shift in the energy. Note that the negative mass term $-\frac{1}{4}\widetilde m^2\phi^2$ in \eqref{eq:phi4_ham_M} results in the canonical mass term $+\frac{1}{2}\widetilde m^2\tilde \phi^2$ in \eqref{eq:phi3}.

To determine the $\widetilde m$ that makes \eqref{eq:phi4_ham} and \eqref{eq:phi4_ham_M} equal, recall the following re-normal-ordering identities 
\begin{equation}\label{eq:renormal}
\begin{aligned}
    N_m\left[\frac{1}{2}\pi^2+\frac{1}{2}(\partial_x\phi)^2\right]&=N_{\widetilde m}\left[\frac{1}{2}\pi^2+\frac{1}{2}(\partial_x\phi)^2\right]+\text{const.}\,,\\
    N_m[\phi^2]&=N_{\widetilde m}[\phi^2]+\text{const.}\,,\\
    N_m[\phi^4]&=N_{\widetilde m}\left[\phi^4+\frac{3}{2 \pi}\log\left(\frac{m^2}{\widetilde m^2}\right)\phi^2\right]+\text{const.}\,.
\end{aligned}
\end{equation}
Equating the Hamiltonian densities \eqref{eq:phi4_ham} and \eqref{eq:phi4_ham_M} (up to a constant shift) and using \eqref{eq:renormal} yields the relation
\begin{equation}\label{eq:logeq}
    \frac{6}{\pi}\log(Y)+\frac{1}{Y}=\frac{6 }{\pi}\log(X)-\frac{2}{X}\,,
\end{equation}
where we defined the effective dimensionless couplings $X\equiv\frac{g}{m^2}$ and $Y\equiv\frac{g}{\widetilde m^2}$. The relation \eqref{eq:logeq} is plotted in Figure~\ref{fig:changdual}.
For $X
> X_*\approx2.26$, \eqref{eq:logeq} has two real solutions, $Y_1(X)$ and $Y_2(X)$, with $Y_1(X) \leq Y_2(X)$.\footnote{The explicit form of the two branches can be written in terms of the Lambert $W$-function
\begin{equation*}
    Y_1(X)=-\frac{\pi }{6 W_{-1}\left(-\frac{\pi  }{6
   X}\exp{\frac{\pi }{3 X}}\right)}\,,\qquad Y_2(X)=-\frac{\pi }{6 W_{0}\left(-\frac{\pi }{6X}\exp{\frac{\pi }{3 X}}\right)}\,.
\end{equation*}} As demonstrated in Figure~\ref{fig:changdual}, the lower branch $Y_1(X)$ sends large dimensionless coupling $X$ to small $Y$, corresponding to a strong/weak duality. At large $X$, this solution has the leading strong-coupling behavior
\begin{equation}
    Y_1(X) \approx \frac{\pi}{6}\frac{1}{\log X}\,.
\end{equation}
Furthermore, since $X_*< \frac{g_c}{m^2}$, the dual formulation \eqref{eq:phi4_ham_M} can also access part of the symmetry preserving phase. The upper branch $Y_2(X)$ sends strong coupling to strong coupling and is not of interest for our purposes, so in the following discussion we take $Y = Y_1(X)$.

\begin{figure}[t]
    \centering
        \includegraphics[width=4in]{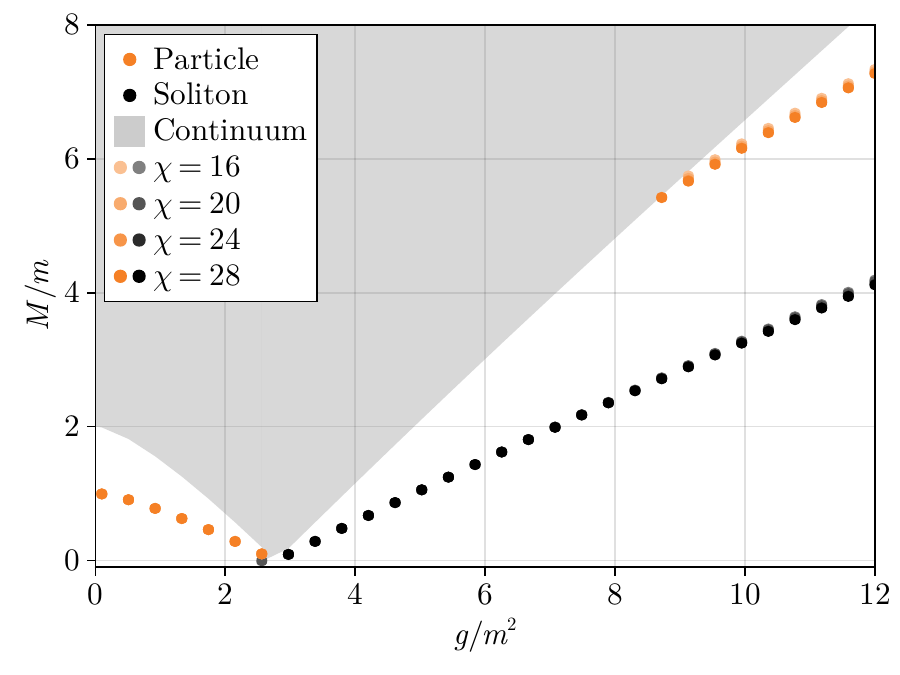}
        \caption{Excitation spectrum of the $\phi^4$ theory obtained using the Hamiltonian density \eqref{eq:phi4_ham}.  The masses of the particle excitations are plotted in orange, and the soliton masses are plotted in black.    The multi-particle continuum is drawn as a shaded gray area and represents the two-particle continuum below the phase transition at $g_c/m^2 \approx 2.77$ and the two-soliton continuum above this transition.}
        \label{fig:phi4}
\end{figure}

The strong coupling region $X=\frac{g}{m^2}\gg1$ can therefore be probed perturbatively using~\eqref{eq:phi3}, with $Y=\frac{g}{\widetilde m^2}\ll1$ obtained as the lower branch solution of \eqref{eq:logeq}.  In particular, the cubic interaction is attractive and, for sufficiently small $Y$, it can dominate over the quartic interaction and give rise to bound states. In \cite{Bajnok:2015bgw, Mussardo:2006iv}, it was argued using semi-classics  that a second stable particle (a bound state) indeed exists at large $X$.  The Hamiltonian truncation studies \cite{Coser:2014lla,Rychkov:2015vap,Bajnok:2015bgw} provide strong evidence for this scenario.  As mentioned above, the $\mathbb Z_2$ symmetry is broken for $g>g_c$, and, in addition to the particle spectrum of either of the two equivalent vacua, one can consider solitons interpolating between these vacua.  The lowest stable particle can be thought of as a bound state of the soliton and anti-soliton.

One can compute the masses of the lowest particle and of the bound state of two such particles in perturbation theory. The correction to the mass $M_1$ of the lowest particle can be determined by considering the self-energy diagrams of the elementary excitation, which was done in \cite{Serone:2019szm} up to four loops, giving
\begin{equation}\label{eq:m1_pert}
    \frac{M_1}{\widetilde m} = 1- \sqrt{3}Y-3.5765(9)Y^2-13.638(15)Y^3-55.3(5)Y^4+\mathcal{O}(Y^5)\,,
\end{equation}
where the uncertainties are due to Monte Carlo integration. The bound state mass $M_2$ can be estimated by considering $2\rightarrow2$ scattering and matching the amplitude to the Born approximation to determine an effective interaction potential in a non-relativistic approximation. The leading correction comes from the tree-level amplitude
\begin{equation}\label{eq:amp1}
	i\mathcal M_\text{tree}=\;\begin{tikzpicture}[baseline={(0.5,0.43)}]
		\node[circle, fill, inner sep=1.pt] at (0.5,0.5) {};
		\draw[thick] (0,0) -- (1,1) (0,1) -- (1,0);
	\end{tikzpicture} \; + \;
	\begin{tikzpicture}[baseline={(0.5,0.45)}]
		\node[circle, fill, inner sep=1.pt] at (0.33,0.5) {};
		\node[circle, fill, inner sep=1.pt] at (0.66,0.5) {};
		\draw[thick] (0,1) -- (0.33 ,0.5) -- (0.66,0.5) -- (1,1) (0,0) -- (0.33,0.5)  (0.66,0.5) -- (1,0);
	\end{tikzpicture}\; + \;
	\begin{tikzpicture}[baseline={(0.5,0.45)}]
		\node[circle, fill, inner sep=1.pt] at (0.5,0.66) {};
		\node[circle, fill, inner sep=1.pt] at (0.5,0.33) {};
		\draw[thick] (0,1) -- (0.5 ,0.66) -- (1,1) (0,0) -- (0.5,0.33) -- (1,0) (0.5,0.33) -- (0.5,0.66);
	\end{tikzpicture}\; + \;
	\begin{tikzpicture}[baseline={(0.5,0.45)}]
		\node[circle, fill, inner sep=1.pt] at (0.5,0.66) {};
		\node[circle, fill, inner sep=1.pt] at (0.5,0.33) {};
		\draw[thick] (0,1) -- (0.5 ,0.66) -- (1,0) (0,0) -- (0.5,0.33) -- (1,1) (0.5,0.33) -- (0.5,0.66);
	\end{tikzpicture}\; = 96ig + \mathcal O\left(\frac{p^2}{\widetilde m^2}\right)\,.
\end{equation}
This amplitude can be matched to a delta-function interaction potential yielding the following non-relativistic center-of-momentum Hamiltonian (see Appendix~\ref{sec:EffQM} for details)
\begin{equation}\label{eq:eff_ham}
    H_\text{NR} = \frac{ p^2}{\widetilde m} -\frac{12 g}{\widetilde m^2}\delta(x)\,.
\end{equation}
The ground state energy of the effective quantum mechanics problem is $E_\text{GS} =- 36 Y^2 \widetilde{m}$.  The mass $M_2$ of the bound state can be found by adding the binding energy $E_{\text{GS}}$ to the mass $2M_1$ of the constituents, yielding
\begin{equation}\label{eq:m2_pert}
    \frac{M_2}{\widetilde m} = 2 - 2\sqrt{3}Y- 43.1529(18) Y^2 +\mathcal O(Y^3)\,.
\end{equation}

\subsubsection{Numerical results}

In Figure~\ref{fig:phi4}, we show our results for the spectrum of excitations in the formulation \eqref{eq:phi4_ham}. As expected, for $g<g_c$ we find a single stable particle below the two-particle continuum. As the coupling is tuned to the critical value, the mass gap closes, and immediately beyond the critical point the only stable excitation is the soliton. Linearly extrapolating the $\chi=28$ data points closest to the phase transition to vanishing mass, we obtain the estimate $g_c/m^2\approx2.78$ for the critical coupling.  This value is in agreement with the literature \cite{Vanhecke:2021noi}.  Furthermore, at $\frac{g}{m^2}\approx 8.7$, a stable particle emerges from the two-soliton continuum.

Note that the slope of the particle mass just below the transition and the slope of the soliton mass just above the transition are equal and opposite. To understand this, we start by noticing that the low-energy physics near the critical point $g_c$ is described by the 2d Ising model deformed by the energy operator $\epsilon$, namely $S=S_\text{Ising} +\lambda \int d^2 x\, \epsilon$ with $\lambda \sim (g - g_c)$. By fermionization, this effective description is equivalent to the free Majorana CFT deformed by a mass term, $S=S_\text{Majorana} +\frac{\lambda}{2} \int d^2 x\, \bar\psi\psi$. Thus, the mass gap on either side of the phase transition is due to the massive Majorana fermion and is proportional to $|\lambda|\sim|g - g_c|$. In the bosonic description, the fermionic particle excitation is represented as a particle for $\lambda<0$ and as a soliton for $\lambda>0$.

Near the critical coupling, our numerical results yield solitons of negative mass for small values of the bond dimension $\chi$. This phenomenon can be understood as follows.  For finite bond dimension, the symmetry properties of the RCMPS of lowest energy denisty need not coincide with those of the physical vacuum. In particular, as the critical coupling is approached from the symmetry-preserving side $g<g_c$, the optimal RCMPS of bond dimension $\chi$ breaks the $\mathbb Z_2$ symmetry at a value of the coupling $g_{c,\chi}$ less than the physical $g_c$ \cite{Tilloy:2021yre}. Moreover, as the bond dimension is increased, the observed critical coupling $g_{c,\chi}$ converges to the physical value. The negative soliton mass is a signal that the optimal RCMPS is not the minimum energy state in the full Hilbert space.
\begin{figure}[t]
    \centering
        \includegraphics[width=4in]{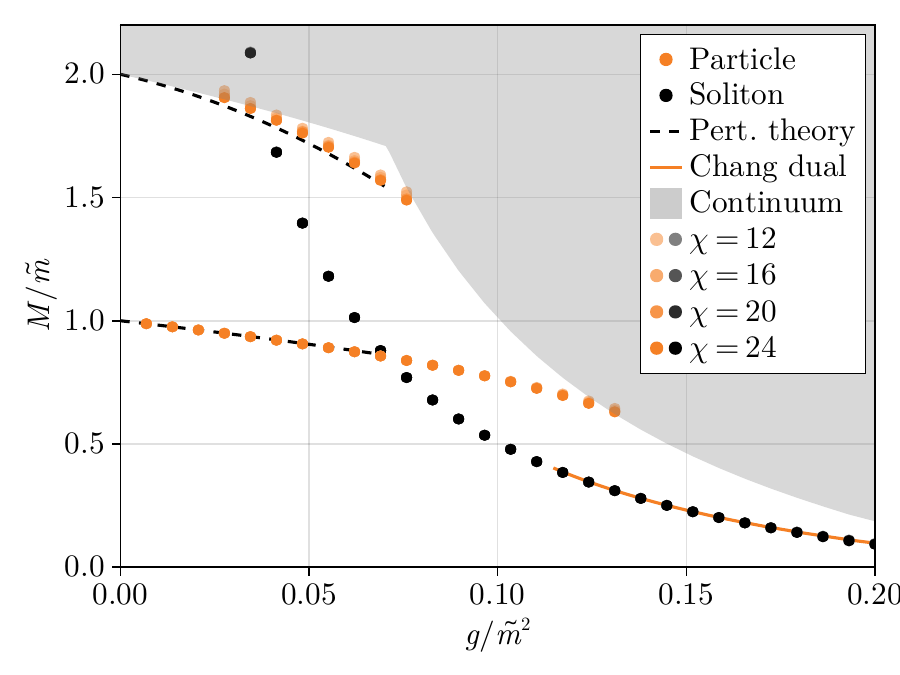}
        \caption{Excitation spectrum of the Hamiltonian \eqref{eq:phi4_ham_M}, with the particle masses in orange and the soliton masses in black. The continuum in the particle sector is shaded in gray, with the threshold on the left side of the plot due to two-particle decay and on the right side due to two-soliton decay. The solid orange line is the soliton mass as determined using \eqref{eq:phi4_ham} at $\chi=28$. The black dashed lines are the perturbative results \eqref{eq:m1_pert} and \eqref{eq:m2_pert}. Note that the quantities are in units of $\widetilde m$ rather than $m$ (see \eqref{eq:mscale}).}
        \label{fig:phi4_chang}
\end{figure}

As mentioned above, in order to study the strong coupling limit of the Hamiltonian density \eqref{eq:phi4_ham}, it is convenient to use Chang duality to map our theory to the weakly-coupled description with Hamiltonian density \eqref{eq:phi4_ham_M}. Recall that 
\begin{equation}\label{eq:mscale}
    \frac{m}{\widetilde m}= \sqrt{\frac{Y_1(X)}{X}}\,,
\end{equation}  
with $X = \frac{g}{m^2}$, $Y = \frac{g}{\widetilde{m}^2}$, and $Y_1(X)$ is the lower branch of solutions of \eqref{eq:logeq}. In Figure~\ref{fig:phi4_chang}, we show the masses of the lowest two particles and of the soliton obtained by using the description \eqref{eq:phi4_ham_M} for couplings $g/\tilde{m}^2$ corresponding to the symmetry broken phase. The main features are as follows.
 \begin{itemize}
  \item In the entire range of couplings, we see only one soliton state, whose mass is represented by black dots.  In particular, the would-be mass of the second soliton extracted from the RQP falls in the continuum of the lowest soliton and lowest particle and appears to converge to the threshold as the bond dimension is increased.  The soliton masses derived from two duality frames \eqref{eq:phi4_ham} and \eqref{eq:phi4_ham_M} are in excellent agreement with each other.  Indeed, for an easy comparison of our results across the duality, in Figure~\ref{fig:phi4}, we show the soliton masses computed using \eqref{eq:phi4_ham} in solid orange (the corresponding data points are also shown as black dots in Figure~\ref{fig:phi4}). 

  \item In the range $0.128 \lsim \frac{g}{\widetilde m ^2} \lsim 0.26 \approx \frac{g_c}{\tilde m^2}$ (corresponding to $\frac{g_c}{m^2}\approx 2.77 \lsim \frac{g}{m^2} \lsim 8.6$) there is no stable particle.  This range agrees with the results obtained using the original duality frame \eqref{eq:phi4_ham}, for which our results are plotted in Figure~\ref{fig:phi4}.

  \item In the range $0.079 \lsim \frac{g}{\widetilde m ^2} \lsim 0.128$ (corresponding to $8.6 \lsim \frac{g}{ m ^2}\lsim 61$), there is only one stable particle, whose masses are shown as orange dots in Figure~\ref{fig:phi4_chang}.

  \item In the range $ \frac{g}{\widetilde m ^2} \lsim 0.079$ (corresponding to $\frac{g}{ m ^2}\gsim 61$) there are two stable particles, whose masses are shown as orange dots in Figure~\ref{fig:phi4_chang}.  In this range, we can compare our numerical results with the perturbative expansions \eqref{eq:m1_pert} and \eqref{eq:m2_pert} for the masses of these particles. For the mass of the lowest particle, we find excellent agreement with the perturbative result \eqref{eq:m1_pert}. The data for the second particle is not fully converged in bond dimension, but it shows reasonable agreement with the perturbative bound state mass \eqref{eq:m2_pert}. The RQP ansatz data clearly distinguishes the second stable particle from the continuum, which was hard to resolve in the earlier works \cite{Coser:2014lla,Rychkov:2015vap} and is in good agreement with the results obtained in \cite{Bajnok:2015bgw}.

 \end{itemize}

\begin{figure}[t]
    \centering
        \includegraphics[width=4in]{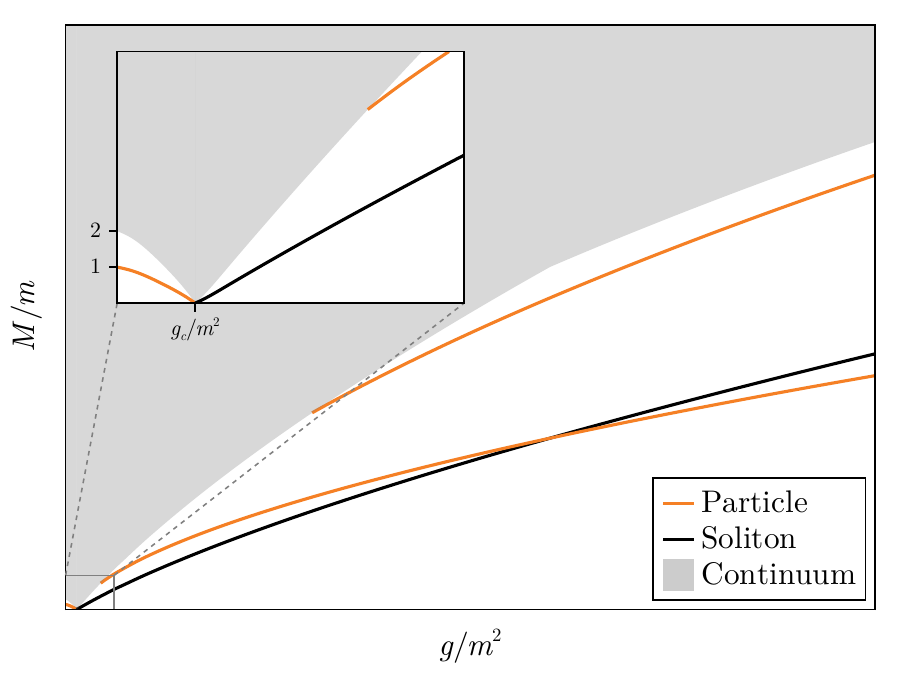}
        \caption{A sketch of the spectrum of excitations in $\phi^4$ theory, plotted in terms of $g/m^2$, with particles in orange and solitons in black. The continuum shaded in gray is due to decay into solitons between teh critical point $g_c /m^2$ and the level crossing between the lowest particle and the lowest soliton, and it is due to decay into particles above that level crossing or below $g_c/m^2$.
        }
        \label{fig:phi4_sketch}
\end{figure}

Finally, we sketch the full spectrum of particles and solitons in Figure \ref{fig:phi4_sketch} for all values of the coupling using the units of the original Hamiltonian \eqref{eq:phi4_ham}. Notice that for large $\frac{g}{m^2}$ the gap between the second particle and the continuum in units of $m$ grows as $|E_\text{GS}|/m = 36Y_1(X)^{3/2}X^{1/2}\sim \frac{X^{1/2}}{\log(X)^{3/2}}$.

\section{Discussion}

In this paper, we developed the framework for the RQP ansatz for relativistic field theories in $(1+1)$d.  This framework allows one to determine not just the spectrum of low-lying particle excitations, but also of solitons between distinct degenerate vacua. Building on the RCMPS framework, the RQP ansatz is defined directly in the continuum, at infinite volume, and without an explicit UV cutoff. We presented the necessary formalism for an efficient numerical implementation of the RQP ansatz, and we showed that the large momentum limit of the ansatz reproduces an ultra-relativistic dispersion. To demonstrate the validity of the RQP ansatz, we studied the sine-Gordon model and numerically reproduced the breather and soliton masses along with their relativistic dispersions. Finally, we studied the $\phi^4$ theory and determined the particle and soliton spectrum for all values of the coupling.

There are various future directions that are worth mentioning.  An interesting scalar field theory that fits within the framework developed in this paper is the Landau-Ginzburg description of the tri-critical Ising model deformed by some relevant operator in terms of a sextic interaction potential. In \cite{Cordova:2024vsq}, it was argued that certain deformations of the tri-critical Ising model give rise to degenerate vacua exchanged by non-invertible symmetries. Furthermore, for certain deformations, the non-invertible symmetries group the soliton (and anti-soliton) between the vacua and a particle above one of the vacua into a degenerate multiplet. Using the RQP ansatz to study a sextic scalar field theory, one could directly probe this multiplet. However, in order to preserve non-invertible symmetry, one would need to simultaneously fine-tune multiple parameters.

The RQP ansatz can naturally be adapted to other studies using RCMPS\@. In particular, one could study the excitation spectrum in the multi-scalar field theories of \cite{Tiwana:2025zhj}. With multiple species, it is necessary to impose further regularity conditions on the RCMPS ansatz to obtain a finite energy density. These bosonic regularity conditions were studied in \cite{Tiwana:2025zhj} and also extended to a characterization at the level of the tangent space. It would be interesting to understand how the tangent space restriction relates to the finiteness of the quasiparticle excitation energy. Similarly, one could also study an RQP based on RCMPSs for Dirac fermions, which effectively has two species of fermions for which one also needs regularity conditions to yield finite energy density  \cite{Haegeman:2012uk}.

Another avenue is to adapt methods from discrete MPS that have not previously been studied using CMPS\@. One such MPS construction is the two-particle scattering ansatz of~\cite{Vanderstraeten:2013xda}, which allowed the authors to directly study the S-matrix for 2-to-2 scattering of the quasiparticles. The two-particle ansatz is constructed by summing over two local perturbations of the vacuum state with distinct variational parameters as a function of the distance between the insertions.  To construct an RCMPS analog of this ansatz, one would, in particular, need to parameterize these variational parameters as a function of continuous distance, unlike for MPS where the lattice offers a natural discretization of the distance.

\section*{Acknowledgments}

We thank Antoine Tilloy for useful discussions.  This work was supported in part by the Simons Foundation Grant No.~917464 (Simons Collaboration on Confinement and QCD Strings) and by the US Department of Energy under Award No.~DE-SC0007968.

\appendix

\section{Formulas for RCMPS}
In this appendix, we provide proofs of various identities utilized in Section \ref{sec:rcmps}.
\subsection{Overlap of RCMPS ansätze}
\label{app:Norm}

In this Appendix we derive the overlap formula \eqref{eq:norm_fin} for the states \eqref{eq:RCMPSxDep}.  Note that under the path ordering symbol we can freely change the order of the operators, so we can equivalently write the state \eqref{eq:RCMPSxDep} as
\begin{equation}\label{eq:RCMPSxDep2}
    \ket{{\bf Q},{\bf R}} = V_{+\infty}^\dagger \mathcal P  \left( e^{\int_{-\infty}^\infty dx\, {\bf Q}(x) \otimes \mathds{1} } e^{\int_{-\infty}^\infty dx\, {\bf R}(x)\otimes a^\dagger(x)} \right)  V_{-\infty}\ket0 \,.
\end{equation}
Expanding the second exponential gives
\begin{align}
    \ket{\mathbf Q,\mathbf R} &= \sum_{n=0}^\infty \int\limits_{-\infty \leq x_1\leq \cdots \leq x_n\leq \infty}\hspace{-1cm}dx_1\cdots dx_n \, V_{+\infty}^\dagger\mathcal P e^{\int_{x_n}^{\infty}dz \, \mathbf Q(z)}\mathbf R(x_n)\cdots \mathcal P e^{\int_{x_{1}}^{x_2}dz \, \mathbf Q(z)}\mathbf R(x_1)\mathcal P e^{\int_{-\infty}^{x_1}dz \, \mathbf Q(z)} V_{-\infty}\nonumber \\
    &{}\times a^\dagger(x_n)\cdots a^\dagger(x_1)\ket{0}\,.
\end{align}
To determine the overlap between two RCMPSs $\ket{\mathbf Q,\mathbf R}$ and $\ket{\mathbf Q',\mathbf R'}$, we can make use of the fact that whenever $x_1 < x_2 < \cdots < x_n$ and $y_1 < y_2< \cdots < y_n$, we have 
\begin{equation}\label{eq:simple_wick}
    \braket{0|a(y_1)\cdots a(y_n)a^\dagger(x_1)\cdots a^\dagger(x_n)|0} = \delta(x_1-y_1)\cdots\delta(x_n-y_n)\,.
\end{equation}
Using this identity, the overlap can be written as
\begin{align}
    \braket{\mathbf Q',\mathbf R'|\mathbf Q,\mathbf R} &= \sum_{n=0}^\infty \int\limits_{-\infty \leq x_1\leq \cdots \leq x_n\leq \infty}\hspace{-1cm}dx_1\cdots dx_n \, V_{+\infty}^\dagger\mathcal P e^{\int_{x_n}^{\infty}dz \, \mathbf Q(z)}\mathbf R(x_n)\cdots \mathcal P e^{\int_{x_{1}}^{x_2}dz \, \mathbf Q(z)}\mathbf R(x_1)\mathcal P e^{\int_{-\infty}^{x_1}dz \, \mathbf Q(z)} V_{-\infty}\nonumber \\
    &{}\times V_{+\infty}^{\prime T}\mathcal P e^{\int_{x_n}^{\infty}dz \, \bar{\mathbf Q}'(z)}\bar{\mathbf R}'(x_n)\cdots \mathcal P e^{\int_{x_{1}}^{x_2}dz \, \bar{\mathbf Q}'(z)}\bar{\mathbf R}'(x_1)\mathcal P e^{\int_{-\infty}^{x_1}dz \, \bar {\mathbf Q}'(z)} \bar V'_{-\infty}\,,
\end{align}
where bars denote complex conjugation. The product can be rearranged using the tensor product identities $(u_1^\dagger A_1v_1)(u_2^\dagger A_2v_2)= (u_1\otimes u_2)^\dagger (A_1\otimes A_2)(v_1\otimes v_2)$, $(A_1B_1)\otimes(A_2B_2) = (A_1\otimes A_2)(B_1\otimes B_2)$ and $\exp A\otimes \exp B = \exp (A\otimes \mathds 1+\mathds 1 \otimes B)$:
\begin{align}\label{eq:norm_ref}
    \braket{\mathbf Q',\mathbf R'|\mathbf Q,\mathbf R} &= \sum_{n=0}^\infty \int\limits_{-\infty \leq x_1\leq \cdots \leq x_n\leq \infty}\hspace{-1cm}dx_1\cdots dx_n \, (V_{+\infty}^\dagger\otimes V_{+\infty}^{\prime T})\mathcal P e^{\int_{x_n}^{\infty}dz \, [\mathbf Q(z)\otimes \mathds 1+\mathds 1 \otimes \bar{\mathbf Q}'(z)]}(\mathbf R(x_n)\otimes \bar{\mathbf R}'(x_n))\nonumber\\
    &\cdots \mathcal P e^{\int_{x_{1}}^{x_2}dz \, [\mathbf Q(z)\otimes \mathds 1+\mathds 1 \otimes \bar{\mathbf Q}'(z)]}(\mathbf R(x_1)\otimes \bar{\mathbf R}'(x_1))\mathcal P e^{\int_{-\infty}^{x_1}dz \, [\mathbf Q(z)\otimes \mathds 1+\mathds 1 \otimes \bar{\mathbf Q}'(z)]} (V_{-\infty}\otimes \bar v'_{-\infty})\,.
\end{align}
This expression can also be interpreted as a single path-ordered exponential given by \eqref{eq:norm_fin}.

\subsection{Free term}
\label{app:free_term}

We will now prove that the expectation value of the free Hamiltonian ${\cal H}_\text{free}(x)$ defined in \eqref{eq:opList} takes the form \eqref{eq:HfreeExp} when evaluated in a translationally invariant state.  Using the mode expansions \eqref{eq:phi,pi} for $\phi(x)$ and $\pi(x)$ in terms of $a_k,a_k^\dagger$, we find
\begin{equation} \label{eq:HfmExp}
 \begin{split}
    \langle \mathcal{H}_\text{free}(x) \rangle 
    &= \frac{1}{2}\int \frac{dk\,dk'}{(2 \pi)^2} \, \frac{-k k' + \mu^2 - \omega_k \omega_{k'}}{2\sqrt{\omega_k \omega_{k'}}}
       \left( \langle a_k a_{k'} \rangle \, e^{i(k+k')x} + \langle a_k^\dagger a_{k'}^\dagger \rangle\, e^{-i(k+k')x} \right) \\
    &{}+ \frac{1}{2}\int \frac{dk\,dk'}{(2 \pi)^2}\, \frac{k k' + \mu^2 + \omega_k \omega_{k'}}{2\sqrt{\omega_k \omega_{k'}}}
       \left( \langle a_{k'}^\dagger a_k \rangle \, e^{i(k-k')x} +  \langle a_{k}^\dagger a_{k'} \rangle\, e^{-i(k-k')x} \right)\,,
\end{split}
\end{equation}
where $\omega_k=\sqrt{k^2+\mu^2}$.  For translationally invariant states, the expression \eqref{eq:HfmExp} can be simplified using the fact that $\langle a_k a_{k'} \rangle$ and $\langle a_k^\dagger a_{k'}^\dagger \rangle$ are proportional to $\delta (k + k')$, and $\langle a_k^\dagger a_{k'} \rangle$ and $\langle a_k a_{k'}^\dagger \rangle$ are proportional to $\delta(k-k')$.  Using the restrictions of the delta functions the pre-factors can be written as
\begin{equation}\label{eq:deltacond}
\begin{aligned}
  \left. \frac{-k k' + \mu^2 - \omega_k \omega_{k'}}{2\sqrt{\omega_k \omega_{k'}}} e^{\pm i (k + k') x}\right|_{k + k'=0}
&= 0 \,,\\
  \left.\frac{k k' + \mu^2 + \omega_k \omega_{k'}}{2\sqrt{\omega_k \omega_{k'}}} e^{\pm i (k - k') x} \right|_{k - k'=0}
   &= \frac{(\mu + ik)(\mu - i k')}{\sqrt{\omega_k \omega_{k'}}}\,,
\end{aligned}
\end{equation}
Applying \eqref{eq:deltacond}, the first line of \eqref{eq:HfmExp} vanishes and the two terms in the second line contribute the same as they differ only by a relabeling of the integration variables which leads to
 \begin{equation}\label{eq:app_hfree}
   \langle \mathcal{H}_\text{free}(x) \rangle 
    = \int \frac{dk\,dk'}{(2 \pi)^2}\, \frac{(\mu + i k)(\mu - i k')}{\sqrt{\omega_k \omega_{k'}}}
        \langle a_{k'}^\dagger a_k \rangle \,.
 \end{equation}
As the delta function restrictions also hold for matrix elements between momentum eigenstates (with identical eigenvalue), the expression \eqref{eq:app_hfree} can also be used for the RQP matrix elements in Section \ref{sec:HeffFree}.

\section{ODEs for matrix element of \texorpdfstring{${\cal H}_\text{free}(0)$}{Hfree(0)}}\label{urulAppendix}
To determine the differential equations for $\bigl|u_{\alpha H, \beta H}^r(x))$ and $\bigl(u_{\alpha H, \beta H}^l(x)|$ needed in Section~\ref{sec:HeffFree}, we take derivatives of \eqref{eq:ODEur} and \eqref{eq:ODEul} with respect to $\alpha$ and $\beta$. Then
\begin{equation}\label{eq:urODE}
\begin{split}
  \frac{d\bigl|u^{r\,(1,1)}(x)\bigr)}{dx}
    &= \left(\mathbb T - ik\right)\bigl|u^{r\,(1,1)}(x)\bigr)
     + J(x)\left(\mu R + [R,Q]\right)\otimes\mathds1\,\bigl|u^{r\,(1,0)}(x)\bigr) \\
    &\quad + J(x)\,\mathds1\otimes\left(\mu\bar R + [\bar R,\bar Q]\right)\bigl|u^{r\,(0,1)}(x)\bigr) \\
    &\quad + \left(V\otimes\mathds1 + W\otimes\bar R\right)\bigl|\widetilde r^{(1,1)}(x)\bigr)
     + \mathcal W(x)\,\bigl|\widetilde r^{(1,0)}(x)\bigr) \,,\\[4pt]
  \frac{d\bigl|u^{r\,(1,0)}(x)\bigr)}{dx}
    &= \left(\mathbb T - ik\right)\bigl|u^{r\,(1,0)}(x)\bigr)
     + J(x)\,\mathds1\otimes\left(\mu\bar R + [\bar R,\bar Q]\right)\bigl|u^r(x)\bigr) \\
    &\quad + \left(V\otimes\mathds1 + W\otimes\bar R\right)\bigl|\widetilde r^{(1,0)}(x)\bigr) \,,\\[4pt]
  \frac{d\bigl|u^{r\,(0,1)}(x)\bigr)}{dx}
    &= \left(\mathbb T - ik\right)\bigl|u^{r\,(0,1)}(x)\bigr)
     + J(x)\left(\mu R + [R,Q]\right)\otimes\mathds1\,\bigl|u^r(x)\bigr) \\
    &\quad + \left(V\otimes\mathds1 + W\otimes\bar R\right)\bigl|\widetilde r^{(0,1)}(x)\bigr)
     + \mathcal W(x)\,\bigl|r(x)\bigr) \,,
\end{split}
\end{equation}
and
\begin{equation}\label{eq:ulSecondEqs}
\begin{split}
  \frac{d\bigl(u^{l\,(1,1)}(x)\bigr|}{dx}
    &= -\bigl(u^{l\,(1,1)}(x)\bigr|\left(\mathbb T + ik\right)
     - \bigl(u^{l\,(1,0)}(x)\bigr|\,J(x)\left(\mu R + [R,Q]\right)\otimes\mathds1 \\
    &\quad - \bigl(u^{l\,(0,1)}(x)\bigr|\,J(x)\,\mathds1\otimes\left(\mu\bar R + [\bar R,\bar Q]\right) \\
    &\quad - \bigl(\widetilde l^{(1,1)}(x)\bigr|\left(V\otimes\mathds1 + W\otimes\bar R\right)
     - \bigl(\widetilde l^{(1,0)}(x)\bigr|\,\mathcal W(x) \,,\\[4pt]
  \frac{d\bigl(u^{l\,(1,0)}(x)\bigr|}{dx}
    &= -\bigl(u^{l\,(1,0)}(x)\bigr|\left(\mathbb T + ik\right)
     - \bigl(u^l(x)\bigr|\,J(x)\,\mathds1\otimes\left(\mu\bar R + [\bar R,\bar Q]\right) \\
    &\quad - \bigl(\widetilde l^{(1,0)}(x)\bigr|\left(V\otimes\mathds1 + W\otimes\bar R\right) \,,\\[4pt]
  \frac{d\bigl(u^{l\,(0,1)}(x)\bigr|}{dx}
    &= -\bigl(u^{l\,(0,1)}(x)\bigr|\left(\mathbb T + ik\right)
     - \bigl(u^l(x)\bigr|\,J(x)\left(\mu R + [R,Q]\right)\otimes\mathds1 \\
    &\quad - \bigl(\widetilde l^{(0,1)}(x)\bigr|\left(V\otimes\mathds1 + W\otimes\bar R\right)
     - \bigl(\widetilde l(x)\bigr|\,\mathcal W(x) \,,
\end{split}
\end{equation}
with the definition
\begin{equation}\label{eq:W(x)def}
  \mathcal W(x)
    \;\equiv\; \frac{\partial}{\partial \beta}\Big[e^{F}U(x)e^{-F}\Big]_{\alpha=\beta=0}
    \;=\; J(x)\Bigl([R,V]\otimes\mathds1 \;+\; [R,W]\otimes\bar R\Bigr)
     \;+\; H(x)\,W\otimes\mathds1\,.
\end{equation}
The boundary conditions for these equations are $|u^{r\,(n,m)}(-\infty))=0$ and $(u^{l\,(n,m)}(+\infty)|=0$ for $(n,m)\neq(0,0)$ and correspond to \eqref{eq:u_BC} for $(n,m)=(0,0)$.
\begin{equation}
\begin{split}
    &\bigl|\widehat{r}^{(1,1)}(x) \bigr)=\bigl|u^{r\,(1,1)}(x)\bigr)-J(x)R \otimes \mathds1 \bigl|u^{r\,(1,0)}(x)\bigr)-J(x) \mathds1 \otimes \bar{R}\bigl|u^{r\,(0,1)}(x)\bigr)+J^2(x) R \otimes \bar{R} \bigl|u^{r}(-\infty)\bigr)\,, \\& \bigl|\widehat{r}^{(1,0)}(x) \bigr)=\bigl|u^{r\,(1,0)}(x)\bigr)-J(x) \mathds 1 \otimes \bar{R} \bigl|u^r (-\infty)\bigr)\,, \\&
    \bigl|\widehat{r}^{(0,1)}(x) \bigr)=\bigl|u^{r\,(0,1)}(x)\bigr)-J(x) R \otimes \mathds1 \bigl|u^r (-\infty)\bigr)\,,
\end{split}
\end{equation}
and
\begin{equation}
\begin{split}
    &\bigl(\widehat{l}^{(1,1)}(x) \bigr|=\bigl(u^{l\,(1,1)}(x)\bigr|+J(x)\bigl(u^{l\,(1,0)}(x)\bigr|R \otimes \mathds1 +J(x) \bigl(u^{l\,(0,1)}(x)\bigr| \mathds1 \otimes \bar{R}\,, \\& \bigl(\widehat{l}^{(1,0)}(x) \bigr|=\bigl(u^{l\,(1,0)}(x)\bigr|\,, \\&
    \bigl(\widehat{l}^{(0,1)}(x) \bigr|=\bigl(u^{l\,(0,1)}(x)\bigr|\,.
\end{split}
\end{equation}

\section{Numerical implementation}\label{app:num_impl}
In this appendix, we describe further implementation details for the numerical work presented in Section \ref{sec:application}. All numerical calculations are carried out in the left-gauge \eqref{eq:left_gauge} for the RCMPS ansatz, which enables us to eliminate the redundant components of $Q$ by writing \cite{Haegeman:2012uk}
\begin{equation}
    Q = -i K -\frac{1}{2}R^\dagger R\,,
\end{equation}
where $K$ is a Hermitian matrix.  For the RQP ansatz, we used the tangent left-gauge \eqref{eq:tangent_left}.

The vacuum state is determined by minimizing the expectation value of the Hamiltonian density, as computed in Section \ref{sec:expvalue}, over the space of RCMPSs. This is a non-linear optimization problem and requires one to determine the gradient of the energy density to efficiently solve it. The expression for the gradient of a given expectation value can be derived using the same methods as used for the RQP in Section~\ref{sec:GenFunsec} (see \cite{Tilloy:2021yre}). As an example, we can consider the gradient $M_b$ of the expectation value $\langle V_b \rangle$ for real $b$ defined in the tangent left-gauge as
\begin{equation}
    \delta \langle V_b \rangle = \tr(M_b W+M_b^\dagger W^\dagger)\,,
\end{equation}
which can be shown to take the form
\begin{equation}
    M_b=\int dx\, \left(-r_{bJ,bJ} l^\dagger_{bJ,bJ} R^\dagger +r_{bJ,bJ}R^\dagger l^\dagger_{bJ,bJ}+bJ(x)r_{bJ,bJ}l^\dagger_{bJ,bJ}\right)\,.
\end{equation}
Similar expressions can be derived for $\langle \mathcal{O}_n\rangle$ and $\langle \mathcal{H}_{\text{free}}\rangle$. To determine the direction of steepest descent one needs to define a metric on the tangent vectors. There are two natural options: the standard Euclidean metric $\tr(W^\dagger W)$ and the Hilbert space metric \eqref{eq:inner_simple}. In left-gauge ($l=\mathds 1$), these metrics give rise to the following steepest descent directions
\begin{equation}\label{eq:steps}
\begin{aligned}
    \text{Euclidean metric:} && W&=-M_b^\dagger\,,\\
    \text{Hilbert space metric:} && W&=-M_b^\dagger r^{-1}\,.
\end{aligned}
\end{equation}
The numerical minimization is performed using the limited-memory Broyden-Fletcher-Goldfarb-Shanno (LBFGS) algorithm, as implemented in \texttt{OptimKit.jl}, loosely following the implementation in \cite{Tang:2025spn}. In particular, we use the standard Euclidean metric for LBFGS and implement the non-trivial Hilbert space metric \eqref{eq:steps} as a preconditioner for the descent step. The matrix inversion $r^{-1}$ in the preconditioner can be ill-conditioned and is therefore regularized as $(r^2+\epsilon^2)^{-1/2}$, with the regulator $\epsilon$ proportional to the norm of the gradient. Furthermore, following \cite{Tang:2025spn}, we prepare the initial state by running LBFGS for~$\sim50-100$ iterations without the preconditioner.

Once the vacuum RCMPS is determined, we solve for the low-lying eigenvalues of the effective Hamiltonian \eqref{eq:eigval}. To compute the soliton spectrum, we need to determine two RCMPSs representing the two degenerate vacua between which the soliton interpolates. From a given vacuum, we can determine the other degenerate vacuum by explicitly acting with the spontaneously broken symmetry operators on the RCMPS ansatz (see Appendix~\ref{app:sym_op} for details). This procedure also fixes the momentum ambiguity described below \eqref{eq:rqp_sol}, because the left and right vacua are related, and therefore cannot be rephased independently. After the two degenerate vacua have been found, the soliton masses are again extracted by diagonalizing the effective Hamiltonian.

\section{Symmetries and RCMPSs}
\label{app:sym_op}
In this appendix, we will derive the action of symmetry transformations on the RCMPS ansatz relevant for the sine-Gordon and $\phi^4$ theories.

\subsection{\texorpdfstring{$\mathbb Z_2$}{ℤ₂} reflection symmetry}

In both sine-Gordon and $\phi^4$ theories, $\mathbb Z_2$ reflection is implemented by a unitary operator $U$ that acts on the operators $\phi,\pi$ as 
\begin{equation}
    U^\dagger\phi(x) U=-\phi(x)\,,\qquad U^\dagger\pi(x) U=-\pi(x)\,.
\end{equation}
The definition \eqref{eq:defa} implies that $U^\dagger a^\dagger(x) U=-a^\dagger(x)$, and so the action of $\mathbb Z_2$ reflection on an RCMPS is
\begin{equation}
\begin{aligned}
    U\ket{Q,R} &= v_L^\dagger U \mathcal P \exp \left[\int_{-\infty}^\infty dx\, Q\otimes \mathds{1}+R\otimes a^\dagger(x)\right]U^\dagger v_R U\ket0 = \ket{Q,-R}\,,
\end{aligned}
\end{equation}
where we used that the Fock vacuum is invariant under $\mathbb Z_2$ reflection $U\ket 0=\ket 0$.
It is easily seen that the $\mathbb Z_2$ reflection transformation
\begin{equation}\label{eq:parity_effective}
    (Q,R) \mapsto(Q,-R)
\end{equation}
preserves the left-gauge condition \eqref{eq:left_gauge}. Consequently, one does not have to regauge the RCMPS after applying \eqref{eq:parity_effective}.

\subsection{Shift symmetry}

The sine-Gordon theory \eqref{eq:sG_ham} is also invariant under discrete shifts of $\phi$.  The shift transformations by some amount $\alpha$ act as
\begin{equation}
    U^\dagger_\alpha \phi(x) U_\alpha =\phi(x) + \alpha\,,\qquad U^\dagger_\alpha \pi(x) U_\alpha =\pi(x)\,,
\end{equation}
and it is explicitly realized by
\begin{equation}\label{eq:shift_op}
    U_\alpha\equiv e^{-i\alpha \int dy \, \pi(y)} = e^{\Delta \int dy\, (a^\dagger(y)-a(y))}\,,
\end{equation}
where we defined $\Delta\equiv\alpha {\sqrt{\frac{\mu}{2}}}$ and in the second equality used \eqref{eq:phi,pi} and \eqref{eq:defa}. To compute~$U_\alpha \ket{Q,R}$, we need the action of $U_\alpha$ on $a^\dagger(x)$ and $\ket{0}$.
For the Fock vacuum, we use~\eqref{eq:shift_op} and split the exponent using  the Baker-Campbell-Hausdorff formula
\begin{equation}\label{eq:Uinterm}
 U_\alpha \ket{0}=e^{- \int dx \frac{\Delta^2}{2}}e^{ \Delta\int dx\, a^\dagger(x)}\ket{0}\,.
\end{equation}
For the operator $a^\dagger(x)$, we have the commutation relation $[\int dy\, (a^\dagger(y)-a(y)),a^\dagger(x)]=-1$ which implies that
\begin{equation}\label{eq:atransform}
    U^\dagger_\alpha a^\dagger(x) U_\alpha=a^\dagger(x)+\Delta\,,
\end{equation}
Putting together \eqref{eq:Uinterm} and \eqref{eq:atransform}, we deduce the action of the shift operator on the RCMPS ansatz
\begin{equation}
\begin{aligned}
    U_\alpha\ket{Q,R} = \ket{Q-\Delta R -\frac{\Delta^2}{2}\mathds 1, R+\Delta \mathds 1}\,.
\end{aligned}
\end{equation}
This transformation of $(Q,R)$ also preserves the left-gauge condition \eqref{eq:left_gauge}. 

\section{Effective quantum mechanics}\label{sec:EffQM}
In this appendix, we give further details on the matching of the amplitude \eqref{eq:amp1} and the effective non-relativistic Hamiltonian by considering $2\rightarrow2$ scattering of the elementary excitation. 

The scattering amplitude corresponding to the tree-level diagrams \eqref{eq:amp1} computed using the Feynman rules derived from \eqref{eq:phi3} is
\begin{equation}
	i\mathcal M_\text{tree}=  -24ig - 72ig\widetilde m^2\left(\frac{1}{s-\widetilde m^2}+\frac{1}{t-\widetilde m^2}+\frac{1}{u-\widetilde m^2}\right)\,,
\end{equation}
where $s,t,u$ are the Mandelstam variables. We will consider the scattering in the center-of-momentum frame, with incoming spatial momentum $p$ and outgoing spatial momentum $q$, with $p=\pm q$.  The Mandelstam variables are
\begin{equation}
    s= 4(\tilde m^2+p^2)\,,\qquad t=-2(p^2-pq)\,,\qquad u=-2(p^2+pq)\,.
\end{equation}
Expanding the amplitude for small momenta, we find the result~\eqref{eq:amp1} quoted in the main text:
\begin{equation}\label{eq:amp2}
	i\mathcal M_\text{tree}=  96ig + \mathcal O\left(\frac{p^2}{\widetilde m^2}\right)\, .
\end{equation}
We want to relate this amplitude to the amplitude computed in an effective quantum mechanics problem that governs the relative motion of the particles in the center-of-momentum frame. In particular, we want to match a Hamiltonian of the form
\begin{equation}\label{eq:HNR1}
    H_\text{NR} = \frac{p^2}{\widetilde m} + V(x)\,,
\end{equation}
where $\widetilde m/2$ is the reduced mass, and $V(x)$ is the two-particle interaction potential, which we need to determine. To deduce $V(x)$, we start from the relation between the S-matrix and the invariant amplitude~\cite{Peskin:1995ev}
\begin{equation}\label{eq:SMQFT}
    \bra{q_1,q_2}S-\mathds1\ket{p_1,p_2}_{\text{QFT}}=(2\pi)^2\delta(\omega_{q_1}+\omega_{q_2}-\omega_{p_1}-\omega_{p_2})\,\delta(q_1+q_2-p_1-p_2)\,i\mathcal{M}\,,
\end{equation}
with the relativistic normalization
\begin{equation}
    \braket{q_1,q_2|p_1,p_2}_\text{QFT} = (2\omega_{p_1})(2\omega_{p_2})(2\pi)^2\left[\delta(q_1-p_1)\delta(q_2-p_2)+\delta(q_1-p_2)\delta(q_2-p_1)\right]\,.
\end{equation}
Here, the two sets of delta functions are due to Bose symmetry. In the center-of-momentum frame, $p_1=-p_2=p$ and $q_1=-q_2=q$, so \eqref{eq:SMQFT} becomes
\begin{equation}\label{eq:QFTSmat}
    \bra{q,-q}S-\mathds1\ket{p,-p}_{\text{QFT}}=(2\pi)^2\delta(0)\,\delta(2\omega_q-2\omega_p)\,i\mathcal{M}\,.
\end{equation}
We compare the tree-level approximation of \eqref{eq:QFTSmat} with the lowest-order (Born approximation) S-matrix element of the effective quantum mechanics, obtained from the Lippmann-Schwinger equation \cite{Sakurai:2011zz},
\begin{equation}\label{eq:NRamp}
    \braket{q,-q|S-\mathds 1|p,-p}_\text{NR} = -i(2\pi)^2\delta(0)\,\delta(E_q-E_p)\left(\braket{q|V|p}+\braket{q|V|-p}\right),
\end{equation}
where $E_p=p^2/\widetilde m$ is the non-relativistic energy from \eqref{eq:HNR1}, and the second term is the exchange contribution arising from the Bose-symmetric normalization
\begin{equation}
    \braket{q_1,q_2|p_1,p_2}_\text{NR} = (2\pi)^2\left[\delta(q_1-p_1)\delta(q_2-p_2)+\delta(q_1-p_2)\delta(q_2-p_1)\right].
\end{equation}
The matching condition is therefore
\begin{equation}\label{eq:NR-R}
\begin{aligned}
	\frac{\braket{q,-q|S-\mathds 1|p,-p}_\text{QFT}}{4\omega_p^2}
	&=(2\pi)^2\delta(0)\,\delta(2\omega_q-2\omega_p)\,\frac{i\mathcal M_\text{tree}}{4\omega_p^2}\\
	&\overset{!}{=}-i(2\pi)^2\delta(0)\,\delta(E_q-E_p)\left(\braket{q|V|p}+\braket{q|V|-p}\right).
\end{aligned}
\end{equation}
For $p,q\ll\widetilde m$, we have $\omega_p\approx\widetilde m$ and $2\omega_q-2\omega_p\approx E_q-E_p$, and the delta functions cancel. With $\mathcal M_\text{tree}=96g$ as in \eqref{eq:amp2}, this gives
\begin{equation}
    \braket{q|V|p}+\braket{q|V|-p} = -\frac{24g}{\widetilde m^2}\,,\qquad V(x) = -\frac{12g}{\widetilde m^2}\,\delta(x)\,.
\end{equation}
For $g>0$, the potential is attractive and has a single bound state with energy $-\frac{36g^2}{\widetilde m^3}$. Moreover, its wavefunction is symmetric, as required by Bose symmetry.

\bibliographystyle{ssg}
\bibliography{refs}

\end{document}